\documentclass[11pt]{article}

\usepackage[utf8]{inputenc} 
\usepackage[T1]{fontenc} 
\usepackage{textcomp}
\usepackage{lmodern}
\usepackage{lastpage} 
\usepackage{siunitx} 
\usepackage{multirow}
\usepackage{textcomp}
\usepackage{pdflscape}
\usepackage{pgfplots}
\pgfplotsset{compat=1.12}
\usepackage[framed,numbered,autolinebreaks,useliterate]{mcode}
\usepackage{media9}
\usepackage{blindtext}
\usepackage[utf8]{inputenc}
\usepackage{amsmath}
\usepackage{amssymb}
\usepackage{float}
\usepackage{graphicx}
\usepackage{setspace}
\usepackage{verbatim}
\usepackage{parskip} 
\usepackage{acronym}
\usepackage{etoolbox}
\usepackage{pgfplotstable} 
\usepackage{float} 
\usepackage{xcolor}
\usepackage{multicol}
\usepackage{enumitem}
\usepackage{etoolbox}
\usepackage{graphicx}
\usepackage{comment}
\usepackage{lipsum}
\usepackage{lineno} 
\usepackage{algorithm}
\usepackage[noend]{algpseudocode}
\usepackage[margin=25mm]{geometry}
\usepackage{amsfonts}
\usepackage{verbatim}
\usepackage{subfigure}
\usepackage{multirow}
\usepackage[bookmarksnumbered=true]{hyperref} 
\hypersetup{
     colorlinks = true,
     linkcolor = blue,
     anchorcolor = blue,
     citecolor = blue,
     filecolor = blue,
     urlcolor = blue
     }

\newcommand{\ep}{\varepsilon}

\newcommand{\ord}{\mathcal{O}}

\newcommand{\der}[2] {\frac{\partial {#1} }{\partial {#2} } }

\newcommand {\red} {\color{red} }

\makeatletter
\newcommand*{\rom}[1]{\expandafter\@slowromancap\romannumeral #1@}
\makeatother
\numberwithin{equation}{section}

\providecommand{\keywords}[1]
{
  \small	
  \textbf{\textit{Keywords---}} #1
}

\title{Convergence analysis of a variational data assimilation scheme for bathymetry detection from surface wave observations}

\author{N. K.-R. Kevlahan$^{1}$ and R. A. Khan$^{1}$\footnote{corresponding author: ramsha.khan@math.mcmaster.ca}\\
        \small $^{1}$ Department of Mathematics and Statistics, McMaster University, Hamilton, ON, Canada\\
}

\date{}

\begin{document}
\maketitle

\begin{abstract}
Accurate mapping of ocean bathymetry is a multi-faceted process, needed for safe and efficient navigation on shipping routes and for predicting tsunami waves. Currently available bathymetry data does not always provide the resolution to capture dynamics of such nonlinear waves accurately. However collection of accurate mapping data is difficult, costly, and often a dangerous affair. As an alternative, in this study we implement a variational data assimilation scheme on the one-dimensional shallow water equations to improve estimates of bathymetry, using a finite set of observations of surface wave height to optimise predictions. We show necessary conditions on system parameters for convergence, and implement a low-pass filter for increased regularity of our reconstructed bathymetry. If our objective is to use this to model tsunami propagation, we observe that a relatively higher error in the optimal reconstruction of the bathymetry still yields a highly accurate prediction of the surface wave, suggesting low sensitivity of surface waves to small scale effects in the bathymetry. These conclusions are based on numerical experiments for both Gaussian and sandbar profile bathymetry, and with different observation operators. These extension of these results to realistic models can potentially have a significant impact, as  computational cost can be minimised through a priori knowledge of sufficient error tolerances needed for accurate tsunami prediction.
\end{abstract}\hspace{10pt}


\section{Introduction}

The processs of bserving ocean bathymetry using evolution of the surface waves is characterized by ill-posedness, often exhibiting sensitivity to small amounts of noise in the system, and susceptible to instability inherent in the inversion process \cite{ozisik}.  Data assimilation is one such inversion process, where observations of a true state are combined with a mathematical model in order to recover missing data governing the system evolution. Subsequently, the implementation of a variational adjoint-based scheme as presented in this study can be challenging, especially with the absence of analytical solutions of the governing nonlinear shallow water system  with non-zero bathymetry. Despite this, the effect of bathymetry is unarguably significant when it comes to predicting rogue waves or tsunamis, capable of effectively modifying wave speed as well as stability of the propagating wave \cite{craig_sulem_99}.

The high number of degrees of freedom in this problem makes it difficult to construct a fundamental set of criteria for optimal bathymetry reconstruction across the entire domain of system inputs; We must operate within the necessary condition of the shallow water system governing tsunami propagation, consider parameters such as bathymetry amplitude and form, we well as  that of the initial condition. This is in addition to calibration of the optimization scheme, and deriving optimal configurations of the observation network.  In this study we aim to quantify some key relationships between parameters in the system, the observation operator, as well as discussing the consequences of convergence error on the resulting surface wave, for example using the reconstructed bathymetry in predictive simulations for tsunami waves. Additionally we implement a Sobolev gradient smoothing technique (effectively a low-pass filter), within our optimization scheme and illustrate its efficacy in counteracting noise present in the bathymetry reconstruction. 

Section \ref{sec2} provides a review of efforts to date to map ocean bathymetry, highlighting empirical, numerical and theoretical approaches. In section \rom{3}, we provide a concise overview of the shallow water system and derivation of the first order adjoint data assimilation scheme using principles of optimal control theory, and a summary of the algorithm. Section \rom{4} gives preliminary result across different choices of initial condition and exact bathymetry for the data assimilation scheme and highlights the presence of noise in the optimal reconstruction. To counter this, section \rom{5} introduces the derivation of a low pass filter, which effectively removes higher frequencies in our reconstruction by increasing the regularity of our estimate at each iteration, taking it from the space $L^2(\mathbb{R})$ to $H^2(\mathbb{R})$. We provide results of the smoothed optimisation scheme and illustrate the removal of noise in the reconstructed bathymetry in multiple cases.  Section \rom{6} contains analyses of the relationships between the amplitudes of the initial condition and bathymetry relative to the average depth, and attempts to formulate a relationship summarising certain necessary conditions for convergence. We also analyse the number and placement of the observation points on the optimal reconstruction. Finally we provide a sensitivity analysis of the surface wave to errors in the bathymetry reconstruction. The key results and conclusions of this work are summarised in section \rom{7}, where we provide insights for further work, such as further sensitivity analyses and extension of the data assimilation scheme to a full two-dimensional model utilising actual observed data for ocean bathymetry from the global topography database ETOPO2\cite{etopo2}. 
	
\section{Review of bathymetry detection efforts}
\label{sec2}

To date, efforts to create an accurate map of oceanic bathymetry have been made by direct measurements, or using information of propagating surface waves. Direct measurement includes platforms like ship-based high frequency radars, however many of these methods are either too costly or have poor spatial resolution.  Often it is easier to measure waves propagating on the free surface, and use this information to create a map of the bottom topography from classical wave theory. This is the classical {\it inverse problem}.

A longstanding approach to solving this inverse problem uses the dispersion relation of surface waves. Earlier works such as Lubard et al.(1980) \cite{lubard_80} utilised  measurements of the frequency-wavenumber spectrum made via optical images, obtained using cameras mounted on an oceanographic research tower. Since then, various methods using dispersion relations have been investigated, where bathymetry is measured by fitting the theoretical dispersion relation for gravity waves, where depth is a system parameter, and derived using inversion formulas.

More recent works include Dugan (1997) \cite{dugan_97}, extended by Williams and Dugan (2002) \cite{dugan_02}, where image sequences of shoaling ocean waves taken from an aircraft are used to retrieve maps of water depth via the linear dispersion relation. The accuracy of this method were found to within 5\% if waves are reasonably linear. however, Grilli (1998)\cite{grilli_98} builds on the research conducted by Dugan et al. \cite{dugan_97}, arguing that the latter is limited due to neglect of amplitude dispersion effects, which accumulate through increasing nonlinearity as waves approach breaking in shallow water. He compares the linear frequency dispersion to a third order polynomial relationship between wave speed $c$ as a function of wavenumber $k$ and depth $h$, showing that due to amplitude dispersion effects, linear wave theory may greatly under estimate $c$, and lead to poorer estimates of bathymetry inversion formulas based on a linear dispersion relation.

These inversion algorithms are calibrated based on results of simulated periodic waves over mild slopes in a two-dimensional `numerical wave tank'. This FNPF (Fully Nonlinear based on Potential Flow). Numerical wave tank methodology was developed by Grilli and Subramanya (1996) \cite{Grill_Sub_96} with wave generation and absorption methods, to calculate speed and height variation for a number of shoaling waves over slopes ranging from $1:35$ to $1:70$, as shown in \cite{grilli_98}.

Tsai and Yue (1996) \cite{Tsai_Yu_96} also demonstrated how FNPF numerical tanks allow calculation of ``numerically exact" properties of shoaling wave sup to breaking point, and can provide accurate representation of surface waves independent of nonlinearity parameters.

While these works are based on an empirical formula for the nonlinear inversion problem, Nichols and Taber (2009) \cite{Nichols_09} derive an inversion formula for bathymetry analytically, using the nonlinearity of the governing equations to detect bathymetry information. The governing equations for the surface wave are expressed as a Hamiltonian system, and a Dirichlet-to-Neumann Operator (DNO) is applied to the system in order to remove some implicit dependencies. The result is a single equation of the wave height at the surface in terms of the bathymetry, and subsequently an inversion is derived. However because of the linear order of their inversion formula, they are required to assume a small amplitude for the bathymetry.

Jang et al. (2010) \cite{JANG2010146} take a similar approach to the related problem of measuring a sudden shift in the sea floor, potentially due to seismic activity, using measurements of the surface waves. Their inversion formula however, is based on the same approach as Dugan\cite{dugan_97}; using the linear dispersion relation with bathymetry as a parameter, and using transforms to show that the problem becomes one of solving an integral equation involving the known surface wave data. The uniqueness of this solution is demonstrated, and analysis concludes that there is a lack of stability in the measurement of the bottom displacement, and a  question as to whether it depends continuously on the wave elevation. They overcome this using  regularisation methods iteratively as a stabilisation technique, and show numerical convergence to the integral solution.

Each of the aforementioned methodologies has its strengths and weaknesses. Dugan \cite{dugan_02} and Grilli \cite{grilli_98} both discuss practical measurement techniques of surface waves, whereas  theoretical approaches such as Nicholls \& Taber \cite{Nichols_09}, and Jang et al. \cite{JANG2010146} show high degree of convergence, but assume full knowledge of the surface wave, and do not address the complexities involved in obtaining accurate measurements in a real world scenario. Future considerations aim to research an affective way of extracting wave parcels from full observational data. In summary, while the theoretical results from such models are promising, their applicability to real-world measurements has not been established. In contrast, in this study we assume that we have a finite set of observations, and so the question of optimal placement of sensors is a significant one, and not considered in detail in reviewed works. 

Another issue is the difference between inversion formulas derived for linear and nonlinear systems. Grilli, and Nicholls \& Taber give results accounting for nonlinearity and resulting dispersion effects, whereas Dugan, and Jang et al. are restricted to the linear dispersion relations for gravity waves. However, there remains the question of whether the empirical formulation of the inversion as derived by Grilli is as rigorous as the analytically derived solution of Nicholls \& Taber. Both assume periodic waves, but the practical limitations on the accuracy of the free surface data make it difficult to asses the relative efficacy of these two methods. Ultimately true comparison will depend on research which systematically compares inversion formulas based on the nonlinear governing equations for free surface wave propagation, as well as addresses the practical issues of collecting realistic field wave data.

In the present study, we approach the bathymetry estimation problem from a variational data assimilation perspective, with the goal of formulating an algorithm for the full nonlinear shallow water system that predicts bathymetry from a small set of observation.  The conclusions of the relatively simple case could indeed prove a significant foundation to extending these methodologies to real world settings, and accurate tsunami prediction.

\section{Derivation of adjoint based data assimilation scheme}
\label{sec3}

The shallow water equations are a coupled system of equations for non-dispersive travelling waves. They are derived from the two-dimensional Euler quations, under the assumption  that the wavelength $\lambda$ is much larger than the total ocean depth $h$ from the sea floor, allowing us to average over the vertical dimension, and our fluid column height becomes $h = H + \eta(x,t) - \beta(x)$, where $H$ is the average depth, $\eta$ is the perturbation of the fluid surface, and $\beta$ is time-independent sea floor perturbation from zero. Eliminating the vertical dimension gives us the irrotational, incompressible  one-dimensional shallow water equations,

\begin{table}
\centering
\begin{tabular}{l l}
\hline
  & \\ [-1em]
Symbol & Definition \\
  & \\ [-1em]
\hline
   & \\ [-1em]
 $\eta(x,t)$    & General solution for the height perturbation   \\
   & \\ [-1em]
$\phi(x) $ &    General initial condition, i.e., $\phi(x) := \eta(x,0)$       \\
   & \\ [-1em]
$\hat{\eta} $ &    Amplitude of the initial condition $\phi(x)$       \\
   & \\ [-1em]
  $\eta^{(t)}(x,t)$   &  True solution for the height perturbation $\eta(x,t)$         \\
   & \\ [-1em]
 $\beta^{(t)}(x) $    &  True bathymetry        \\
   & \\ [-1em]
$\hat{\beta} $ &    Amplitude of the true bathymetry $\beta^{(t)}(x) $      \\
   & \\ [-1em]
 $\beta^{(g)}(x) $ & Starting guess for bathymetry  \\
   & \\ [-1em]
   $\beta^{(n)}(x)$ & Approximate bathymetry at iteration n of the assimilation algorithm, i.e., $\beta^{(0)}:= \beta^{(g)}$\\
   & \\ [-1em]  
  $\beta^{(b)}(x)$ & Best approximation to the bathymetry (e.g., fixed point of iterations) \\ 
   & \\ [-1em] 
$y^{(o)}(t)$ & Observations of the true height perturbation at positions $\{ x_j\}, j=1,...,N_{obs}$ \\
   & \\ [-1em] 
 $\eta^{(f)}(x,t)$  & Approximate (``forecast'') solution generated by approximate bathymetry \\
   & \\ [-1em] 
 $\mathcal{J}^{(n)}$ & Cost function at iteration n \\
  & \\ [-1em] 
  $(\cdot)^{*}$ &  Adjoint \\
   & \\ [-1em] 
\hline
\end{tabular}
\caption{Notation used in the derivation of data assimilation scheme of the SWE to find the optimal bathymetry, using same format as given in \cite{khan_2019}.}
 \label{tab:symbols}
\end{table}
\begin{subequations}
\begin{align}
\frac{\partial \eta}{\partial t} + \frac{\partial }{\partial x} \Big((H + \eta - \beta) u \Big) = 0, \\
\frac{\partial u}{\partial t} + \frac{\partial }{\partial x} \Big( \frac{1}{2} u^2 + g \eta \Big) = 0  ,\\
\eta(x,0) = \ \phi (x) , \ \ u(x, 0) = \ 0. 
\end{align}
\label{swe_2d}
\end{subequations}
We assume that $\phi(x)$ is compactly supported, and that we have periodic boundary conditions on some domain $\Omega = \{x; x \in [-L,L]\}$. Our objective is to implement a variational data assimilation scheme constrained by \eqref{swe_2d} in order to estimate the bathymetry $\beta(x)$, We wish to derive an optimal estimate of the bottom topography using a finite number of observations of the surface wave perturbation, for all time $t$ in our temporal domain $[0, T]$. To simplify further, we normalise the system \eqref{swe_2d} by  the average height $H$ and gravitational acceleration $g$ such that the wave propogation speed $c = \sqrt{gH} = 1$.

We can quantify our objective as the PDE constrained minimisation of some cost function $\mathcal{J}$,
\begin{equation}
\mathcal{J}(\beta) = \frac {1}{2} \int_0^T  \sum_{i=1}^{M} \Big[\eta^{(f)}(x_j,t;\beta) - y_j^{(o)}(t) \Big]^2 \  dt.
\label{cost_func}
\end{equation}
where $y_j^{(o)}(t)$  are the observations of the true height perturbations taken at positions ${x_j}$, $j = 1,...,N_{obs}$,  and  $\eta^{(f)}(x_j,t;\beta)$ is the solution of our system at ${x}_j$ generated by the bathymetry $\beta$. We define the optimal bathymetry $\beta^{(b)}$ by
\begin{equation}
 \beta^{(b)} = \text{argmin}_{\beta \in L^2(\Omega)} \mathcal{J}(\beta)  
\end{equation}

This is equivalent to solving
\begin{equation}
    \nabla \mathcal{J}^{L^2}(\beta^{(b)}) = 0.
\end{equation}
As direct computation of this optimization problem is too computationally expensive, we utilise principles of control theory to formulate a dual adjoint system in terms of some Lagrange multipliers (also called adjoint variables), that allow us to find $\beta^{(b)}$ with more efficiency \cite{nichols94}.

The first variation of $\mathcal{J}$, given some arbitrary perturbation $\beta'$ of scale $\ep$ is given by the Gateaûx derivative,
\begin{align}
\mathcal{J}'(\beta;\beta') =\lim_{\ep \to 0} \frac{J(\beta+ \ep \beta')- J(\beta)}{\ep} .
\label{gat_der}
\end{align}
Expanding the perturbation to $\ord(\ep)$,  we can reformulate \eqref{gat_der} as
\begin{align}
\mathcal{J}'(\beta;\beta') = - \int_0^T \Big(\eta^{(f)}(x_j,t;\beta) - y^{(o)}(t) \Big) \eta'  \ dt ,
\label{gat_2}
\end{align}
where $(\eta', u')$ are the solutions of the perturbed system of \eqref{swe_2d} given $\beta'$, found by linearising about $(\eta, u)$ and extracting the $\ord(\ep)$ system.
As the Gateaux derivative is a directional derivative in the direction of the perturbation $\beta'$, we can express \eqref{gat_2} as the inner product between $\nabla \mathcal{J} $ and $\beta'$,
\begin{equation}
\mathcal{J}'(\beta;\beta') \ = \ \langle \nabla \mathcal{J}, \beta' \rangle_{{L^2}(\Omega)} \ = \ \int_{-L}^{L} \nabla^{L_2} \mathcal{J} \ \beta' \ dx.
\end{equation}
Then the  following forms of $\mathcal{J}(\beta;\beta') $ are equivalent,
\begin{equation}
\label{r_rep}
J'(\beta;\beta') \ = - \int_0^{T} \Big(\eta^{(f)}(x_j,t;\beta) - y^{(o)}(t) \Big) \eta' \ dt =\ \int_{-L}^{L} \nabla^{L^2} \mathcal{J} \ {\beta'} \ dt.  
\end{equation}
We form a Lagrangian of our linearised system for $(\eta',u')$ with some arbitrary adjoint variables $(\eta^*,u^*)$,

\begin{equation}
 \int_{0}^{T} \int_{-L}^L \eta^*(x,t) \Big[ \der {\eta'} t + \der {} x \big( (\eta' - \beta')u + (\eta+1 - \beta) u' \big) \Big] + u^*(x,t) \Big[ \der {u'} t + \der{} x \big(\eta' + u u'\big) \Big] dx \ dt = \ 0.
 \label{dual}
\end{equation}
Integrating by parts in time and space reduces \eqref{dual} to 
\begin{align} 
0 =  &- \int_{0} ^{T}  \int_{0}^L  \eta' \Bigg \{   \der {\eta^*}{t} + u \der {\eta^*}{x} +  \der {u^*} x 
 \Bigg \}  +  u' \Bigg \{\der {u^*}{t} + (\eta + 1 - \beta)  \der {\eta^*}{x} +   u \der {u^*} x 
 \Bigg \} - \beta' u \der{\eta^*} x \ dxdt \nonumber \\
 &+  \int_{ 0} ^{T} \eta^* \big[ (\eta' - \beta')u + (\eta+1-\beta)u' \big] \Bigr|_{-L}^{L} dt  \ +  \int_{ 0} ^{T} u^* \big[ \eta' + uu'\big] \Bigr|_{-L}^{L} dt  \ \nonumber \\
 &+ \int_{-L} ^{L} \eta^*\eta' \Bigr|_{\substack{t=T}} dx \  - \int_{-L} ^{L} \eta^*\eta' \Bigr|_{\substack{t=0}} dx  \nonumber \\
&+ \int_{-L} ^{L} u^*u' \Bigr|_{\substack{t=T}} dx -  \int_{-L} ^{L} u^*u' \Bigr|_{\substack{t=0}} dx . \label{eq_var3}
\end{align}
Due to periodicity our boundary terms vanish, and if we pick $(\eta^*, u^*)$ as the solution to 
\begin{subequations}
\begin{align}
 \der {\eta^*} t + {u \der {\eta^*} x}  + \der {u^*} x = \ \big( \eta^{(f)}(x,t;\beta) - y^{(o)}(t) \big)  \delta(x - x_j)  &,  \\
\der {u^*} t + (1+{ \eta} - \beta) \der {\eta^*} x + u{ \der {u^*} x} = \ 0 &, \\
 \eta^*(x,T) = \  0&, \\
\label{adj_ic}
\ {u^*}(x,T) = \ 0 &,
\end{align}
\label{adj_eq}
\end{subequations}
Then \eqref{eq_var3} is reduced to 
 \begin{equation}
   \int_0^T \int_0^L \Big(\eta^{(f)}(x_j,t;\beta) - y^{(o)}(t) \Big)\eta ' \ dx dt = \int_0^T \int_0^L \beta' \der{u^*} x \ dx dt,
 \end{equation}
 
and \eqref{adj_eq} are called the adjoint equations. Combining this result with the equivalence given by  \eqref{r_rep}, we have

 \begin{equation}
\int_{-L}^{L}  \int_0^{T}   u \der{\eta^*} x \beta' \ dt \ dx  = \int_{-L}^{L}  \nabla^{L^2} \mathcal{J} \ \beta' \  dx,
\label{rrp}
 \end{equation}
and thus since our functional is linear and bounded, and belongs to the space of square-integrable functions, we can use the Riesz representation theorem to extract $\nabla^{L_2} \mathcal{J}$, giving us 
 \begin{equation}
\nabla^{L^2} \mathcal J = \int_0^{T}   u \der{\eta^*} x \ dt.
\label{grad_J}
\end{equation}
Losch and Wunsch (2003) \cite{losch_wunsch_03} also utilise a similar adjoint based minimisation for their bathymetry detection analysis, however they do not consider the infinite-dimensional case as we have here. The benefits of this approach is that our formulation is independent of any discretisation used in the numerical implementation of the data assimilation scheme.

To verify that our formulation for $\nabla^{L^2} \mathcal J$ is correct, we define 

\begin{equation}
    \kappa(\ep) = \lim_{\ep \to 0} \frac{1}{\ep} \frac{J(\beta+ \ep \beta')- J(\beta)}{\langle \nabla \mathcal{J}^{L^2}, \beta' \rangle_{{L^2}(\Omega)} },
    \label{kappa}
\end{equation}

where $\kappa(\ep)$ is the quotient of the two equivalent forms for  the variation $\mathcal{J}(\beta;\beta')$ we used in the above derivation. Given some $\beta'$, if we have correctly defined $\nabla \mathcal{J}^{L^2}$, then as $\ep \rightarrow 0$, we should see $\kappa(\ep) \rightarrow 1$. In the numerical implementation we define this as the kappa test.

We utilise an iterative steepest descent algorithm to find our minimiser $\beta^{(b)}$  yielding $\nabla \mathcal{J}^{L^2} = 0$, given some starting guess $\beta^{(g)}$. Using a line minimisation algorithm to find the optimal step size at each iteration, this can be summarised as

\begin{equation}
 \beta^{(n+1)} = \beta^{(n)} - \tau_n \nabla^{\mathcal{L}_2}\mathcal{J} \big( \beta ^{(n)} \big)   
\end{equation} 
where  
\begin{equation}
\tau_n =  \text{argmin}_{\tau \in \mathbb{R}} \ \mathcal{J}  \Big( \beta^{(n)}(x) - \tau \nabla^{\mathcal{L}_2}\mathcal{J}  \big(\beta^{(n)}(x) \big)  \Big).
\end{equation}

The optimal bathymetry reconstruction $\beta^{(b)}$ is the fixed point of this iterative scheme. The steps for the process are outlined in algorithm \ref{alg:bath1}.

\begin{algorithm}[H]
\caption{Data Assimilation Algorithm for Bathymetry Estimation}
\begin{algorithmic}[1]
\State Pick initial estimate for $\beta^{(g)}$.
\State Solve the initial value problem for $(u,\eta)$ from $t=0$ to $t=T$.
\State Solve adjoint problem for $(u^*,\eta^*)$ backwards in time from $t=T$ to $t=0$ to find $\eta^*(x,t)$.
\State Approximate $\int_0^T u\der{\eta^*} x \ dt$ at every point in  spatial domain $\Omega$.
\State Define $\nabla^{L^2} \mathcal{J} = \int_0^T u\der{\eta^*} x \ dt$.
\State Compute the optimal time step $\tau_n $ through a line minimisation algorithm. 
\State Use a gradient descent algorithmn to compute $\beta^{(n+1)}(x) = \beta^{(n)}(x) - \tau_n \nabla^{L^2}\mathcal{J} \big( \beta ^{(n)}(x) \big)$.
\State Repeat until $\parallel \nabla^{L^2}\mathcal{J} \parallel < \ep$ for some small $\ep$ ($\parallel \int_0^T u\der{\eta^*} x \ dt  \parallel \approx 0 $).
\State Set $\beta^{(b)}(x):= \beta^{(n)}(x)$.
\end{algorithmic}
\label{alg:bath1}
\end{algorithm}

\section{Initial results}
\label{sec4}

\begin{table}[h]
\centering
\begin{tabular}{ |c||c||c| }
 \hline
Case   &  Bathymetry & Initial Condition \\
         \hline
  \rom1  & Gaussian &  Gaussian  \\
  \rom2  & Gaussian &  Sinusoidal  \\
  \rom3  & Sandbar & Gaussian  \\
\hline
\end{tabular}
\caption { Cases considered for data assimilation algorithm \ref{alg:bath1}}
\label{table_case}
\end{table}

\begin{figure}[H]
\centering
\subfigure[Case \rom 1]{\includegraphics[width=0.26\textwidth]{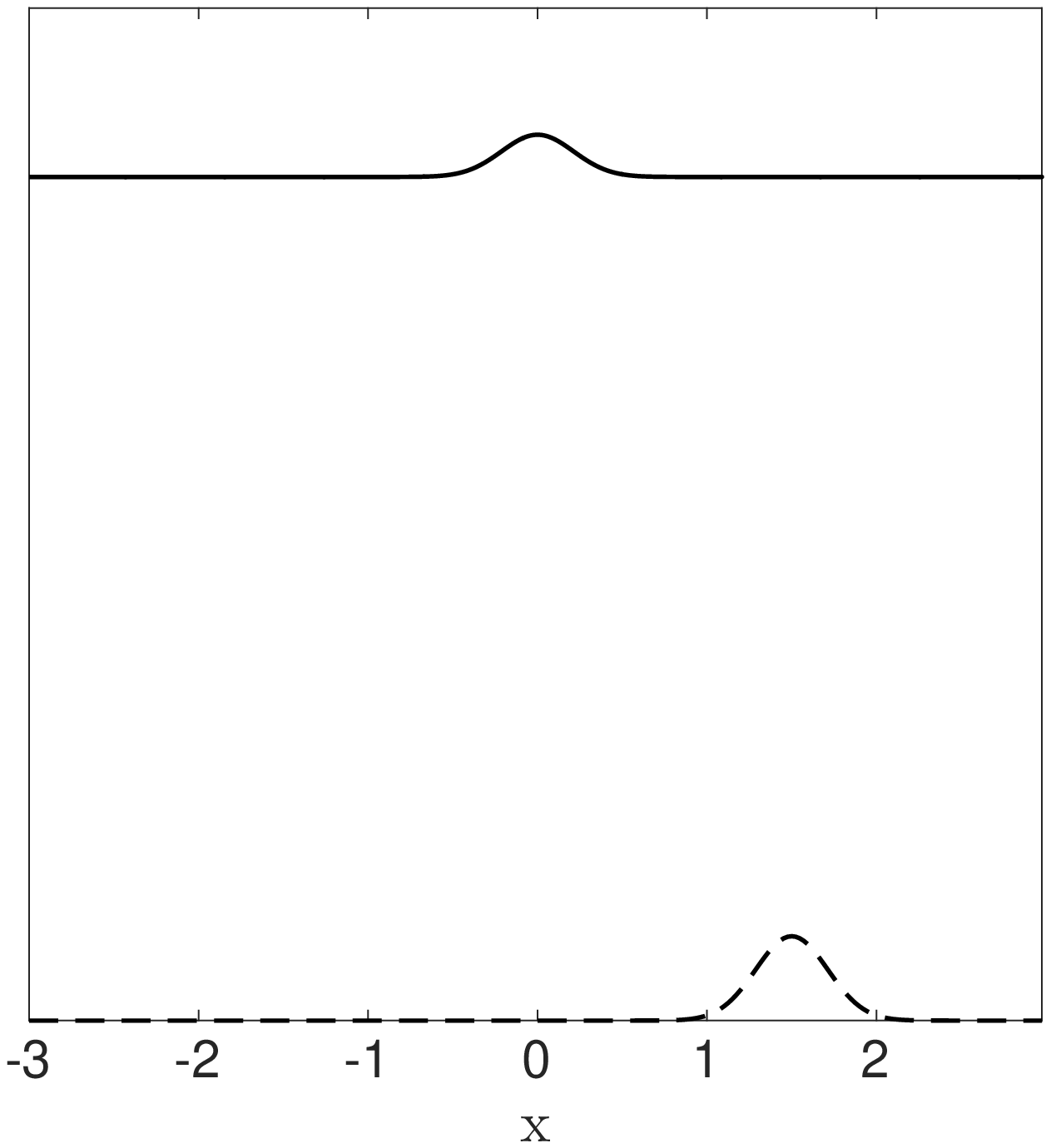}} \hspace{40pt}
\subfigure[Case \rom 2]{\includegraphics[width=0.26\textwidth]{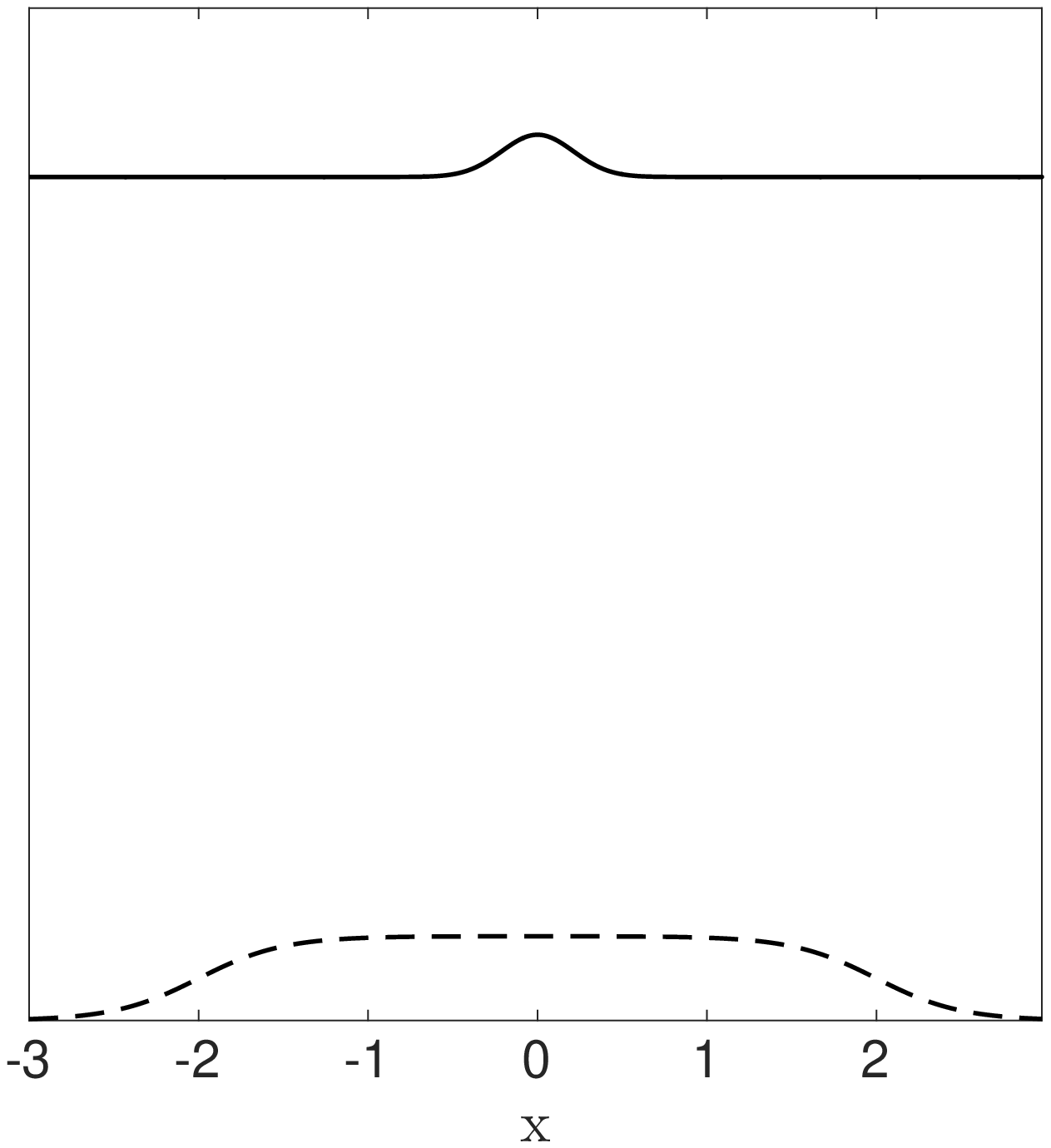}} \hspace{40pt}
\subfigure[Case \rom 3]{\includegraphics[width=0.26\textwidth]{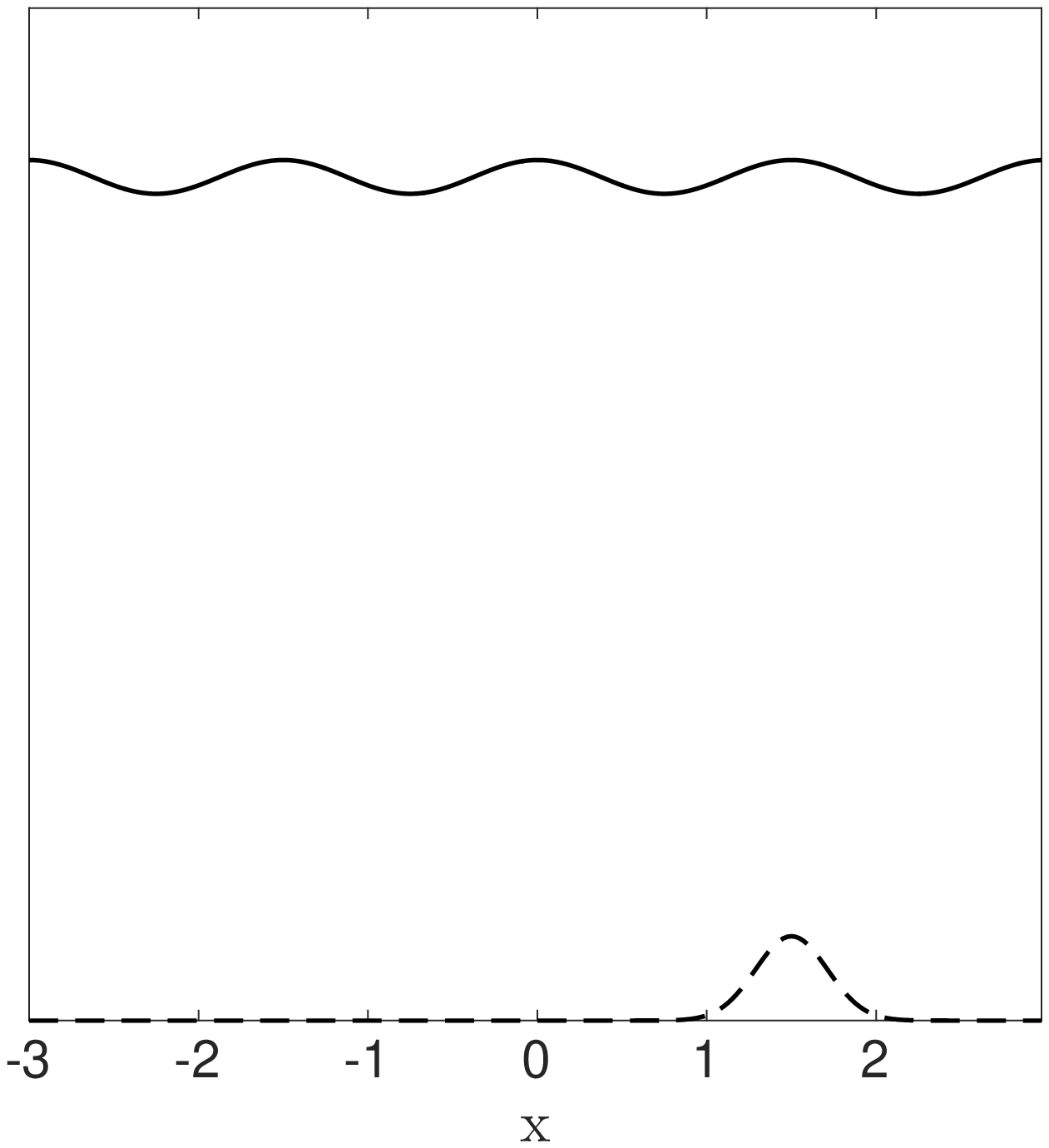}}
\caption{The three test cases for bathymetry $\beta(x)$ and surface perturbation initial condition $\eta(x,0)$ for the data assimilation scheme. Note that while the spatial distribution is correct, amplitude of the initial condition $\hat{\eta}$, amplitude of the bathymetry $\hat{\beta}$, and average depth $H$ are not to scale in these diagrams, as $\hat{\eta}$ was restricted to $1\%$ of $\hat{\beta}$ across most of the numerical analyses.}
\label{fig:cases}
\end{figure}

For the numerical implementation of algorithm \ref{alg:bath1}, we consider three different cases, comprising of different forms of the initial condition $\phi(x)$ and the true bathymetry $\beta^{(t)}(x)$. These cases are represented in Table \ref{table_case}, and a visual representation (not to scale) is given in fig \ref{fig:cases}. These cases were chosen to analyse convergence of the data assimilation scheme in scenarios where the support of $\phi(x)$ and the support of $\beta^{(t)}(x)$ overlap or are disjoint. Additionally we want to consider the effect of a surface wave where $\phi(x)$ is compactly supported, as in case \rom1 and case \rom2, or periodic across the entire domain $\Omega$. We restrict $\beta^{(t)}(x)$ to being compactly supported on some subdomain of $\Omega$, as the inverse problem of bathymetry detection is not well posed and subsequently unlikely to show convergence when the bathymetry is not compactly supported \cite{cobelli_petitjeans_maurel_pagneux_2018}. Ultimately the primary application of this study is intended to be tsunami propagation given some optimal reconstruction of missing bathymetry data, hence we are primarily interested in a non-periodic propagating surface wave, as in case \rom1 and case \rom2. However, including case \rom3 in our analysis should prove insightful on the effects of the observation operator and parameters in the system, on the optimal reconstruction.

We implement these schemes using a second order finite difference approximation in space, and a four stage third order Runge-Kutta scheme (as outlined in \cite{ruuth_spiteri}) in time. The resolution of our spatial grid is $N=512$, and given our domain $ \Omega = \{ x; -L \leq x \leq L \}$ we pick a final time $T < |L|$ such that with a propagation speed $c=1$, there are no boundary effects for $t \in [0, T]$. We integrate the system \eqref{swe_2d} on a staggered grid where $u(x,t)$ values are located at grid edges and $\beta(x)$ and $\eta(x,t)$ are located at grid centres. Periodic boundary conditions are imposed at $x=L$ and $x=-L$. We assume we have no background information for bathymetry a priori, and set $\beta^{(g)}(x) = 0$. 

\begin{figure}[H]
\centering
\subfigure[$|\kappa(\ep)-1|$]{\includegraphics[width=0.26\textwidth]{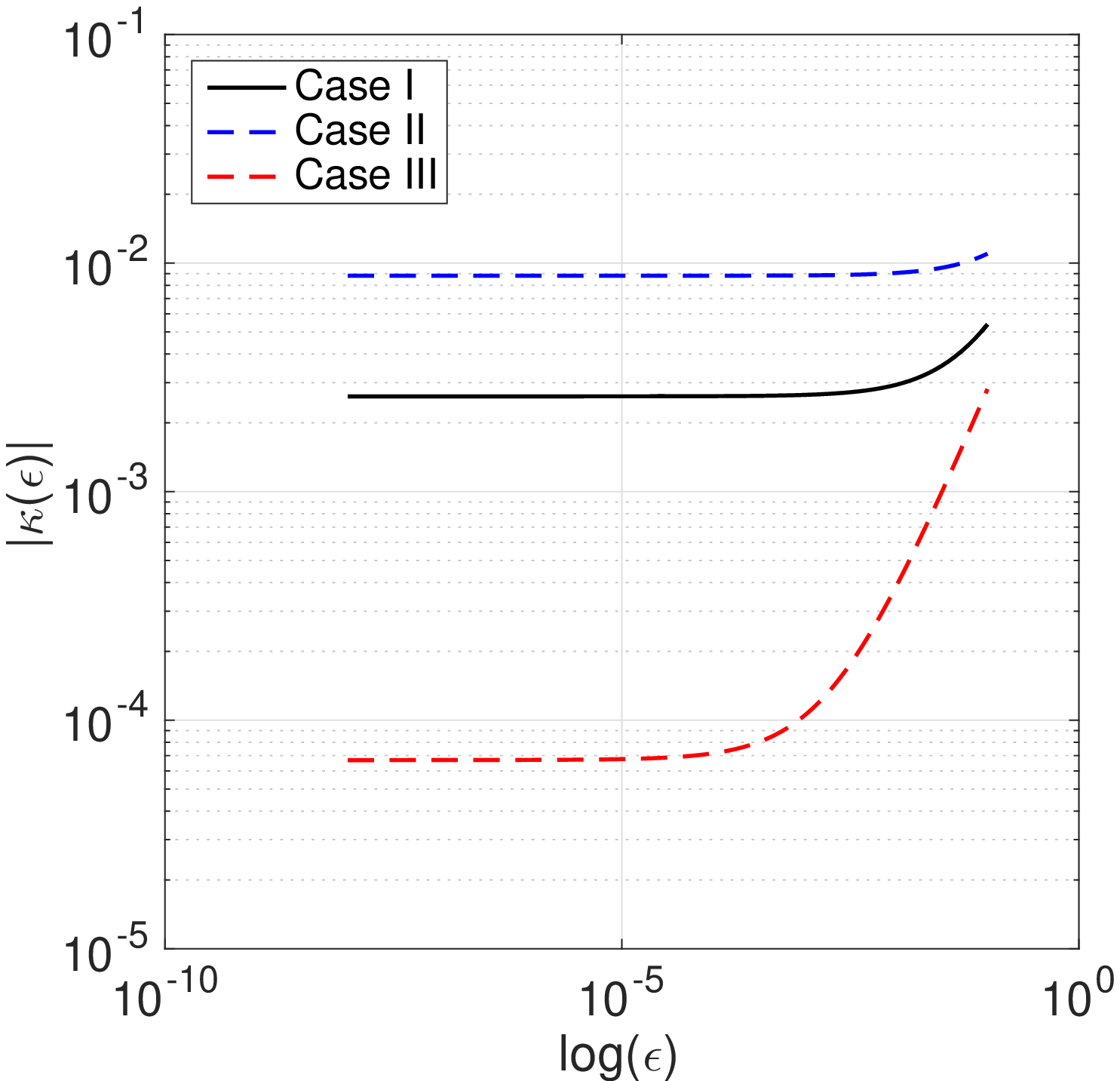}}\hspace{40pt} 
\subfigure[Relative reduction in $\mathcal{J}^{(n)}$]{\includegraphics[width=0.26\textwidth]{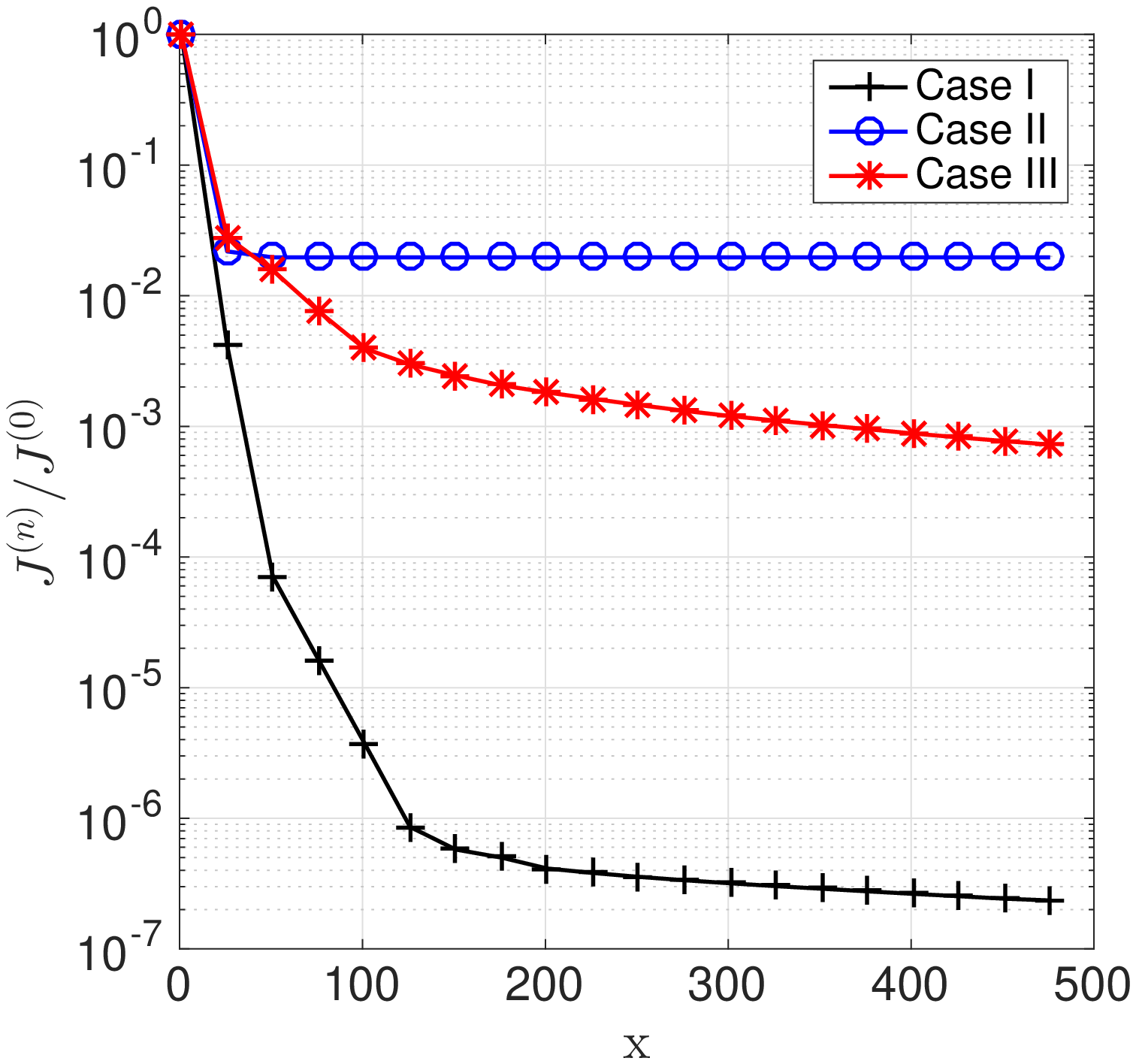}} \hspace{40pt}
\subfigure[$\parallel \beta^{(t)} - \beta ^{(n)} \parallel_{L^2}^2 $/$\parallel \beta^{(t)} \parallel_{L^2}^2$]{\includegraphics[width=0.26\textwidth]{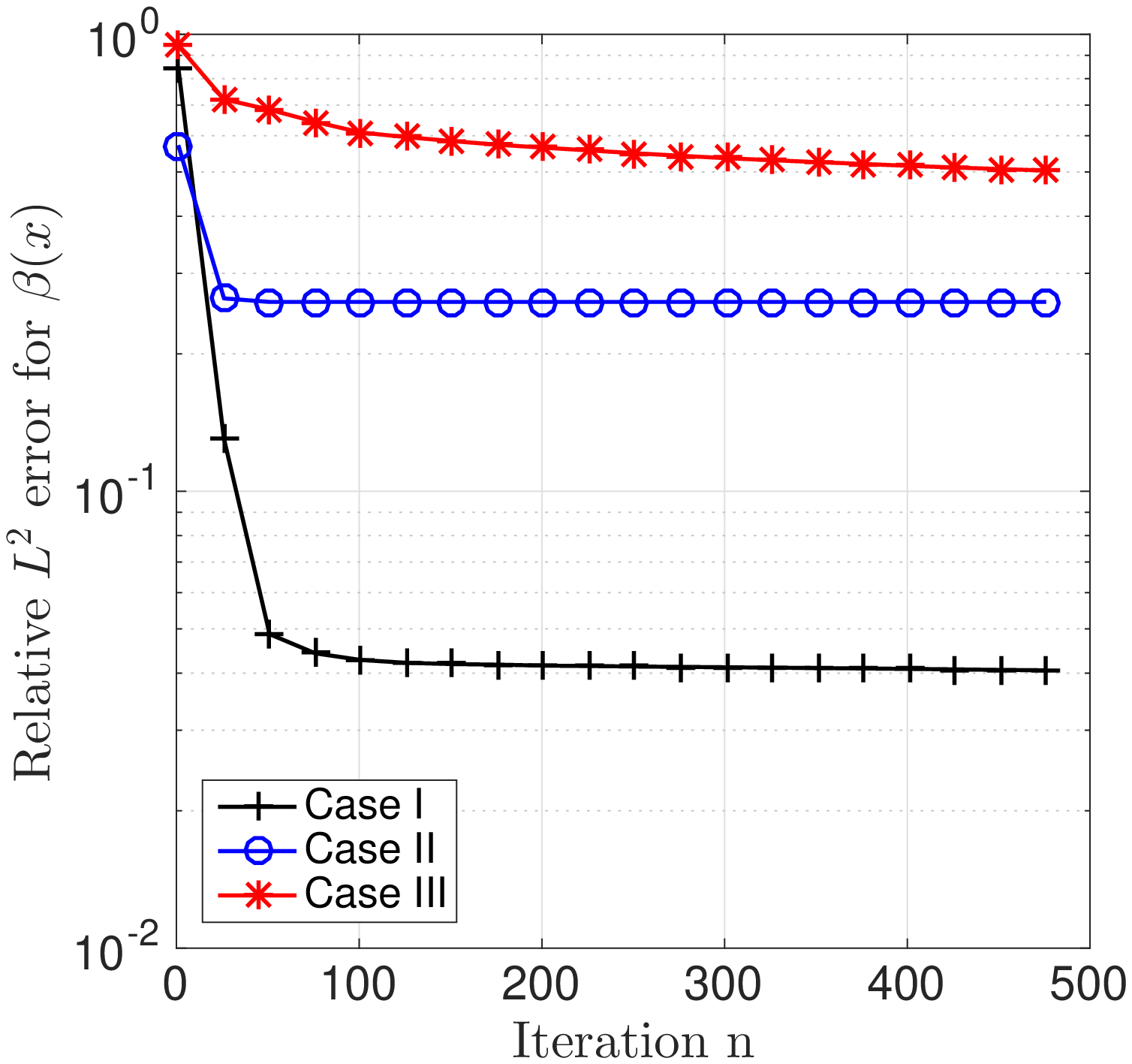}}

\subfigure[$\beta^{(g)}$ for Case \rom 1]{\includegraphics[width=0.26\textwidth]{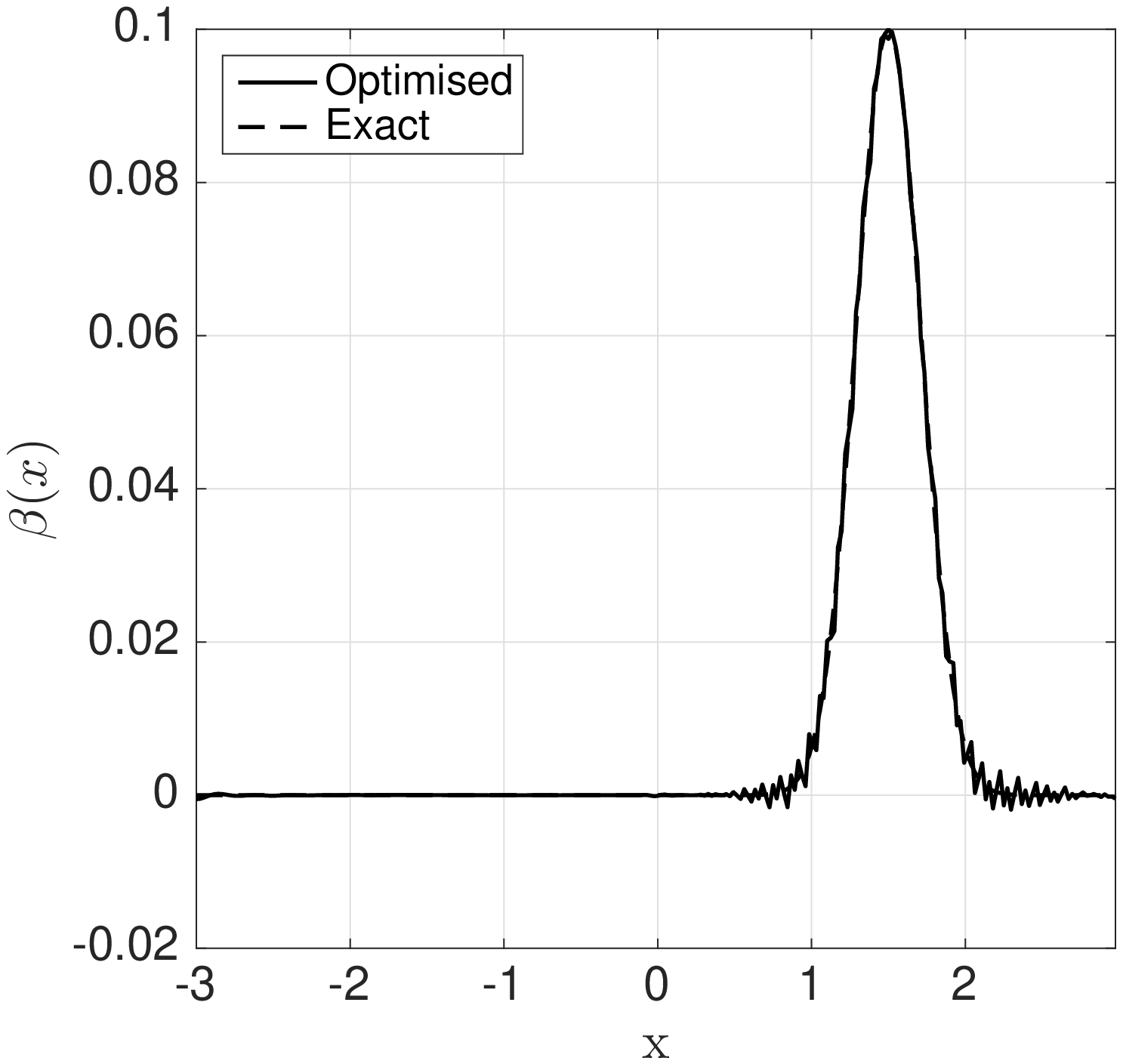}} \hspace{40pt}
\subfigure[$\beta^{(g)}$ for Case \rom 2]{\includegraphics[width=0.26\textwidth]{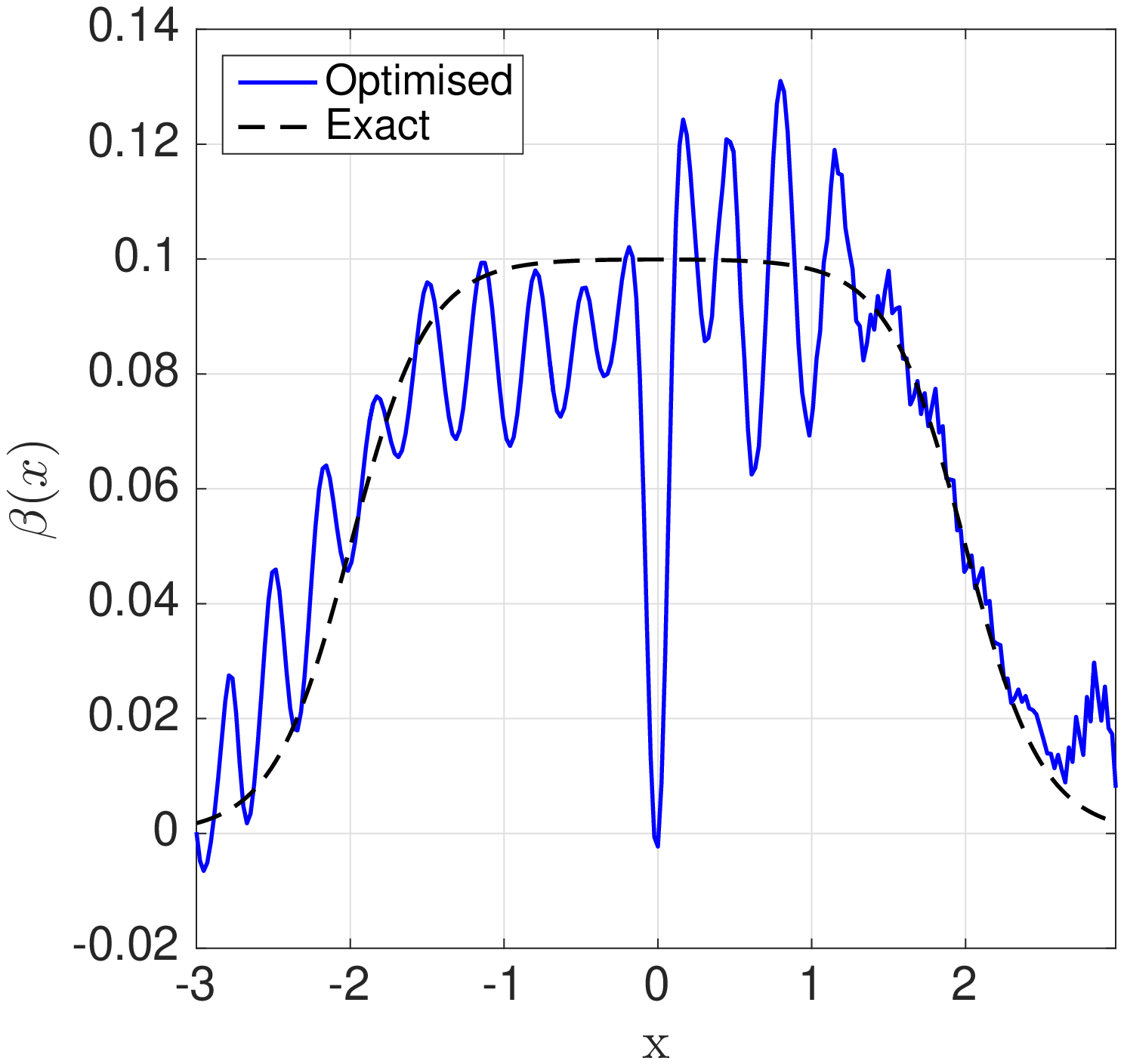}} \hspace{40pt}
\subfigure[$\beta^{(g)}$ for Case \rom 3]{\includegraphics[width=0.26\textwidth]{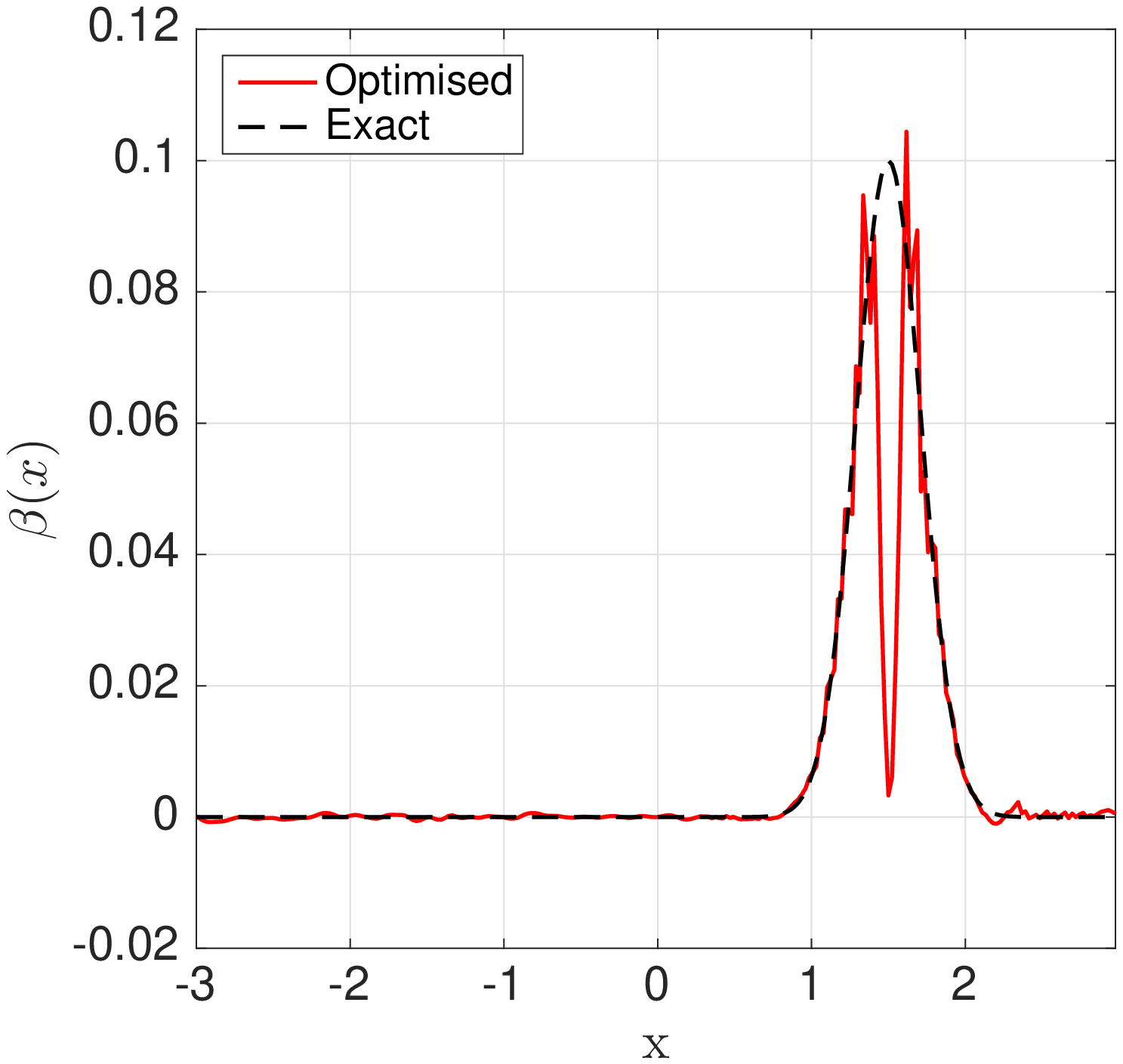}}
\caption{Results for iterative data assimilation scheme outlined in algorithm \ref{alg:bath1}}
\label{fig:no_filt}
\end{figure}

The results in fig \ref{fig:no_filt} illustrate the convergence of the data assimilation scheme for each case outlined in table \ref{table_case}, and the convergence of the kappa test \eqref{kappa}. Let us first consider the convergence of the cost function \eqref{cost_func} to zero; ultimately the explicit purpose of the optimization scheme was to minimise the error between the observations of the true height perturbation $y^{(o)}(t)$ and the approximated solution, $\eta^{(f)}$ given the optimal bathymetry $\beta^{(b)}$; accurate reconstruction of bathymetry was predicted to be a consequence of this, however due to the ill-posed nature of the problem, that may not necessarily be guaranteed. Indeed, the results highlight the difference between these two objectives. We see in fig \ref{fig:no_filt}(b), that the relative decrease in the cost function over $500$ iterations of algorithm \ref{alg:bath1} is greatest for case \rom1, at $\ord(10^{-6})$. From this alone, we can infer that the objective of the iterative scheme was successfully achieved. The relative decrease in $\mathcal{J}^{(n)}$ for case \rom2 and \rom3 does not go below $\ord(10^{-2})$ and $\ord(10^{-4})$ respectively, and to consider if this translates to a greater error in $\beta^{(b)}$, we consider figure \ref{fig:no_filt}(c). 

It is clear that for each case, convergence of the cost function to zero does not generate an equivalent convergence to the true bathymetry $\beta^{(t)}$. The relative error does not go below $\ord(10^{-2})$, even for case \rom1 where the reduction of the cost function was greatest. figures \ref{fig:no_filt}(d), (e) and (f) give the reconstructed bathymetry for each case, and it is immediately clear that the source of the large error is high-frequency noise in the reconstruction. for case \rom 1, we see the peak of the Gaussian is resolved, however there is noise present at the tails. However for case \rom3 where we assume a periodic initial condition $\phi(x)$, the noise is amplified even at the peak, and we deduce that the observability of the bathymetry by sensors measuring a sinusoidal propagating surface wave is significantly lower than that of a travelling Gaussian wavefront, as in case \rom1.

For case \rom2, it is interesting to note that the noise is of a larger scale on the sloping bathymetry for $x<0$ than for $x>0$. As the placement of the observation points is on the right hand side of the Gaussian initial condition $\phi$, this could suggest a lack of observability  of bathymetry at some point $x_0$ by a set of measurements taken at $\{x_j\}$, $j = 1, ..., N_{obs}$, where $x_0 + \delta <x_j$ for some distance $\delta >0$. We give a deeper analysis of the placement of the observations and effects on observability in section \ref{sec6}. 

Considering the convergence of the kappa test in \ref{fig:no_filt}(a), we see that the error magnitude is not optimal, ideally we would like to see convergence to $\kappa(\ep) = 1$ with at most an error of $\ord(10^{-3})$. However, we see that convergence of the kappa test does not proportionally result in convergence to the exact bathymetry; we see in fig \ref{fig:no_filt}(a) that the highest convergence for the kappa test is for case \rom3, whereas case \rom3 also shows the biggest error in the bathymetry reconstruction in \ref{fig:no_filt}(c). 

Before we move onto analysis of the various parameters that are inputs for the optimisation scheme, further insight on whether the regularity of the optimised bathymetry can be improved is necessary. We find that an optimisation scheme where the gradient of the cost function \eqref{grad_J} exists in $L^2(\Omega)$ is not rigorous enough to obtain classical solutions to \eqref{swe_2d}, and thus in section \ref{sec5} we analytically derive a low pass filter by smoothing our gradient such that $\nabla \mathcal{J} \in H^{2}(\Omega)$ and provide results of the numerical implementation.

\section{Smoothing using Sobolev gradients}
\label{sec5}
When considering the system \eqref{swe_2d}, we note that the bathymetry is incorporated into the system via the $(\beta u)_x$ term. Thus if we require a classical solution to this system, then we require smoothness not just on $\beta$, but its derivative. Because of this, we require the gradient \eqref{grad_J} to be in the Sobolev space$H^2(\Omega)$, which imposes smoothness conditions on $ {\beta}_{x}$ as well as $\beta$. The following derivation of the corresponding Sobolev gradient is adapted from Matharu and Protas \cite{matharu2019optimal}.\\

$H^2(\Omega)$ is a Sobolev space equipped with the inner product 
\begin{align}
     \langle  v_1, v_2 \rangle_{{H^2}(\Omega)} 
  = &\ \langle  v_1, v_2 \rangle_{{L^2}(\Omega)}  + {l_1^2} \Big \langle  \der {v_1} s ,\der {v_2} s \Big \rangle_{{L^2}(\Omega)} + {l_2^4} \Big \langle  \der{{}^2 v_1} {s^2}, \der{{}^2 v_2} {s^2} \Big \rangle_{{L^2}(\Omega)}  \nonumber  \\ 
  =& \ \int_{s=a}^{b} \Big[ v_1 v_2  + {l_1^2} \der {v_1} s \der {v_2} s + {l_2^4}\der{{}^2 v_1} {s^2}\der{{}^2 v_2} {s^2} \Big] \ ds,
  \label{h2_ip}
\end{align}

where $v_1, v_2 \in {H^2}(\Omega)$, and $l_1,l_2 \in \mathbb{R}$ are the length scale parameters used to adjust the regularity. As long as $l_1,l_2$ are finite, by the Riesz representation theorem,
\begin{align}
  \ \mathcal{J}'(\beta;\beta')  =& \ \langle \nabla^{L^2} \mathcal{J}, \beta' \rangle_{{L^2}(\Omega)} \nonumber \\
 =&  \ \langle \nabla^{H^2} \mathcal{J}, \beta' \rangle_{{H^2}(\Omega)} \nonumber \\
 =& \ \langle \nabla^{H^2} \mathcal{J}, \beta' \rangle_{{L^2}(\Omega)} + {l_1^2} \Big \langle  \der{\nabla^{H^2} \mathcal{J}}{s} ,  \der {\beta'}{s} \Big \rangle_{{L^2}(\Omega)} + {l_2^4} \Big \langle  \der{{}^2\nabla^{H^2} \mathcal{J} }{s^2} ,  \der {{}^2 \beta'}{s^2} \Big \rangle_{{L^2}(\Omega)}. \label{h2_grad_ibp}
\end{align}

In order to extract the gradient as we did in \eqref{rrp}, we need to isolate the $\beta'$ term, and define an equivalent expression for $\nabla ^{L^2}\mathcal{J}$. And so we utilise integration by parts as before, on the second and third term in \eqref{h2_grad_ibp}. We impose periodic conditions in space to get rid of the resulting boundary terms, and subsequently we have
\begin{align}
   \ \langle \nabla^{L^2} \mathcal{J}, \beta' \rangle_{{L^2}(\Omega)}
 =&  \ \langle \nabla^{H^2} \mathcal{J}, \beta' \rangle_{{H^2}(\Omega)}\nonumber \\ 
  =& \ \int_{s=a}^{b} \Big[ \nabla^{H^2} \mathcal{J} - l_1^2 \der{{}^2\nabla^{H^2} \mathcal{J} }{s^2}  + l_2^4 \der{{}^4\nabla^{H^2} \mathcal{J} }{s^4}  \Big] \beta' \ ds.
\end{align}

Since this holds for every arbitrary $\beta'$, the process of smoothing the gradient from ${L^2}(\Omega)$ to ${H^2}(\Omega)$ is equivalent to solving the inhomogeneous boundary value problem 
\begin{subequations}
\begin{align}
 & \nabla^{L^2} \mathcal{J}(s) =  \nabla^{H^2} \mathcal{J}(s) - l_1^2 \der{{}^2\nabla^{H^2} \mathcal{J}(s) }{s^2}  + l_2^4 \der{{}^4\nabla^{H^2} \mathcal{J}(s) }{s^4}, \\
 & \der{{}^{(2i+1)}\nabla^{H^2} \mathcal{J}(s) }{s^{(2i+1)}  } \Bigr|_{\substack{s=a}} = \der{{}^{(i)}\nabla^{H^2} \mathcal{J}(s) }{s^{(i)}  } \Bigr|_{\substack{s=b}}, \ \ \ i = 0,1.
\end{align}
\label{ellip_bvp}
\end{subequations}

In Fourier space, solving \eqref{ellip_bvp} is simplified to
\begin{equation}
 \widehat{(\nabla^{H^2} \mathcal{J})}_k = \frac{1}{1 + l_1^2k^2 + l_2^4k^4} \widehat{(\nabla^{H^2} \mathcal{J})}_k.
 \label{lowpassf}
\end{equation}

This is effectively a low pass filter applied to the $L^2$ gradient. We can make this as aggressive as needed by ``tuning'' $l_1$ and $l_2$, and the case where $l_1 = l_2 = 0$ simply gives us back our original $L^2$ gradient. 

For our smoothed data assimilation algorithm, we set $l_1 = 0$ and calibrate $l_2$, as this gives us the desired regularity and reduces a further degree of freedom in the problem. We now consider the numerical implementation of this updated optimization scheme, summarised in algorithm \ref{alg:bath2}.

\begin{algorithm}[H]
\caption{Data Assimilation algorithm with low pass filter for bathymetry estimation}
\begin{algorithmic}[1]
\State Pick initial estimate for $\beta^{(g)}$.
\State Solve the initial value problem for $(u,\eta)$ from $t=0$ to $t=T$.
\State Solve adjoint problem for $(u^*,\eta^*)$ backwards in time from $t=T$ to $t=0$ to find $\eta^*(x,t)$.
\State Approximate $\int_0^T u\der{\eta^*} x \ dt$ at every point in  spatial domain $\Omega$.
\State Define $\nabla^{L^2} \mathcal{J} = \int_0^T u\der{\eta^*} x \ dt$.
\State Apply low pass filter \eqref{lowpassf} to $\nabla^{L^2} \mathcal{J}$ to get $\nabla^{H^2} \mathcal{J}$ 
\State Compute the optimal time step $\tau_n $ through a line minimisation algorithm. 
\State Use a gradient descent algorithmn to compute $\beta^{(n+1)}(x) = \beta^{(n)}(x) - \tau_n \nabla^{H^2}\mathcal{J} \big( \beta ^{(n)}(x) \big)$.
\State Repeat until $\parallel \nabla^{H^2}\mathcal{J} \parallel < \ep$ for some small $\ep$ ($\parallel \int_0^T u\der{\eta^*} x \ dt  \parallel \approx 0 $).
\State Set $\beta^{(b)}(x):= \beta^{(n)}(x)$.
\end{algorithmic}
\label{alg:bath2}
\end{algorithm}

\begin{figure}[H]
\centering
\subfigure[Case \rom 1]{\includegraphics[width=0.3\textwidth]{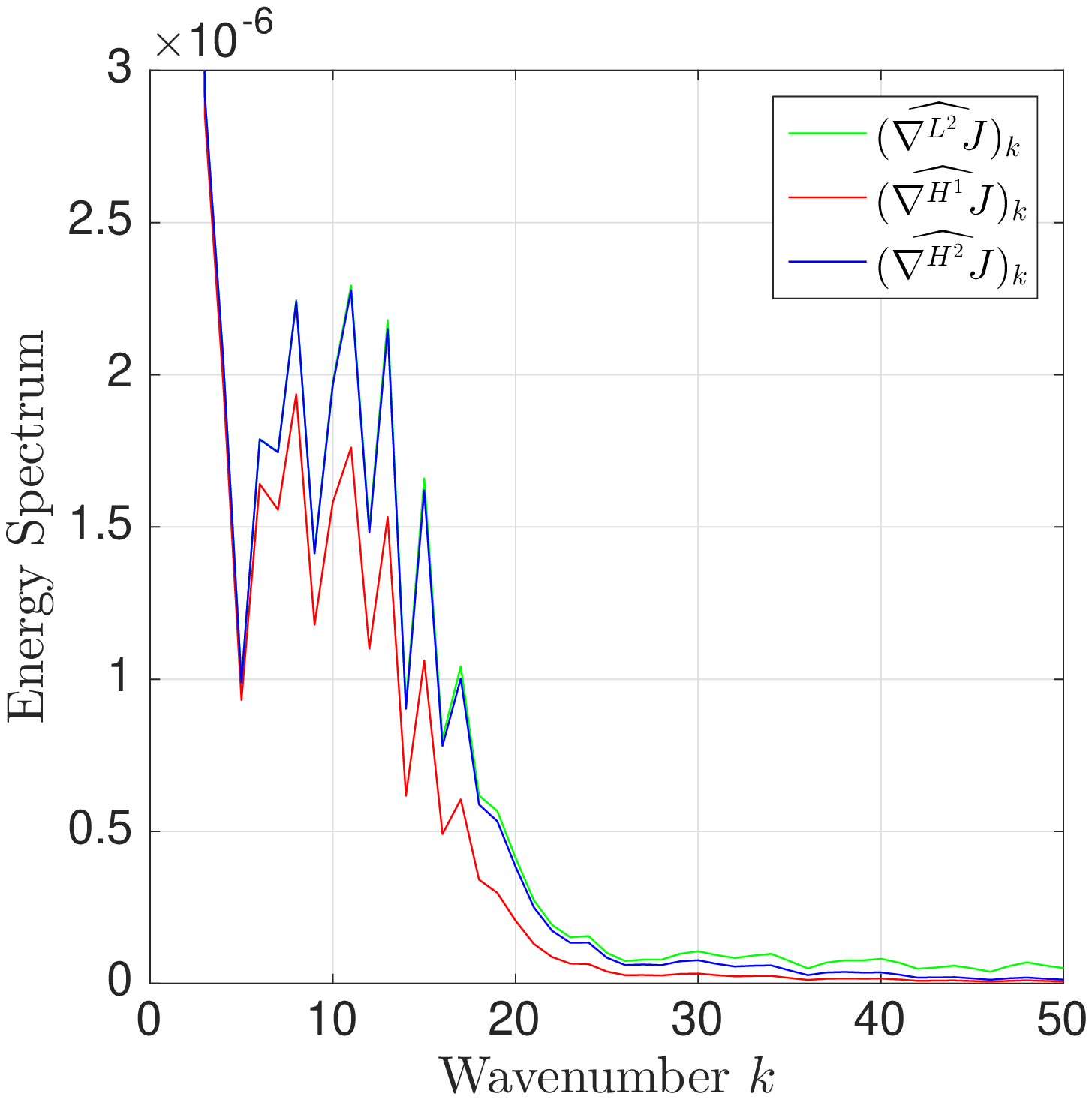}} \hspace{40pt}
\subfigure[Case \rom 2]{\includegraphics[width=0.3\textwidth]{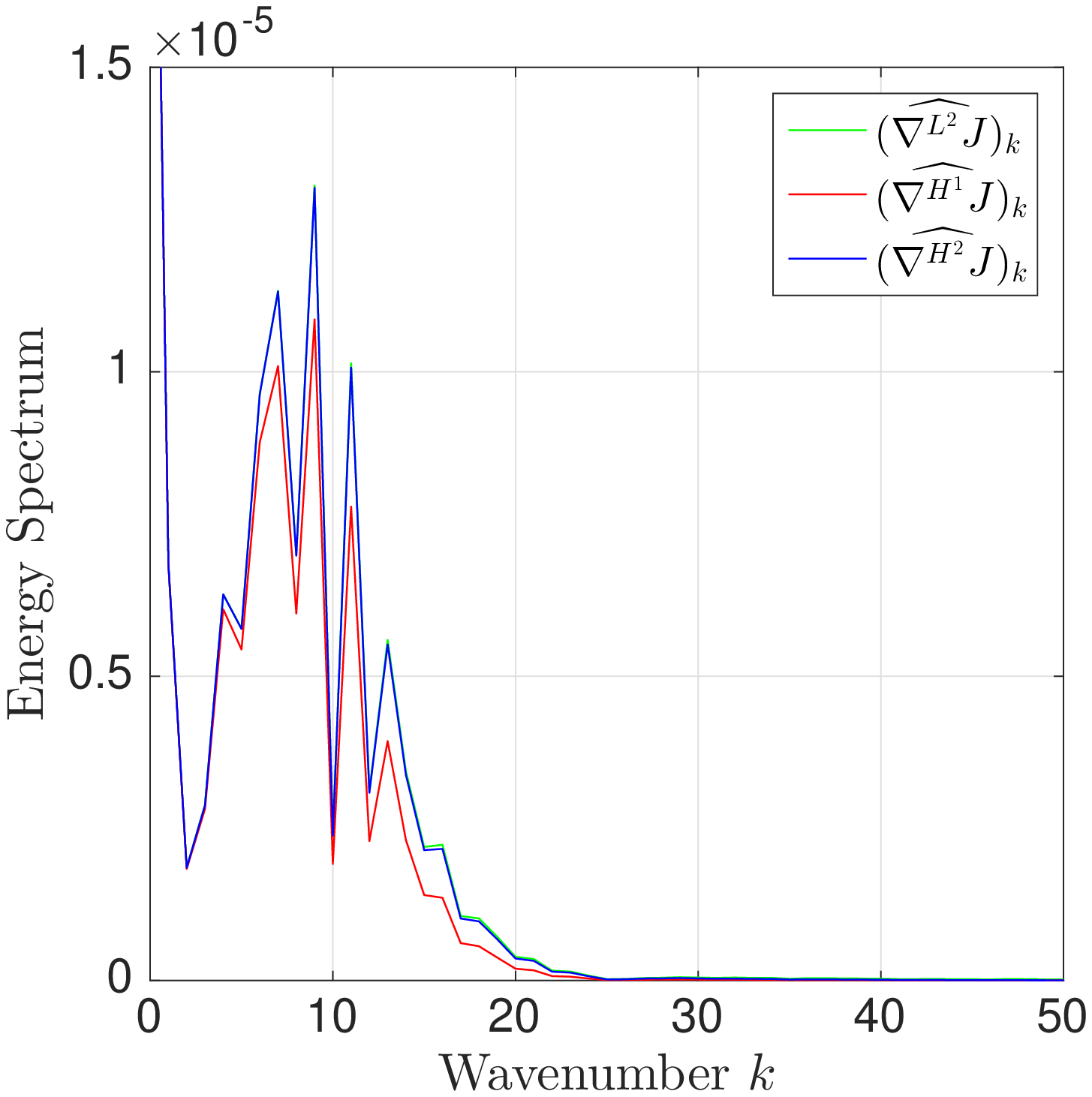}}

\subfigure[Case \rom 3]{\includegraphics[width=0.3\textwidth]{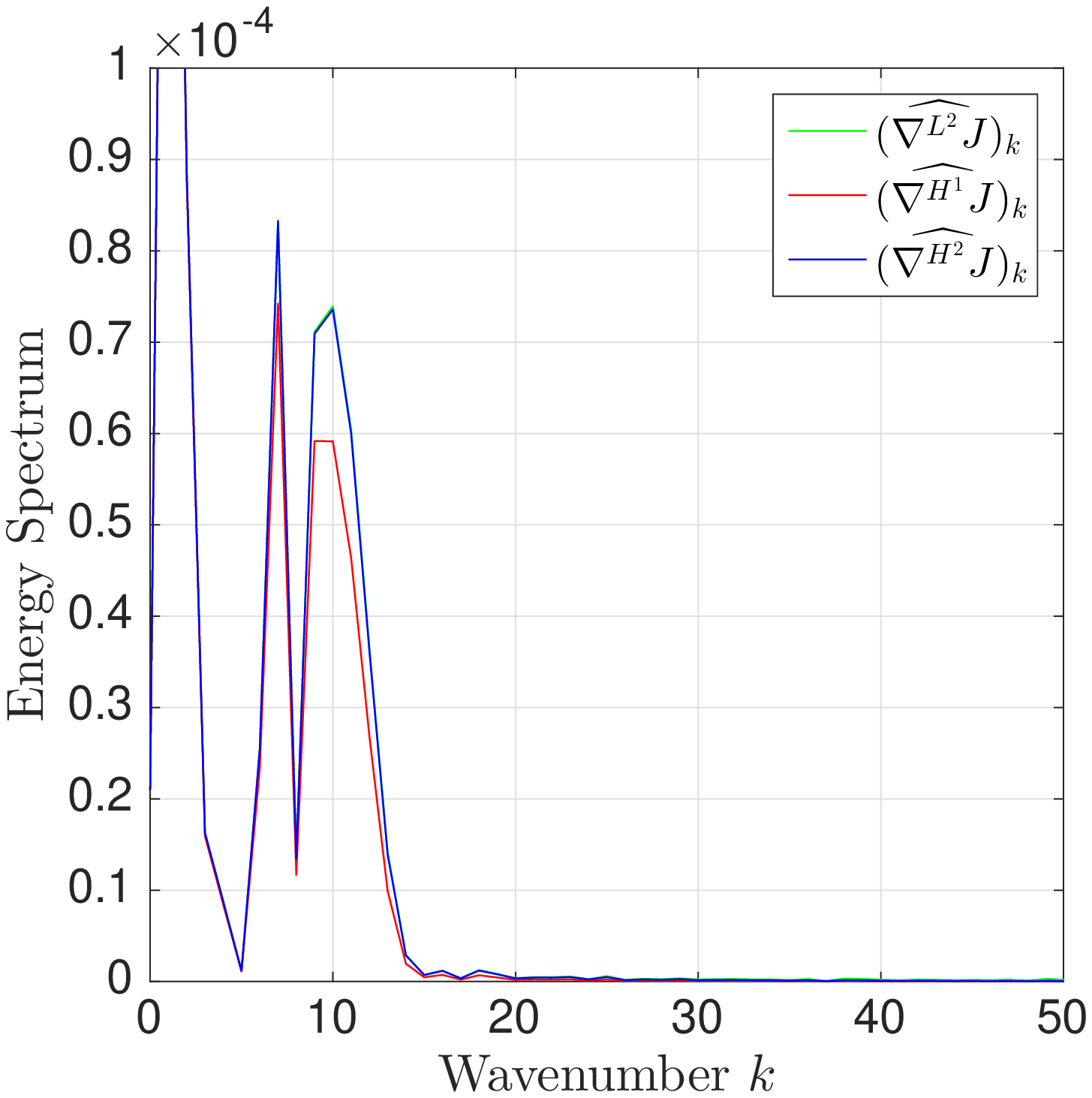}} \hspace{40pt}
\subfigure[$\nabla \mathcal{J}$ for Case \rom 3]{\includegraphics[width=0.25\textwidth]{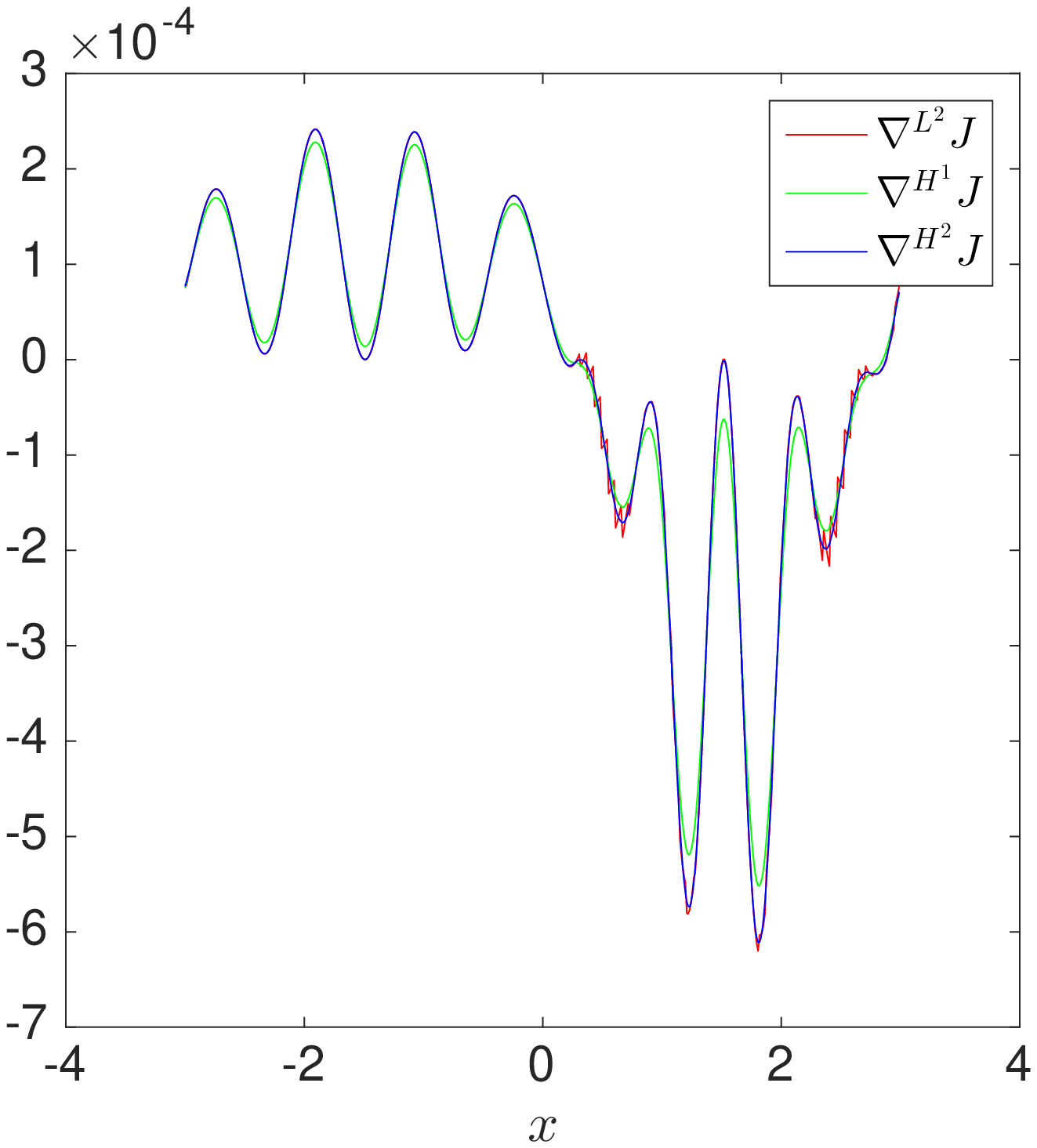}}
\caption{(a), (b) and (c) show the energy spectrum of the gradient of $\mathcal{J}(\beta)$, obtained after one iteration. (d) shows $\mathcal{J}(\beta)$ for case \rom3.}
\label{filt_h1_h2}
\end{figure}

Before we present the results of the data assimilation scheme, we illustrate the efficacy of the filtering in fig \ref{filt_h1_h2}. For each case \rom1, \rom2 and \rom3, we plot the energy spectrum of the gradient of $\mathcal{J}(\beta)$, obtained after one iteration. We compare the spectrum of gradient in $L^2$ with the spectra of the gradient in $H^1$, and $H^2$. The purpose of this comparison is to illustrate that $H^2$ smoothing is more appropriate than having $\nabla \mathcal{J}$ be in $H^1$, where the $H^1$ inner product is equivalent to the $H^2$ inner product \eqref{h2_ip} with $l_2 = 0$. In both cases, we try to choose optimal values of $l_1$ (for $H^1$) and $l_2$ (for $H^2$) to filter out  higher frequencies that contribute to the noise in the bathymetry reconstruction, without also getting rid of necessary information for observing the bathymetry propagated by the lower frequencies.

\begin{figure}[H]
\centering
\subfigure[]{\includegraphics[width=0.26\textwidth]{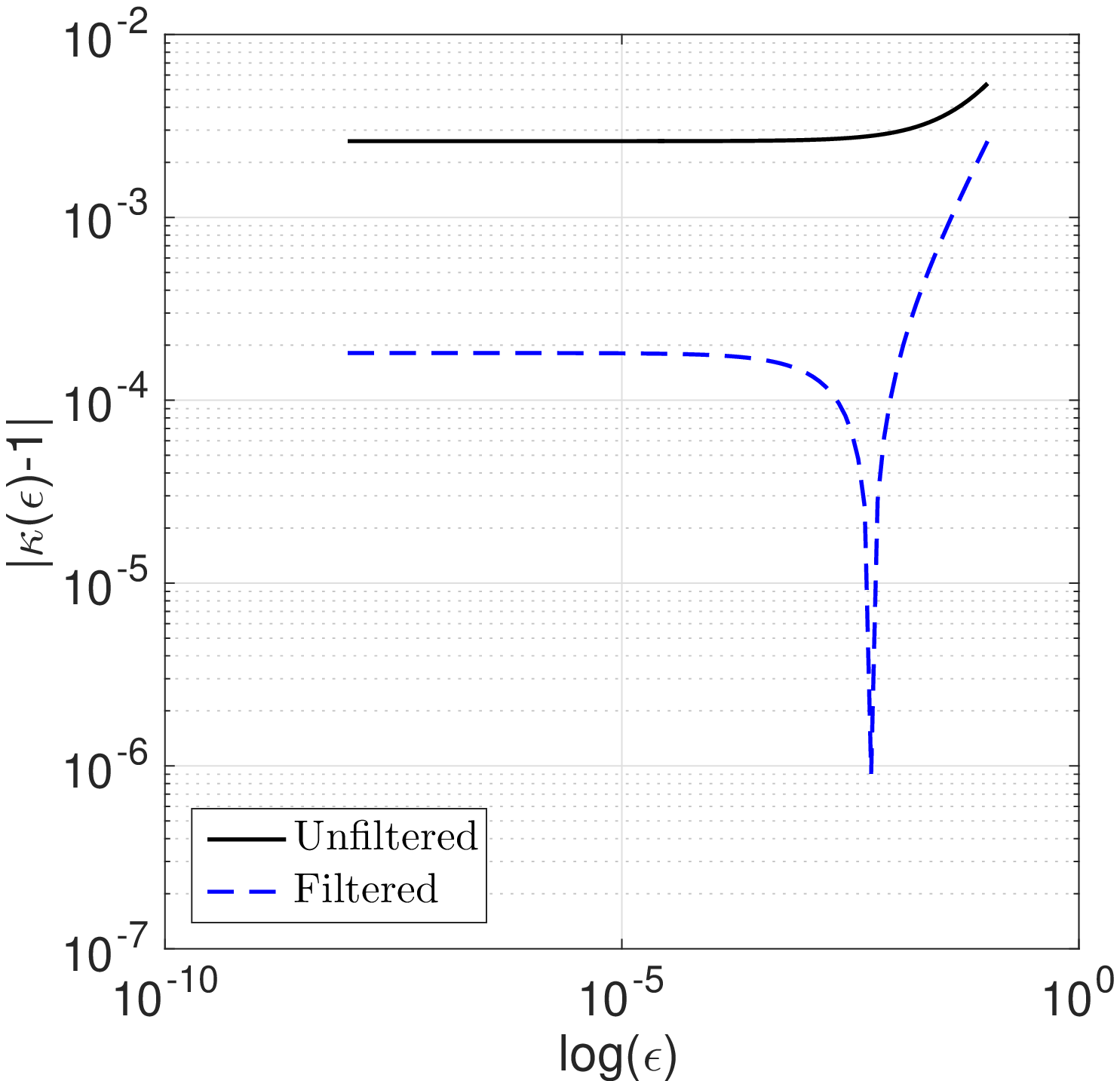}} \hspace{40pt}
\subfigure[]{\includegraphics[width=0.26\textwidth]{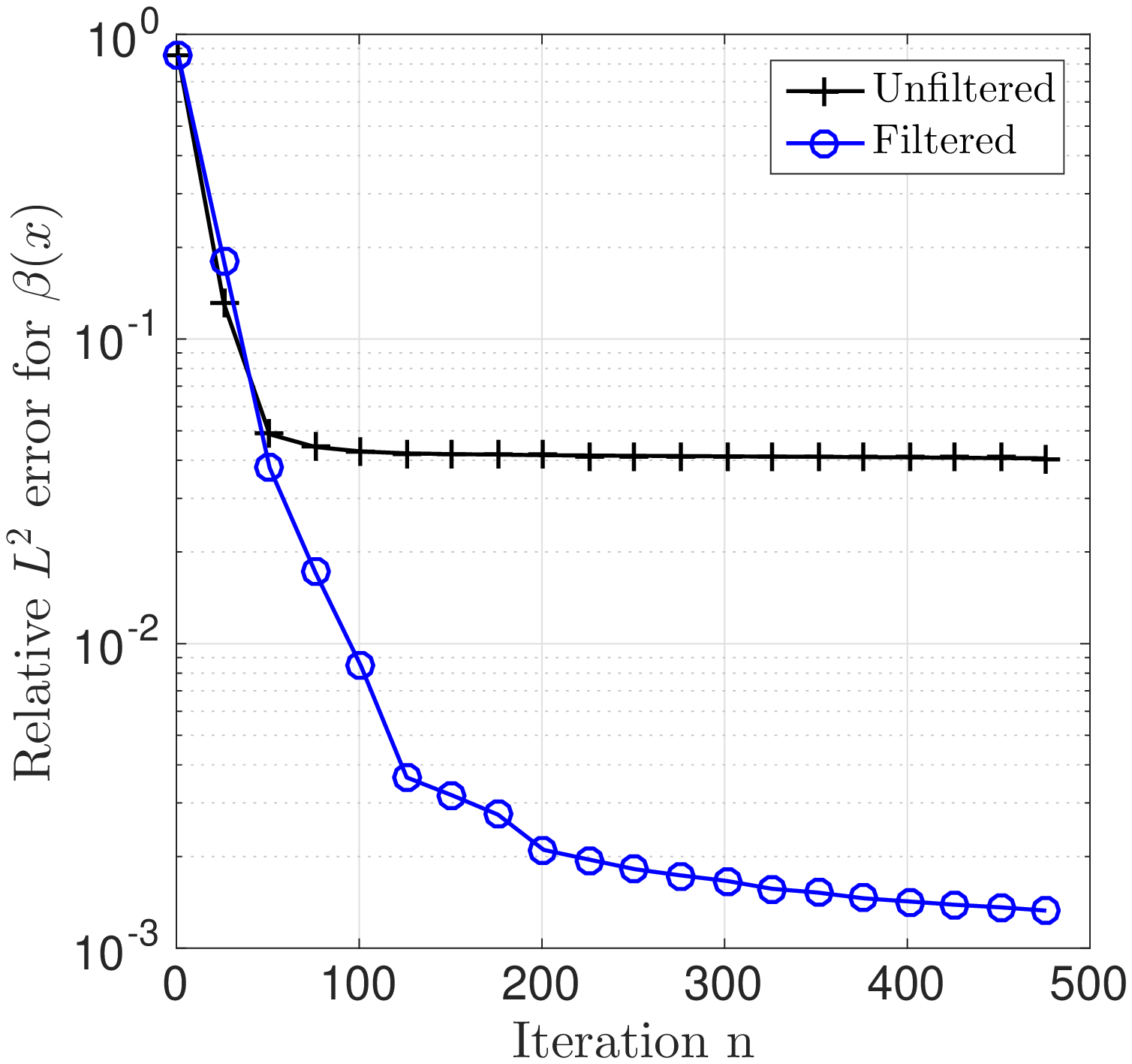}}\hspace{40pt}
\subfigure[]{\includegraphics[width=0.26\textwidth]{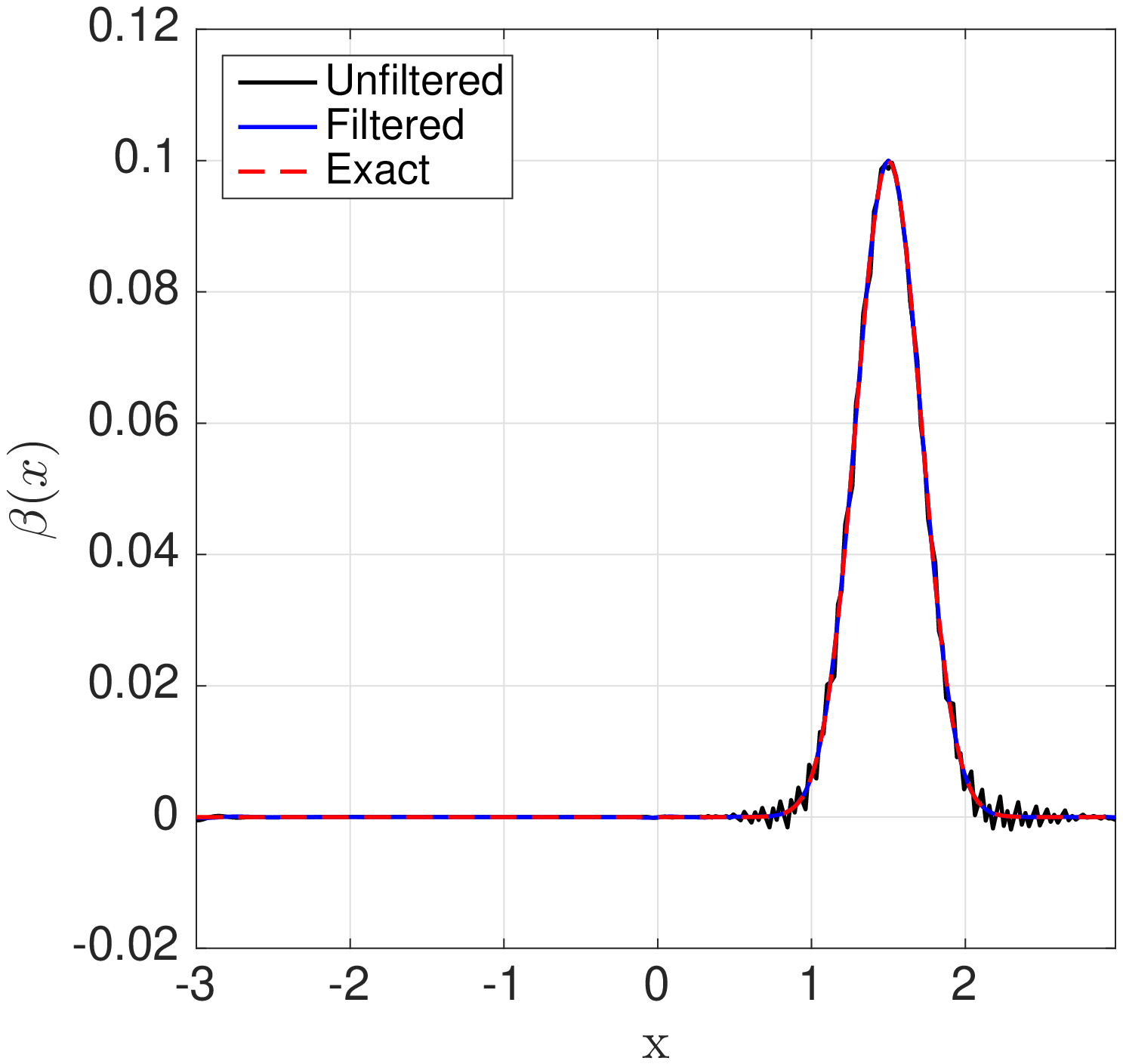}}

\subfigure[]{\includegraphics[width=0.26\textwidth]{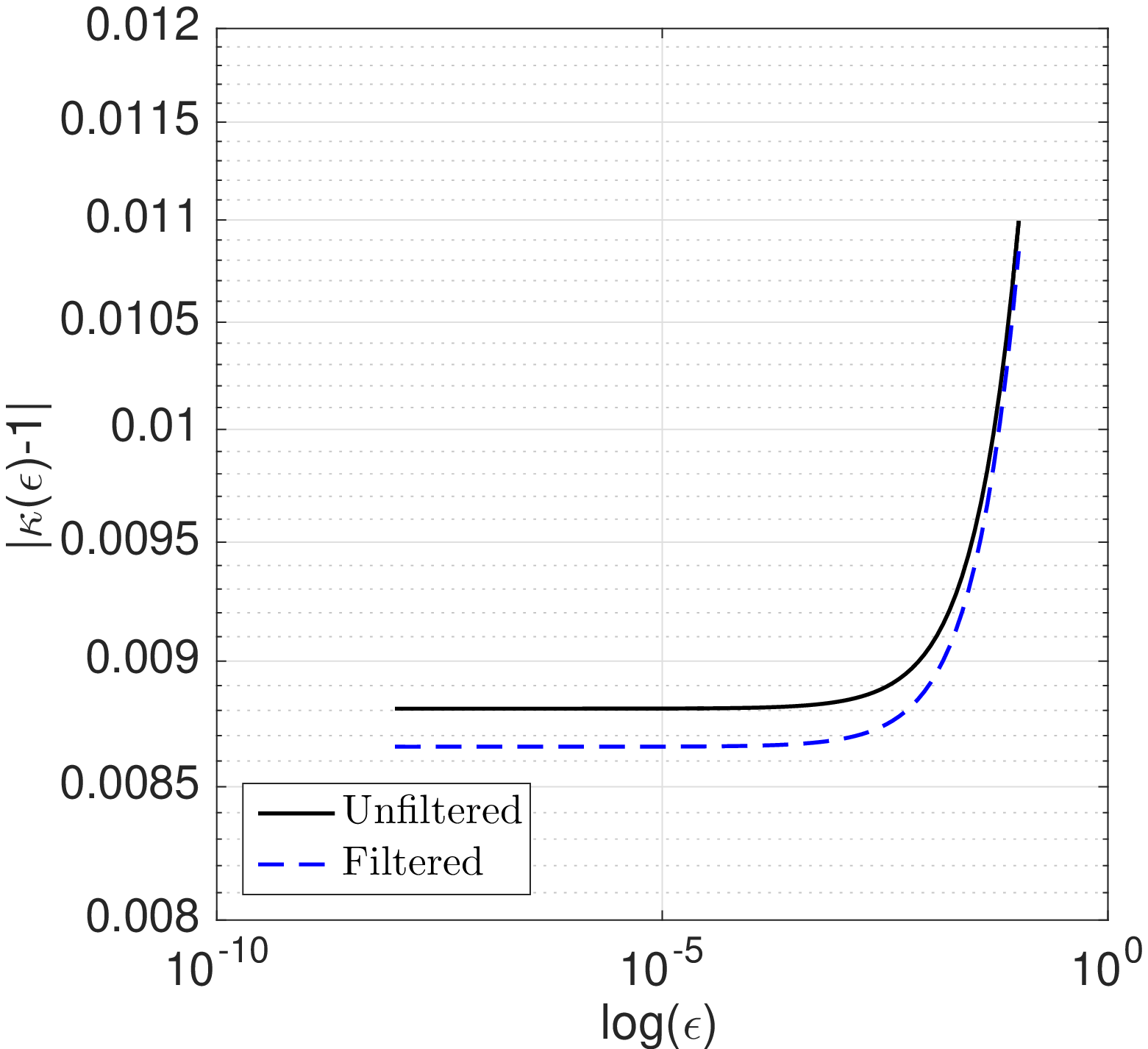}} \hspace{40pt}
\subfigure[]{\includegraphics[width=0.26\textwidth]{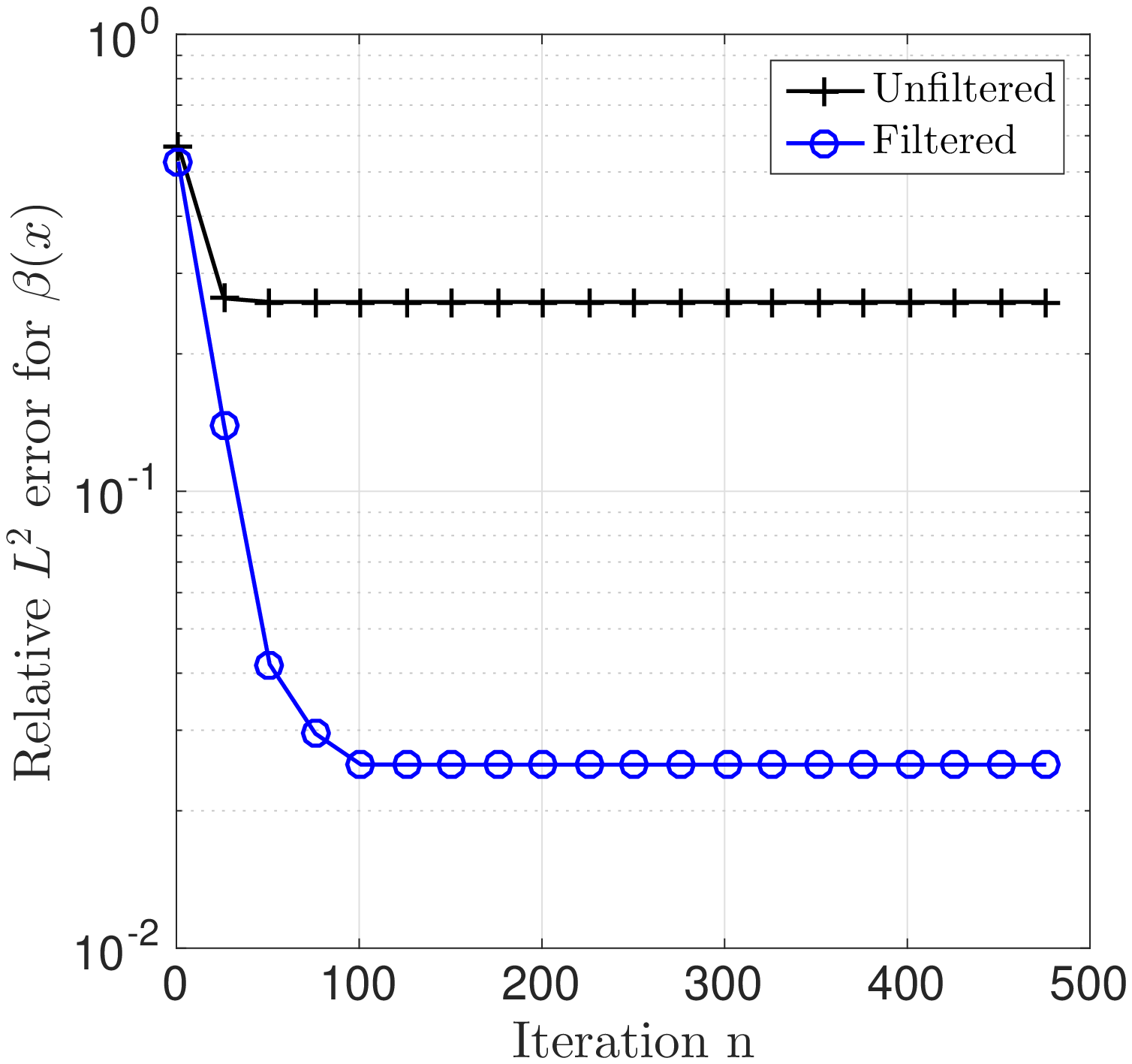}} \hspace{40pt}
\subfigure[]{\includegraphics[width=0.26\textwidth]{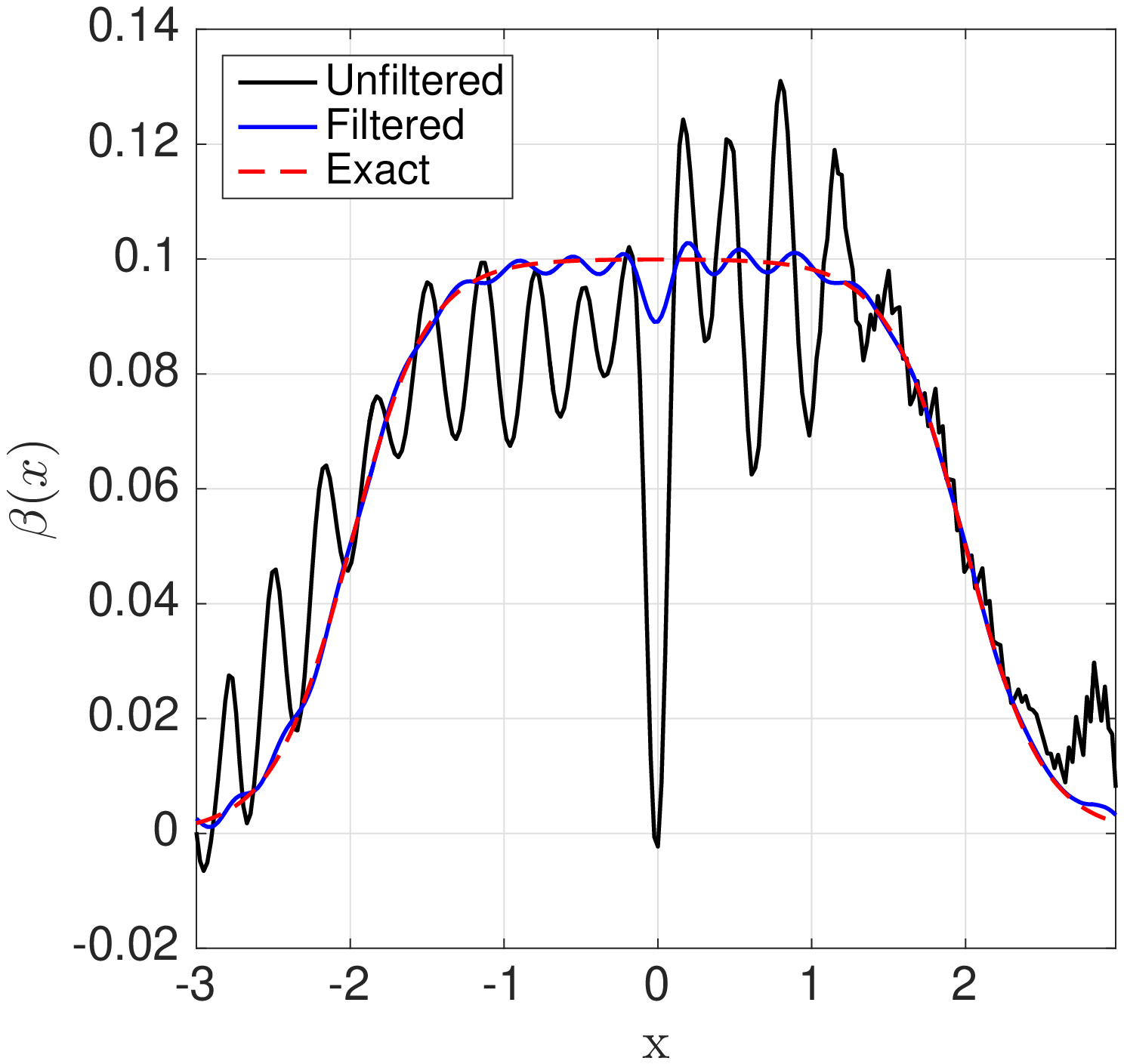}}

\subfigure[]{\includegraphics[width=0.26\textwidth]{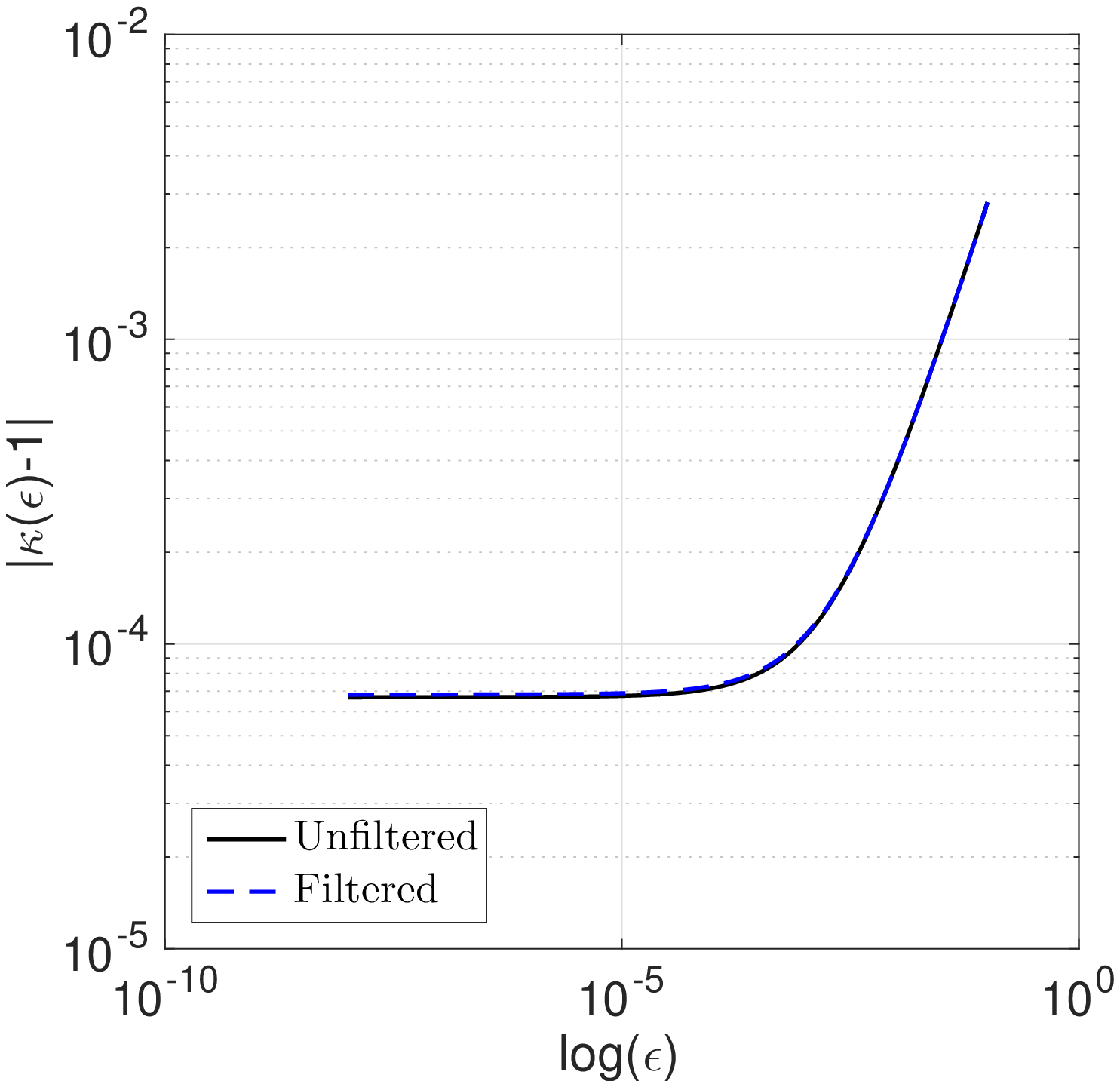}} \hspace{40pt}
\subfigure[]{\includegraphics[width=0.26\textwidth]{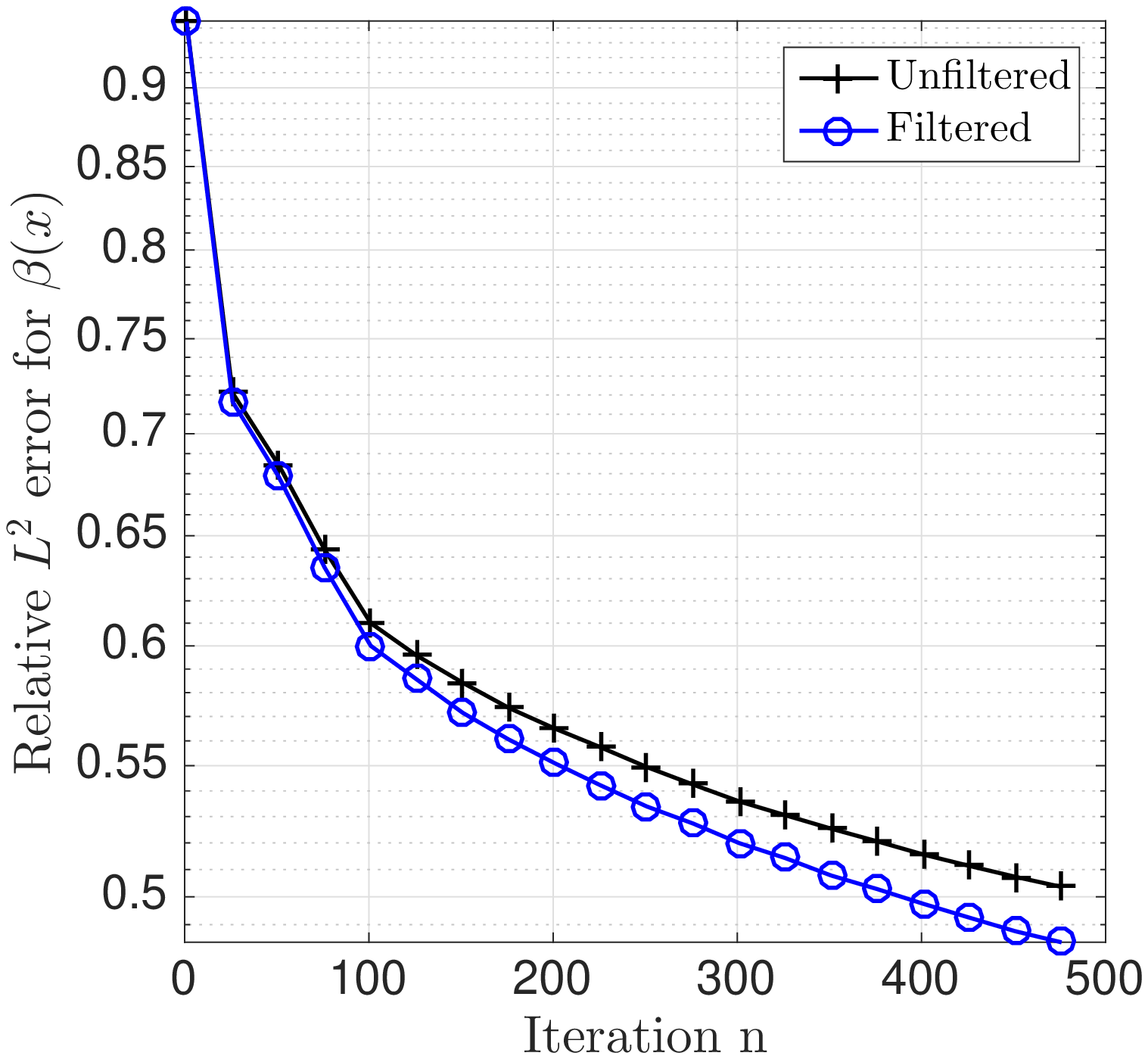}} \hspace{40pt}
\subfigure[]{\includegraphics[width=0.26\textwidth]{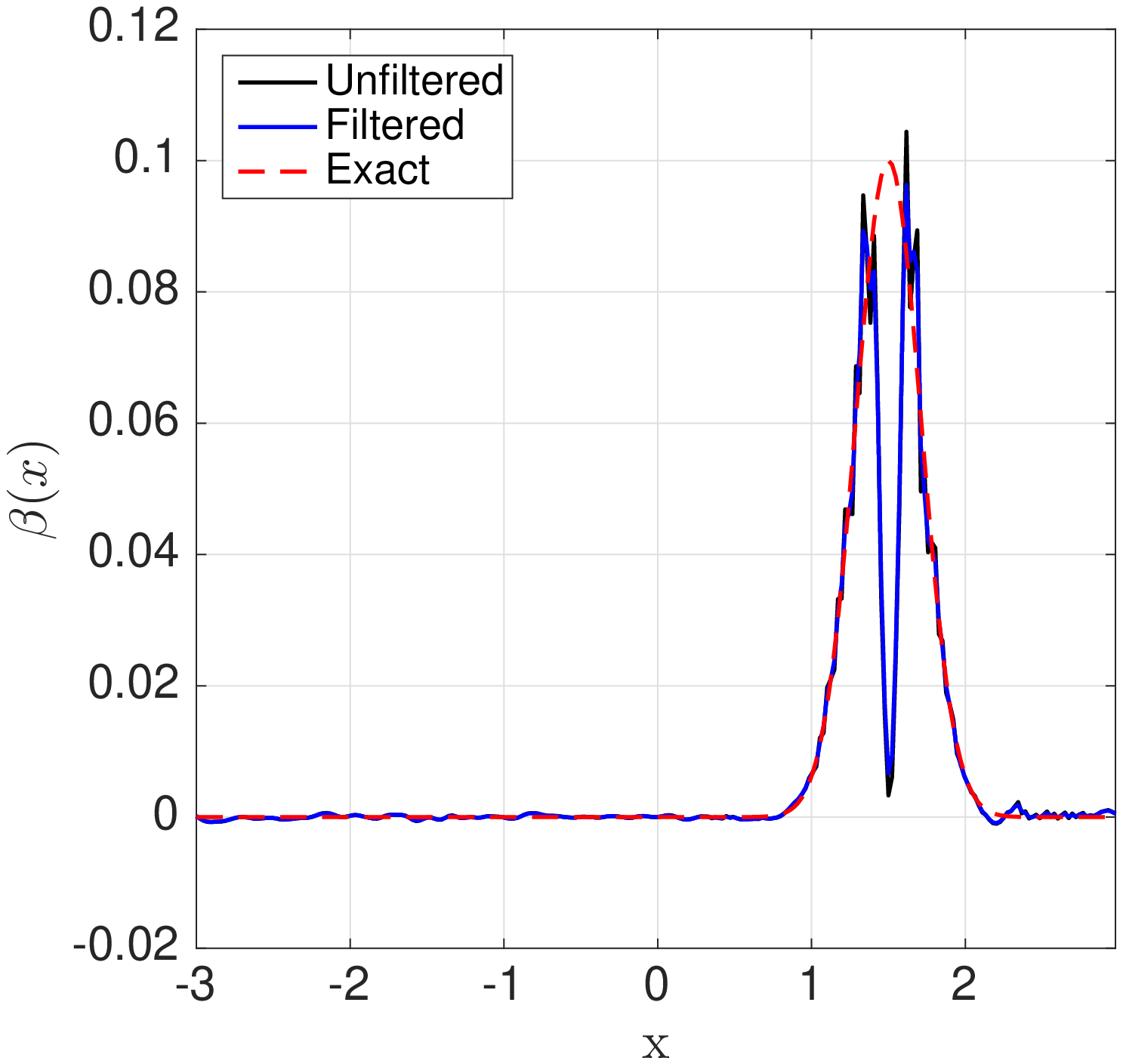}}
\caption{Results for assimilation scheme with $H^2$ smoothing applied to $\nabla^{L^2} J$. For for case \rom1, (a) shows the absolute  error of the kappa test convergence to $1$,  (b) shows the relative $L^2$ error $\parallel \beta^{(t)} - \beta ^{(n)} \parallel_{L^2}^2 $/$\parallel \beta^{(t)} \parallel_{L^2}^2$ between the exact and reconstructed bathymetry at each iteration, and (c) shows the reconstructed bathymetry with filtering, without filtering, and the exact bathymetry. Figures (d) - (f) represent the same simulations for case \rom2, and figures (g) - (i) for case \rom3. }
\label{results_filt}
\end{figure}

For a more detailed view of the spectrum for these lower frequencies, we restrict the plots to wavenumbers $k$ less than $50$, as all noise for higher frequencies in our domain have been suitably filtered out by both $H^1$ and $H^2$ smoothing. This restricted range allows us to see that in all cases, the choice of $l_1$ that smoothed out higher frequencies also damps the lower frequencies of $\widehat{(\nabla^{H^2} \mathcal{J})}_k$ as well, thereby losing information of the true signal. On the other hand, $\widehat{(\nabla^{H^2} \mathcal{J})}_k$ is almost indistinguishable form the unfiltered signal $\widehat{(\nabla^{L^2} \mathcal{J})}_k$ for smaller wavenumbers, while having effectively removed higher frequency noise. We can observe this in fig \ref{filt_h1_h2}(d), which shows the $L^2$, $H^1$, and $H^2$ gradients in the domain $\Omega$. We see that the $H^1$ gradient filters out the noise, but also reduces the amplitude of the signal peaks, whereas the $H^2$ gradient filters the noise and also retains the form of the original signal without further damping. 

Analysis of the results of algorithm \ref{alg:bath2} are given in fig \ref{results_filt}. The plots (b), (e), and (h) show the relative error in the reconstruction for case \rom1, \rom2, and \rom 3 respectively. The first thing we note is that in each case, the error is lower for the filtered scheme as opposed to the unfiltered. Especially with case \rom1 and \rom2, we observe the error decrease by at least an order of magnitude. The reconstructed bathymetry shown in (c) and (f) illustrates  how the noise has been greatly reduced visually, and for case \rom1 is almost negligible. For case \rom 2 we see some noise remaining on the plateau of the sandbar, however it is a drastic improvement from the unfiltered result. This increase is reflected in the kappa test results given in (a) and (d) for case \rom1 and \rom2 respectively, though the increased convergence does not scale proportionally with the error, as we saw before in fig \ref{fig:no_filt}.

However, this improvement does not extend to case \rom3. We see no improved convergence for the kappa test in (g), and negligible improvement in the reconstruction of $\beta(x)$ given in (i). These results did not improve with a more restrictive choice of $l_2$, leading us to consider whether additional factors such as the system parameters and placement of the sensors $y^{(o)}(t)$ effect the observability of the bathymetry. In section \ref{sec6} we attempt to analyse some of these effects.

\section{Necessary conditions on model parameters and the observation operator}
\label{sec6}

Having achieved improved results by smoothing our gradient to $H^2$, we now try to analyse the effect of model parameters and the observation operator on the convergence of our data assimilation scheme. 
\subsection{Necessary Conditions on Parameters}

For the purpose of this study, we restrict our parameter analysis to understanding the relationship between the amplitudes of the initial condition $\hat{\eta}$, the amplitude of the true bathymetry $\hat{\beta}$, and the average depth $H$, which we have normalised to $1$. As this research is focused on improvement in tsunami prediction, our objective is to understand how tsunami waves propagate over sea bathymetry, and these factors play an important part when considering the scales involved. Tsunamis are characterised by their long wavelength, often reaching 100-150km in the deep oceans, and their relatively small amplitude can be between $0.1-1$m, making them often imperceptible; even when approaching coastlines the amplitude of the surface wave can be $20-50$m whereas the wavelength may still be up to $2$km. As the energy flux of the wave speed is dependent on wave speed $c$ and depth $H+ \eta(x,t) -\beta(x)$, bathymetry amplitudes and regularity can have a big impact on the solution \cite{tsunami_ch}. 

Subsequently, we give special consideration to case \rom1, where we have a Gaussian initial condition of the surface wave, generating a travelling wave to the right ( we also have a left-propagating wave in the simulation, however we do not consider it in our analysis as for tsunamis we generally focus on a single direction and thus all sensors are placed at $x>0$), and we introduce a Gaussian bathymetry perturbation to the right, which is observed at some point $t>0$. the results for convergence of the kappa test, and the error in the reconstruction given some magnitude scale for $\frac{\hat{\eta}}{\hat{\beta}}$ and $\frac{\hat{\beta}}{H}$ are given in table \ref{table_amp}. As we can observe, there are cases where the data assimilation scheme \ref{alg:bath2} fails to converge and becomes unstable, and we have highlighted those in red. Additionally, if we require that the error in the kappa test convergence be at most $\ord(10^{-3})$, then some cases fail to meet that criteria, which we have also highlighted in red. That leaves two valid cases remaining, where $\frac{\hat{\eta}}{\hat{\beta}} \approx \ord(10^{-1}), \frac{\hat{\beta}}{H} \approx \ord(10^{-3}) $, and $\frac{\hat{\eta}}{\hat{\beta}} \approx \ord(10^{-3}), \frac{\hat{\beta}}{H} \approx \ord(10^{-1}) $. It is insightful to note that we require either the ratio between the amplitudes to be small, or the ratio between $\hat{\beta}$ and $H$ to be small ($\ord(10^{-3})$), but not both at the same time. This is also true for when both are large ($\ord(10^{-1})$). In fact, the only admissible cases are when the resulting ratio between $\hat{\eta}$ and $H$ is $\ord(10^{-4})$ as a consequence. And so we surmise that for case \rom1, a necessary condition for convergence of the optimisation scheme and observability of the true bathymetry, is that the initial condition of the surface wave be of low amplitude relative to the average sea depth $H$. 

\begin{figure}[H]
\centering
\subfigure[Case \rom 1]{\includegraphics[width=0.26\textwidth]{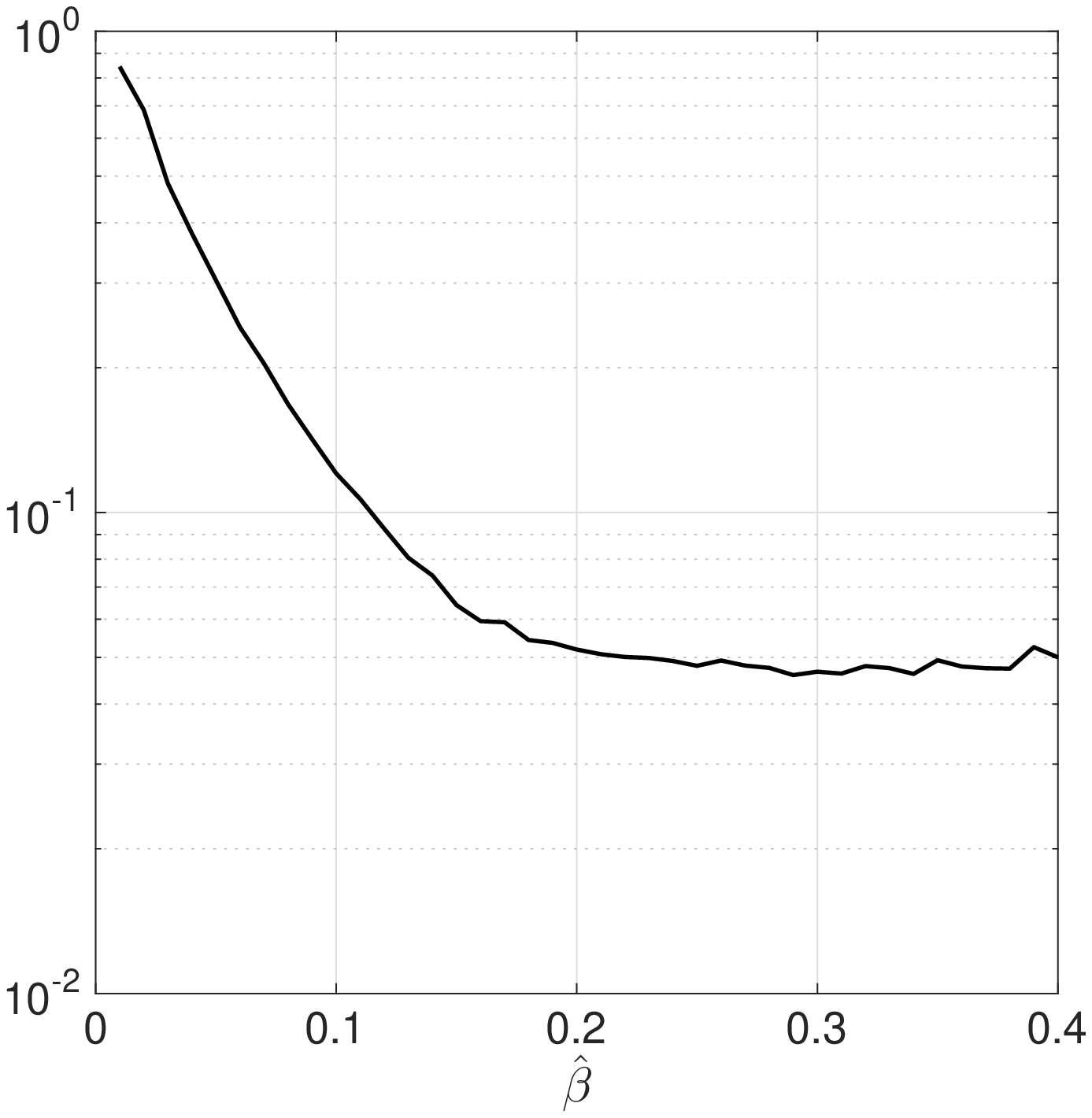}} \hspace{40pt}
\subfigure[Case \rom 2]{\includegraphics[width=0.26\textwidth]{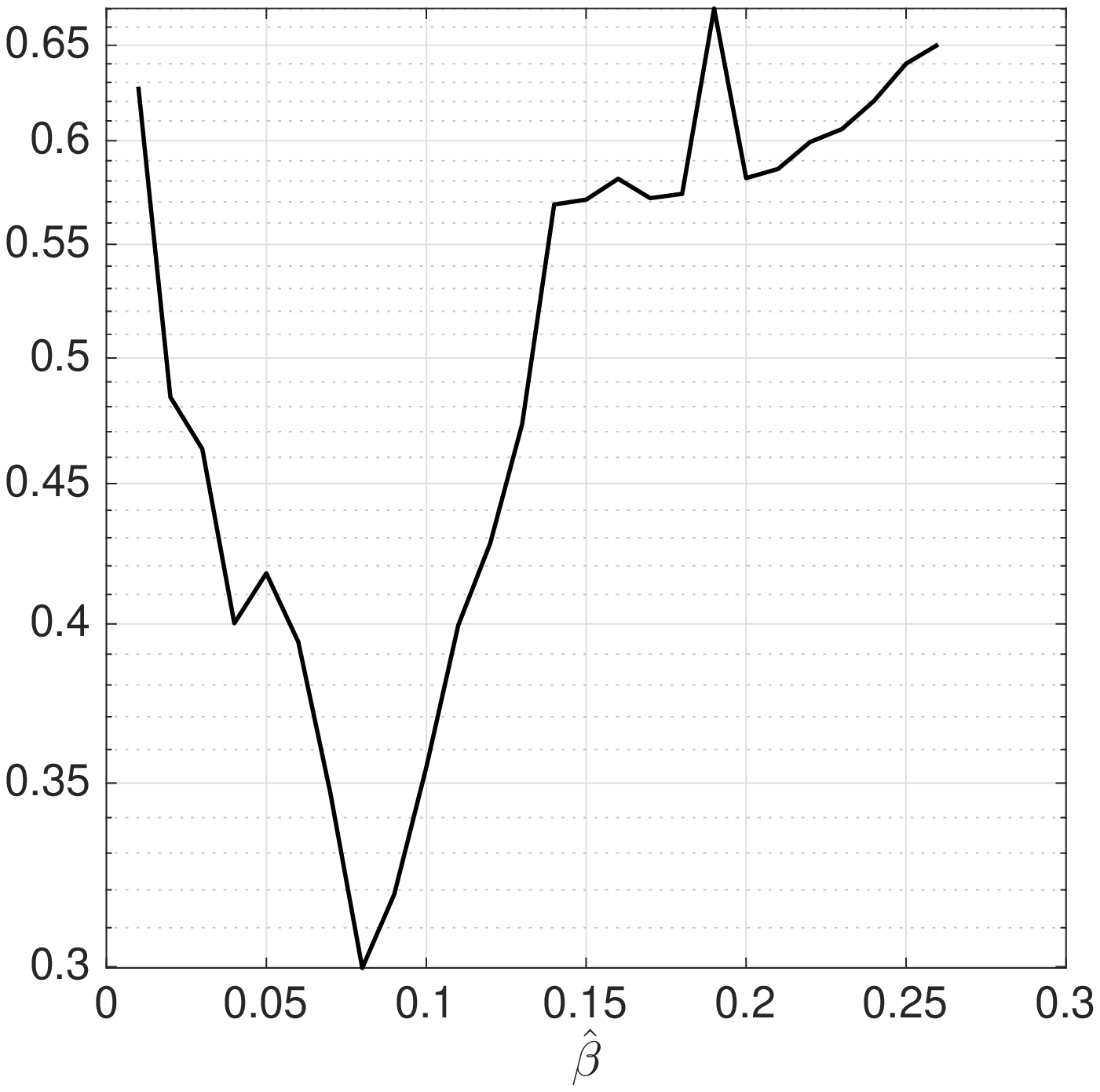}} \hspace{40pt}
\subfigure[Case \rom 3]{\includegraphics[width=0.26\textwidth]{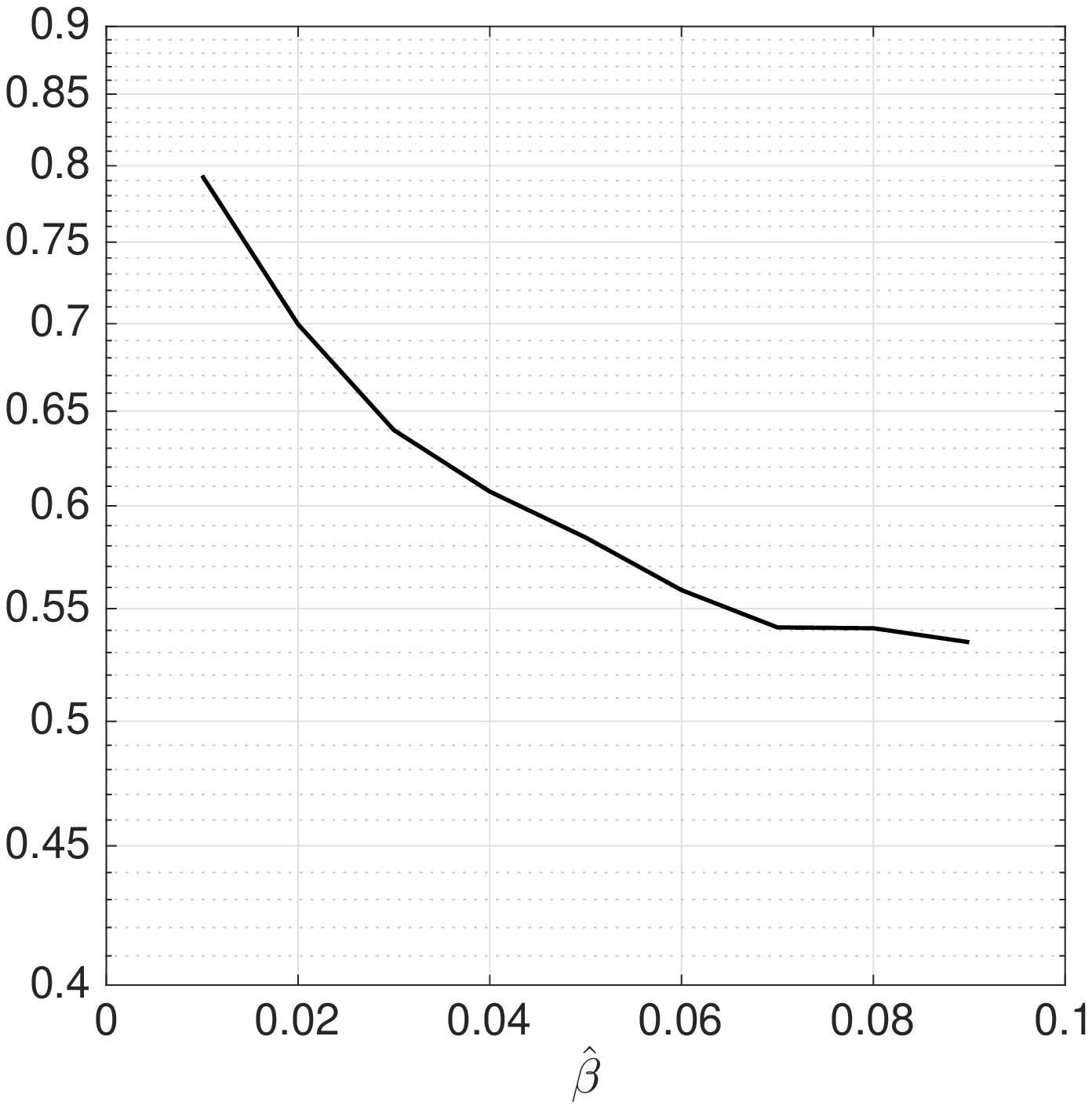}}\hspace{40pt}
\caption{ $\parallel \beta^{(t)} - \beta ^{(b)} \parallel_{L^2(\Omega)}^2 $/$\parallel \beta^{(t)} \parallel_{L^2(\Omega)}^2$ (where $\Omega = [-L,L]$), shown for different amplitudes $\hat{\beta}$, with amplitude of initial condition $\hat{\eta}$ fixed to be $1\%$ of $\hat{\beta}$}
\label{sens_bath}
\end{figure}

To consider whether the amplitude of the bathymetry $\hat{\beta}$ plays a role, we analysed convergence of algorithm \ref{alg:bath2} for amplitudes ranging from $1\%$ of the average deoth to $40 \%$. The results are summarised in fig \ref{sens_bath}, and we have included analysis for cases \rom 2 and \rom 3 for more insight. In all cases we fix $\hat{\eta}$ to be $1\%$ of  $\hat{\beta}$. We see that for case \rom1 and the results corresponding to table \ref{table_amp}, fig \ref{sens_bath}(a) shows that the error is highest when  $\hat{\beta}$ is small, and then remains stable even when it is up to $40\%$ of $H$. However for cases \rom2 and \rom3, the algorithm becomes unstable when  $\hat{\beta}$ gets bigger than approximately $10 \%$ of $H$. And so if we are to generalise to beyond case \rom1, based on the current analysis we state that an additional necessary condition is that the amplitude of the bathymetry  $\hat{\beta}$ be no greater than $\ord(10^{-2})$ for convergence when we have sandbar type profiles, and if we are generalising to surface waves with periodic initial conditions. While we did not highlight the results for $\hat{\eta}$ for case \rom2 and \rom3 in as much detail as case \rom1, in all cases the algorithm only converged for $\frac{\hat{\eta}}{\hat{\beta}} \approx \ord(10^{-1}) $ when $\frac{\hat{\beta}}{H}$ was very small. Hence we conclude that a necessary condition for our data assimilation scheme is that the amplitude of the initial condition be small relative to average depth $H$, and $\hat{\beta}$ be no greater than $10\%$ of $H$. This is not an unreasonable condition, and corroborates the necessary assumptions derived by Nichols and Taber(2009) \cite{Nichols_09} with their dispersion relation inversion. Additionally considering the focus of tsunami predictive models is in generally in the Pacific ocean where the average depth is $4$km, sea floor perturbations of $10\%$  are not insignificant.

\begin{table}[H]
\centering
\begin{tabular}{|c|c|c|c|}
 \hline
  & & &\\[-1em]
 &  $\frac{\hat{\eta}}{\hat{\beta}} \approx \ord(10^{-1})$ 
 & $\frac{\hat{\eta}}{\hat{\beta}} \approx \ord(10^{-2})$  & $\frac{\hat{\eta}}{\hat{\beta}} \approx \ord(10^{-3})$ \\ 
  & & &  \\[-1em] \hline 
  & & &  \\[-1em]
\multirow{2}{*}{$\frac{\hat{\beta}}{H} \approx \ord(10^{-1})$} 
 & $|\kappa(\ep) -1| = \red {3 \times10^{-2}}$  &$|\kappa(\ep) -1| =   {\red 2.7 \times 10^{-2}}$ &  $|\kappa(\ep) -1| = 4.2 \times10^{-4}$ \\
 &  & &\\[-1em]
 & $L^2$ Error = {\red \textbf{Undefined}} & $L^2$ Error $= 1.3 \times 10^{-3}$   &  $L^2$ Error $=  4.6 \times10^{-3}  $  \\
 & &  &\\[-1em]
  \hline
  & &  &\\[-1em]
\multirow{2}{*}{$\frac{\hat{\beta}}{H} \approx \ord(10^{-3})$}  
 & $|\kappa(\ep) -1| = 2.5 \times10^{-3}$      &  $|\kappa(\ep) -1| =  {\red 3 \times 10^{-2}}$  & $|\kappa(\ep) -1| = 2.7 \times 10^{-3}$  \\ 
  & & &  \\[-1em]
 & $L^2$ Error $= 7.4 \times10^{-3} $ & $L^2$ Error = {\red \textbf{Undefined}} & $L^2$ Error = {\red \textbf{Undefined}} \\
\hline
\end{tabular}
\caption { Analysis of kappa test convergence and the Error $\parallel \beta^{(t)} - \beta ^{(b)} \parallel_{L^2(\Omega)}^2 $/$\parallel \beta^{(t)} \parallel_{L^2(\Omega)}^2$ by varying amplitudes of $\phi(x)$ and $\beta^{(t)}$ for Case \rom1}
 \label{table_amp}
\end{table}

\subsection{Observation Operator}

The analysis of the observation operator in our optimisation scheme is significant, as it is a degree of freedom which we have the most direct control over in real world simulations, and subsequently are interested in finding configurations that give us maximum convergence to the exact bathymetry. 

\begin{figure}[H]
\centering
\subfigure[Case \rom1 Cost ]{\includegraphics[width=0.25\textwidth]{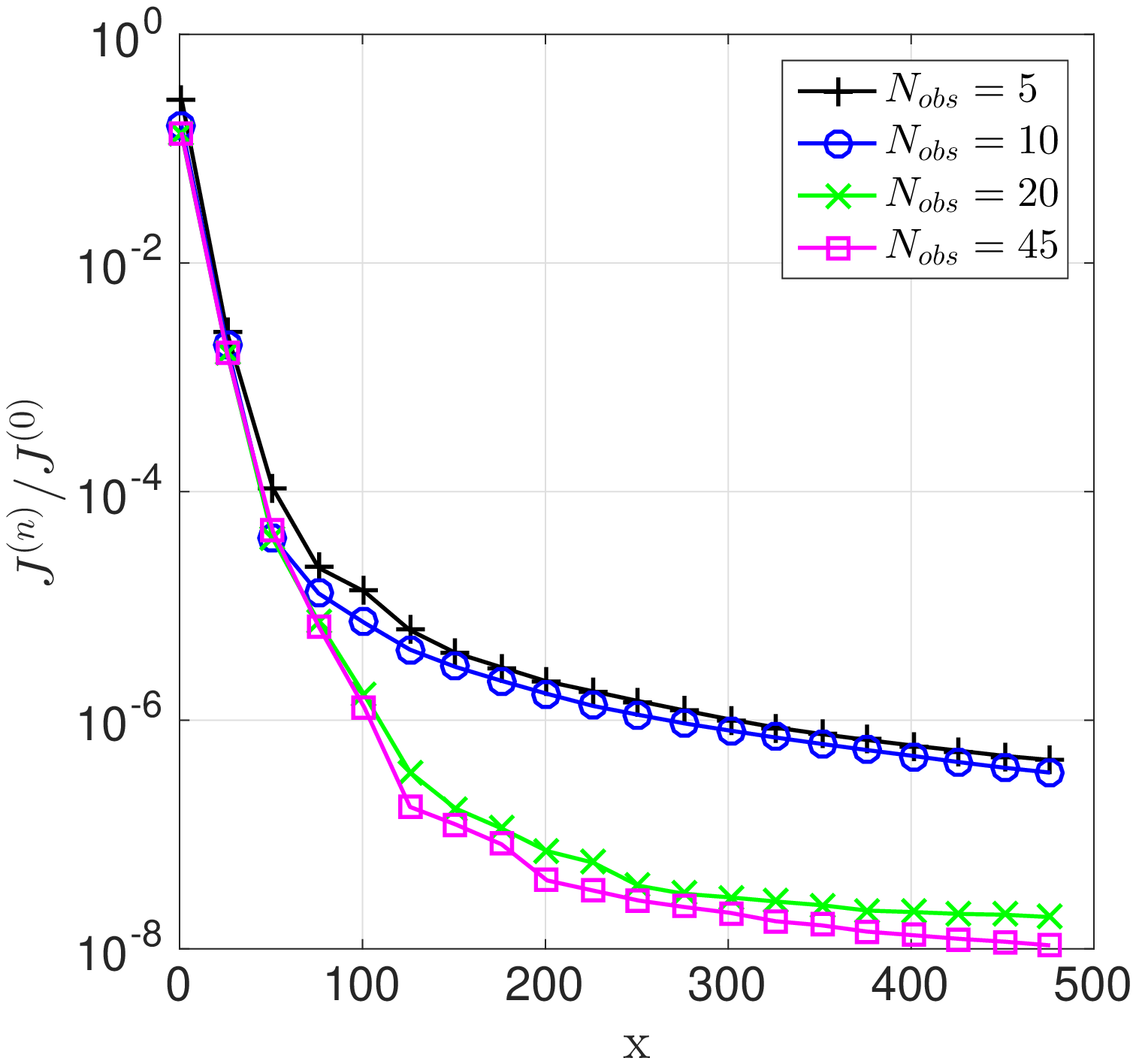}}\hspace{40pt} 
\subfigure[Case \rom2 Cost ]{\includegraphics[width=0.25\textwidth]{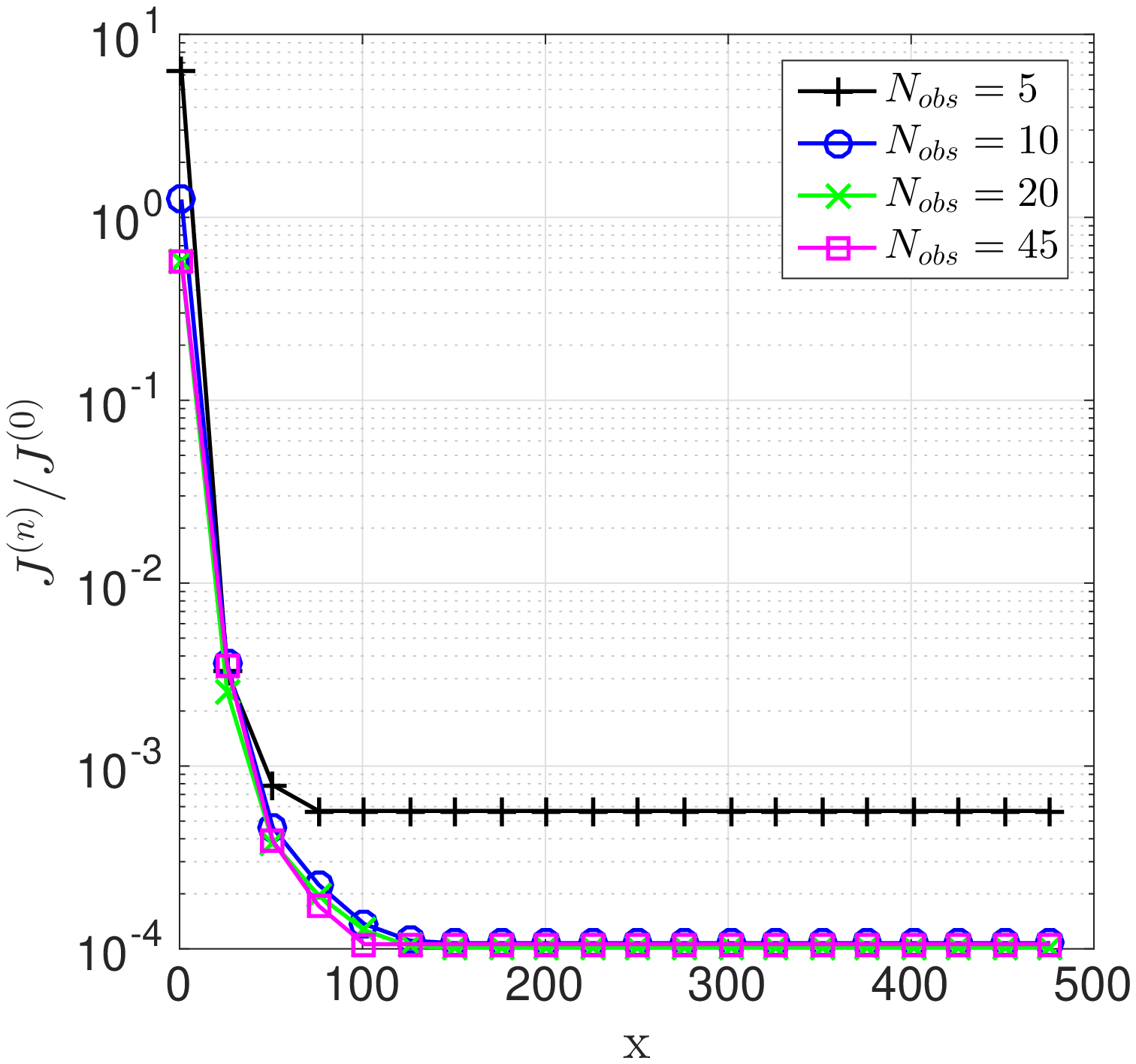}} \hspace{40pt}
\subfigure[Case \rom3 Cost ]{\includegraphics[width=0.25\textwidth]{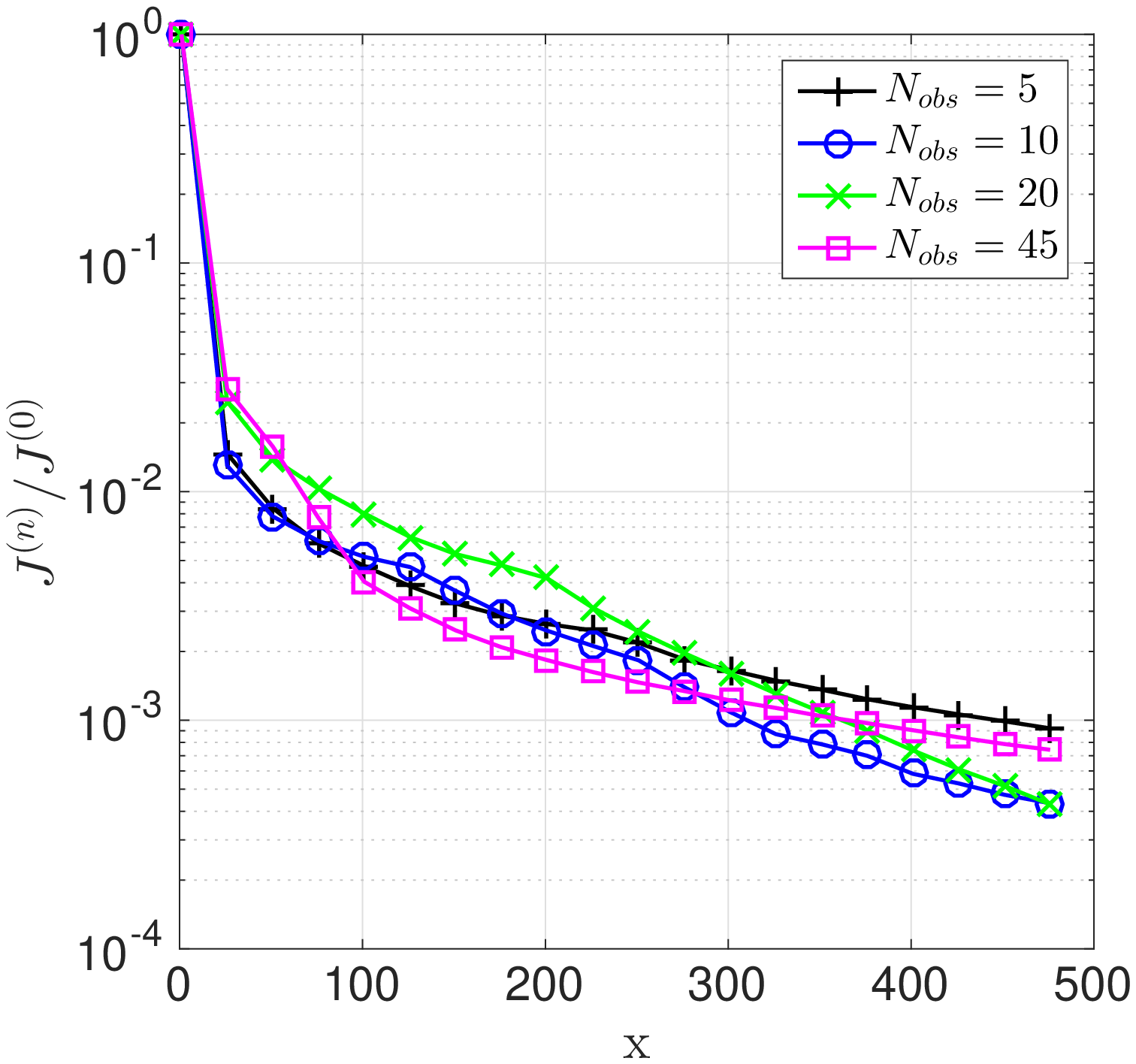}} 
\subfigure[Case \rom1 Error]{\includegraphics[width=0.25\textwidth]{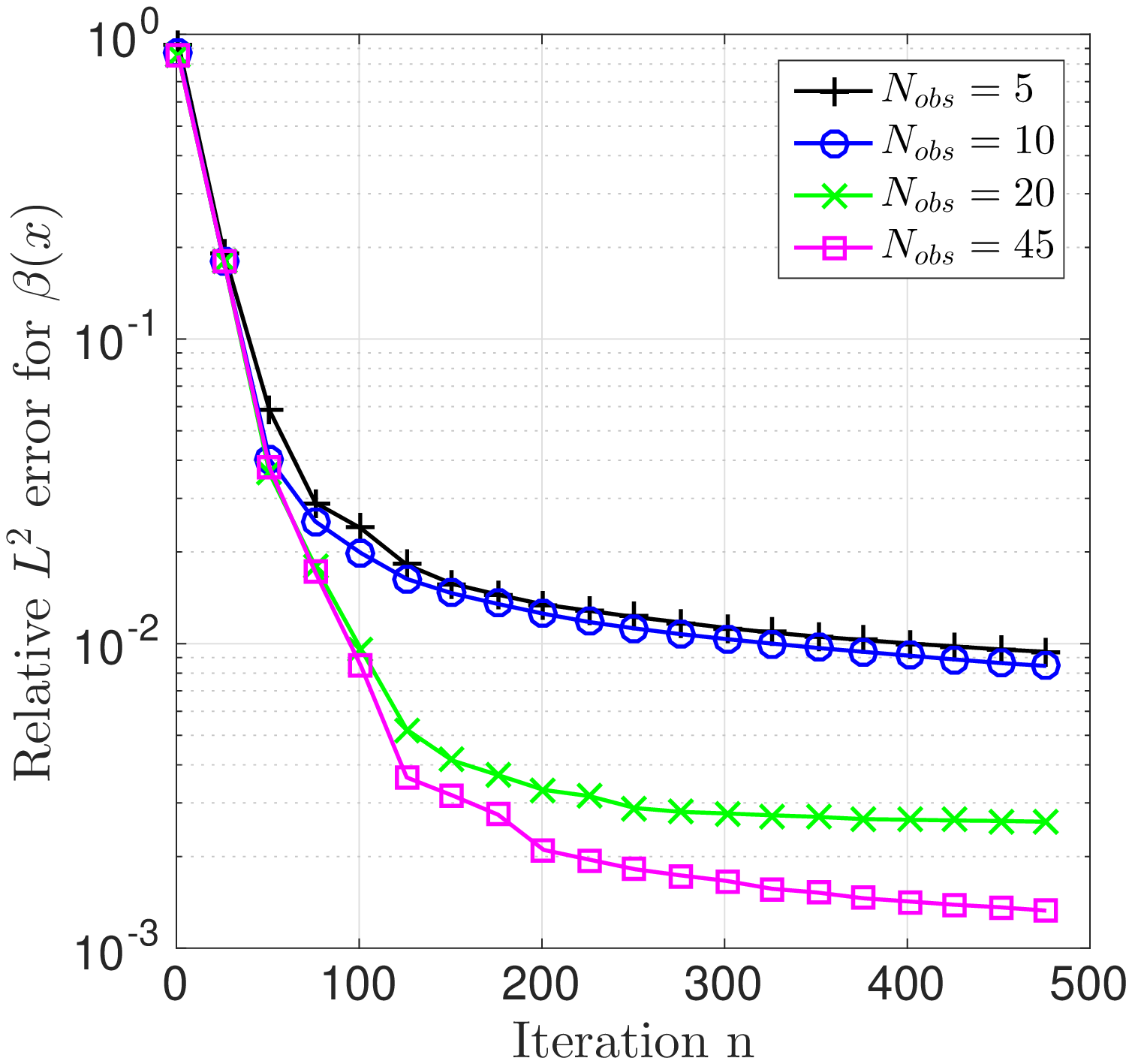}} \hspace{40pt}
\subfigure[Case \rom2 Error]{\includegraphics[width=0.25\textwidth]{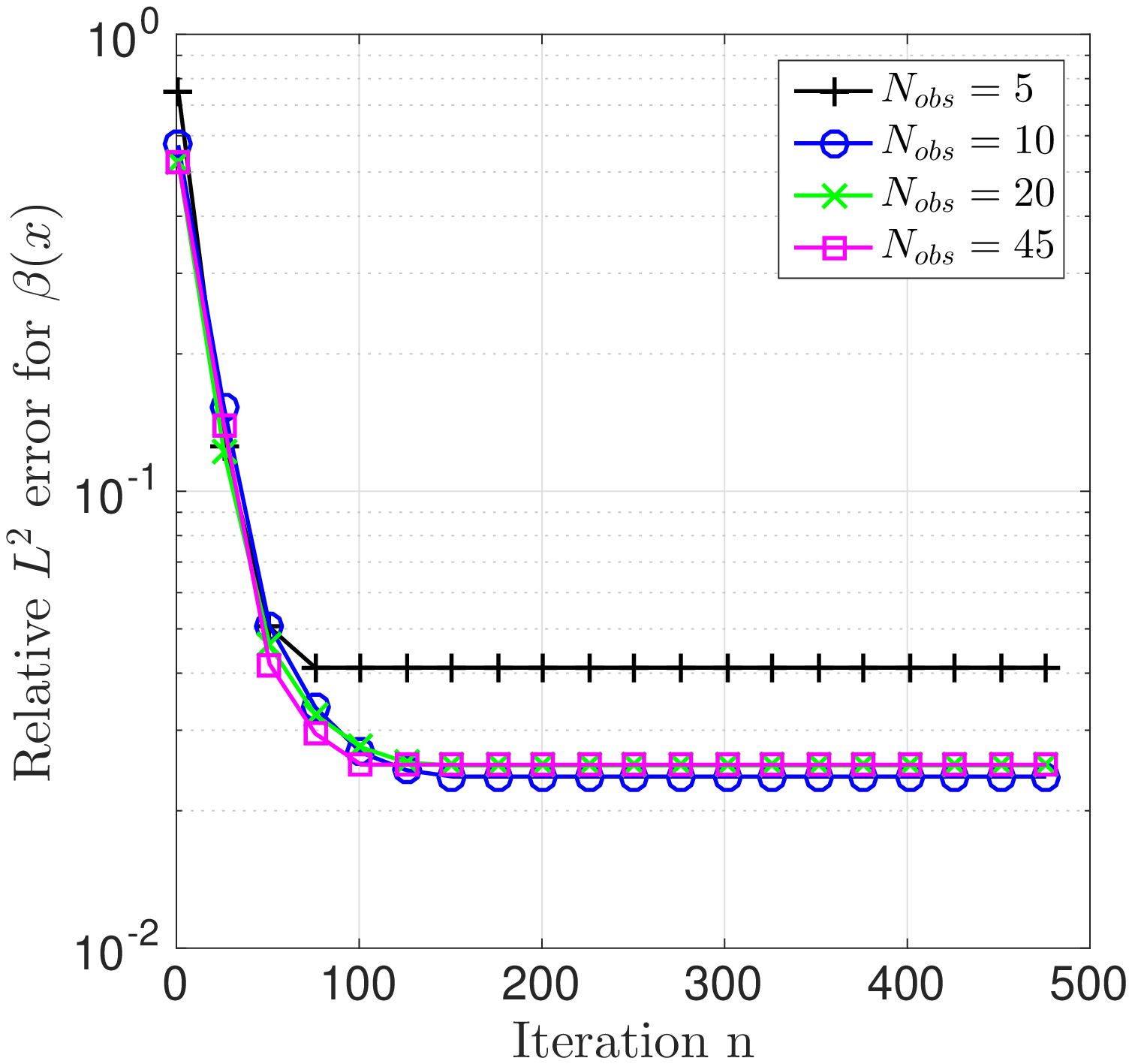}} \hspace{40pt}
\subfigure[Case \rom3 Error]{\includegraphics[width=0.25\textwidth]{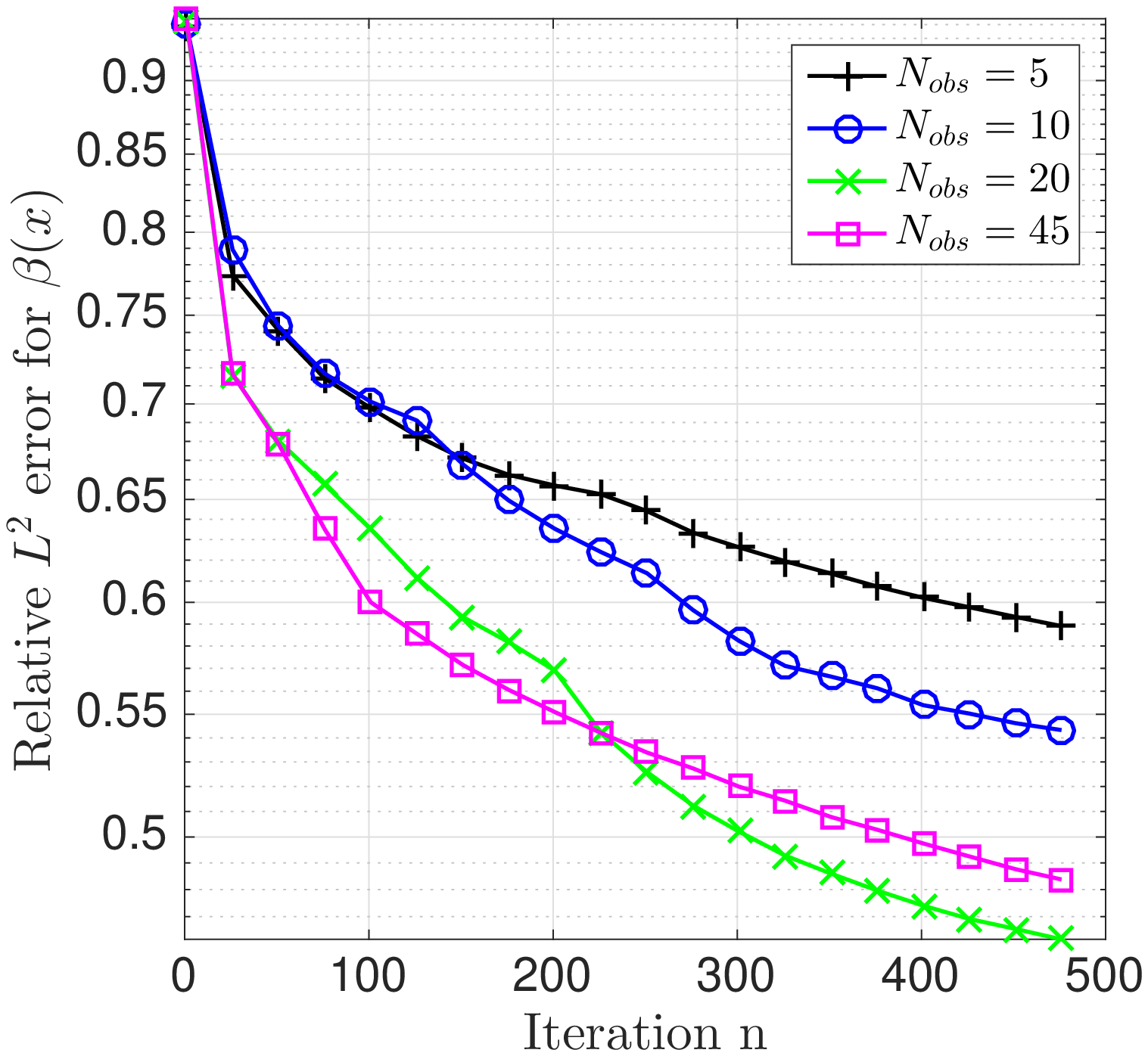}}
\caption{Relative cost function and relative $L^2$ error for different numbers of observation points}
\label{mult_obs}
\end{figure}

Firstly, we consider the effect of the number of observation points on convergence. In fig \ref{mult_obs} we present the results of algorithm \ref{alg:bath2} for $N_{obs} = 5, 10, 20$, and $25$. Each configuration is a set of equidistant points, with the first point placed at $x= L/10$. For case \rom1 and \rom2 this is when the first point is within the support of the Gaussian initial condition. We show the convergence of both the cost function and the relative $L^2$ error, and across all three cases, the general consensus is that more observation points result in better convergence for each. On closer observation, we note that for case \rom2, having $5$ points only is suboptimal, however there is not much difference in convergence for $N_{obs} = 10, 20, 45$ respectively. Additionally, it is interesting to note that for case \rom3 (fig \ref{mult_obs}(f)), having $20$ observation points shows a better convergence for the error than $45$ points. This suggests that there are additional factors influencing observability, which can influence the efficacy of the scheme. 

\begin{figure}[H]
\centering
\subfigure[]{\includegraphics[width=0.255\textwidth]{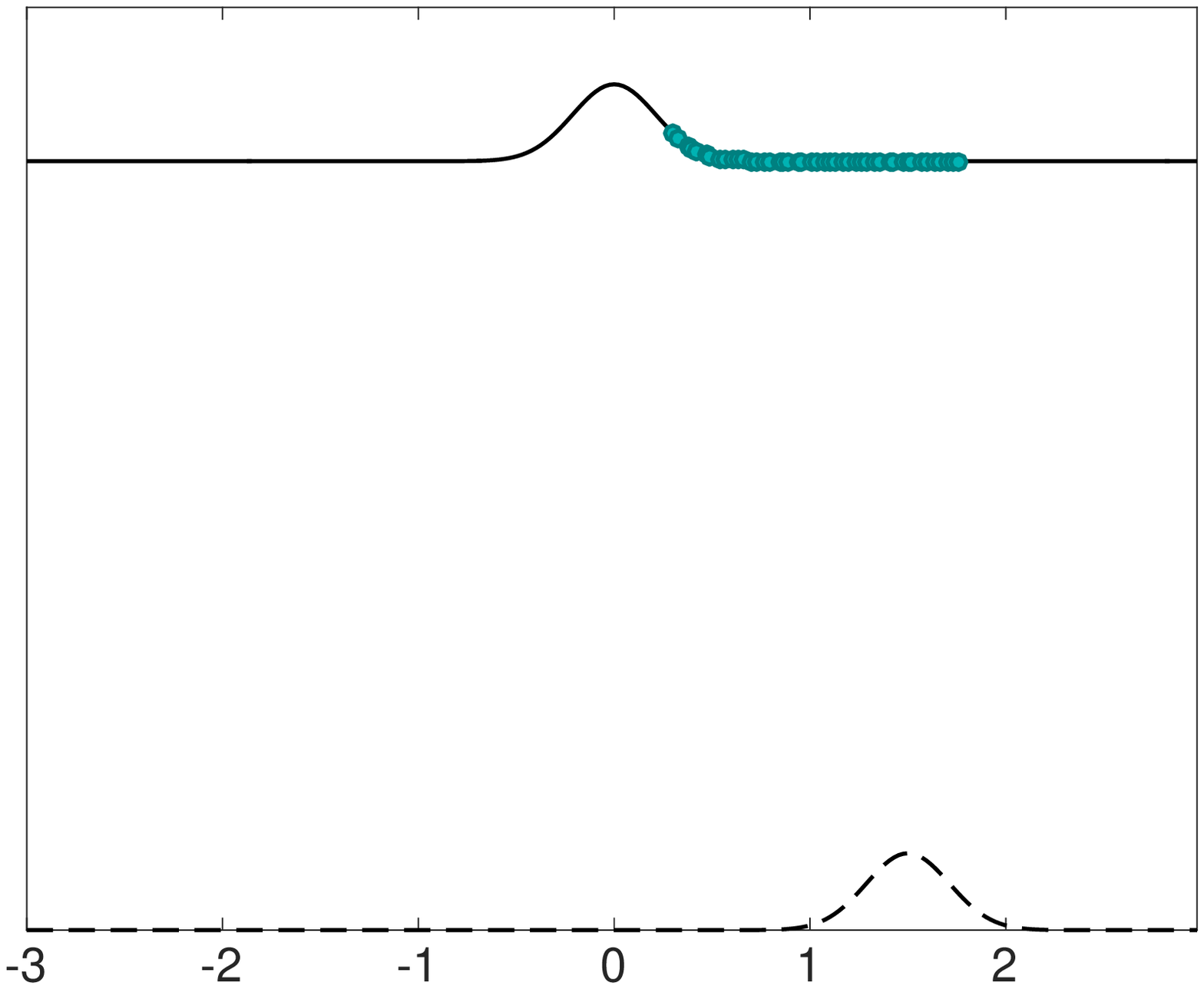}} \hspace{40pt}
\subfigure[]{\includegraphics[width=0.255\textwidth]{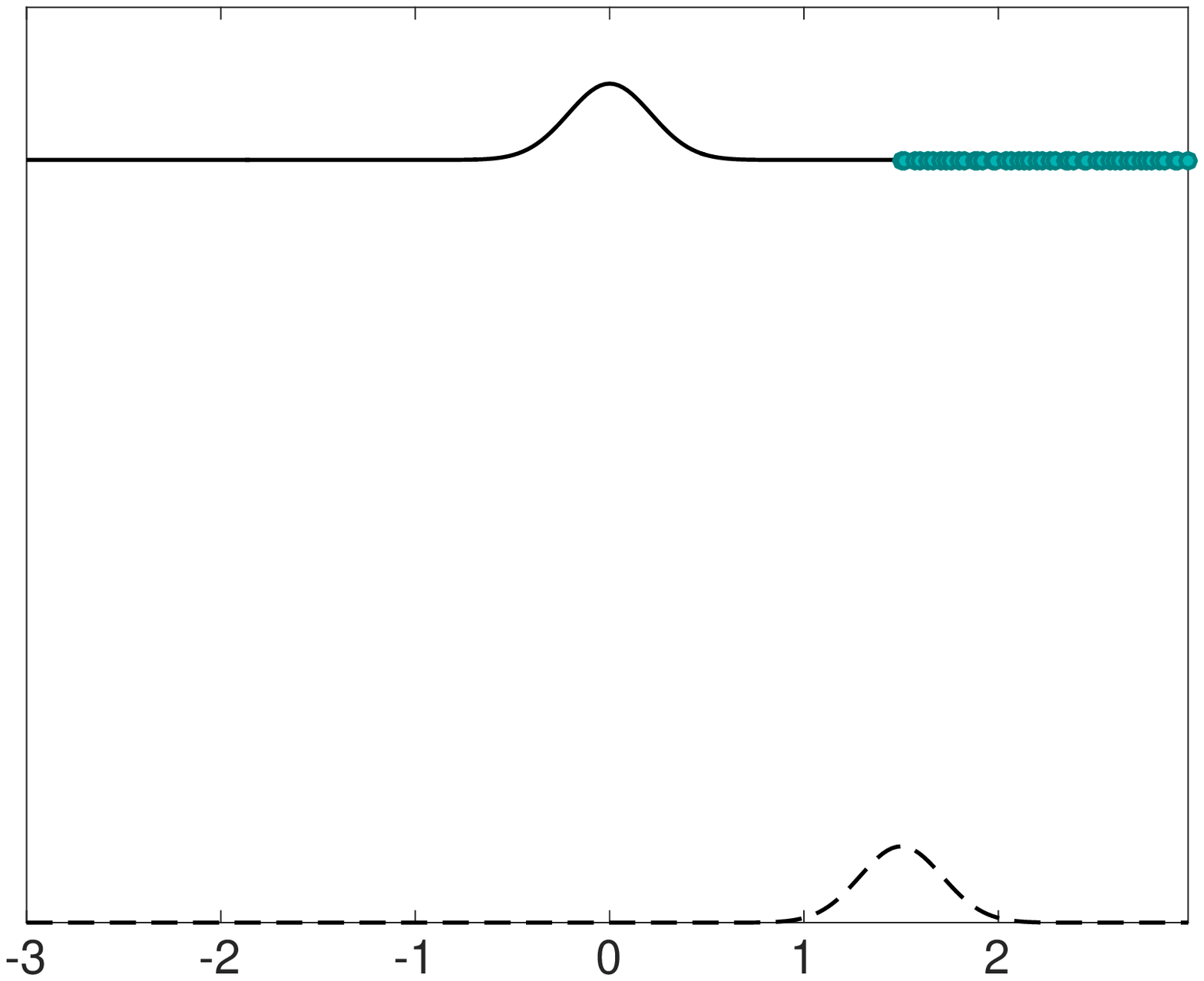}} 

\subfigure[]{\includegraphics[width=0.255\textwidth]{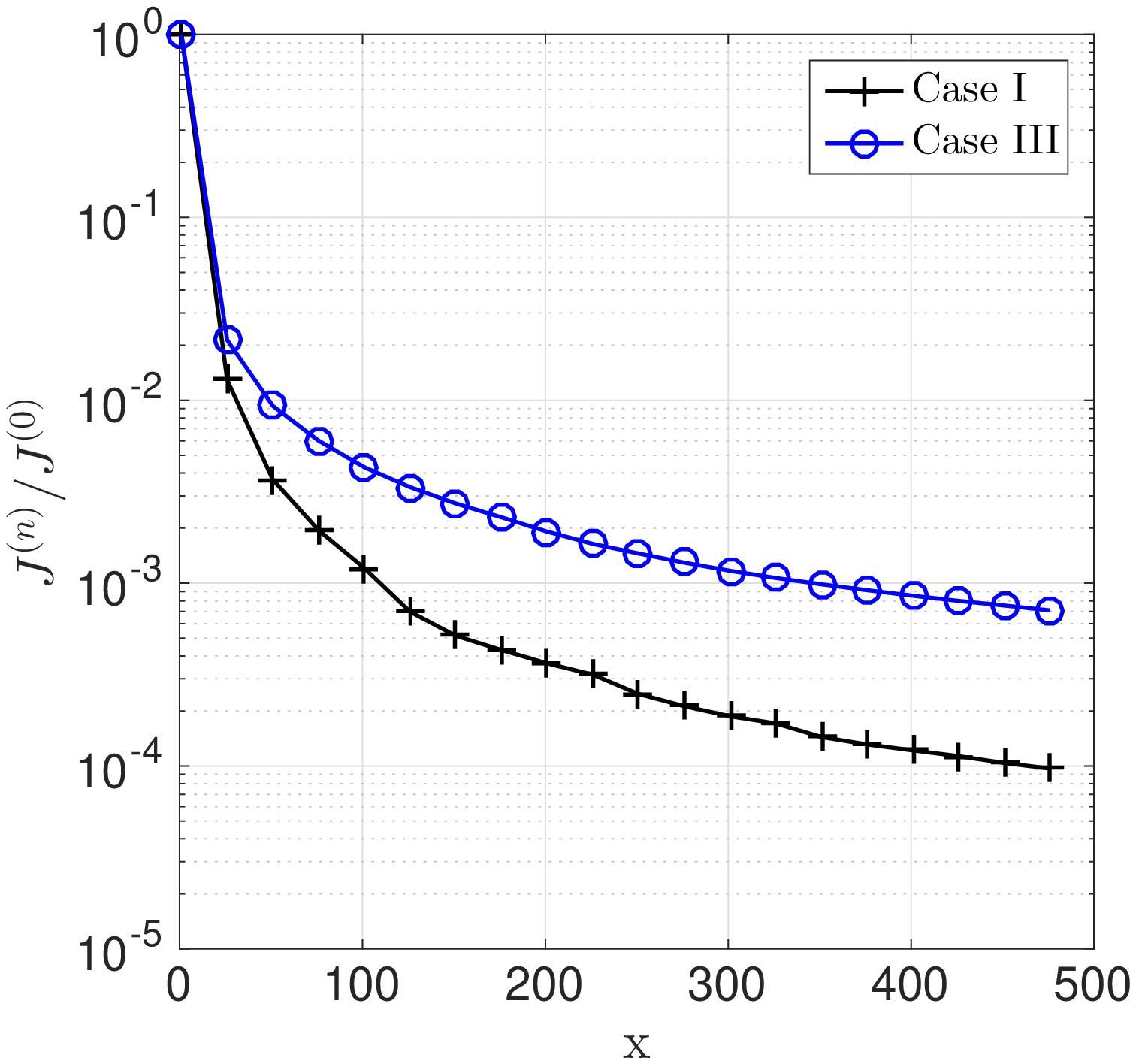}} \hspace{40pt}
\subfigure[]{\includegraphics[width=0.255\textwidth]{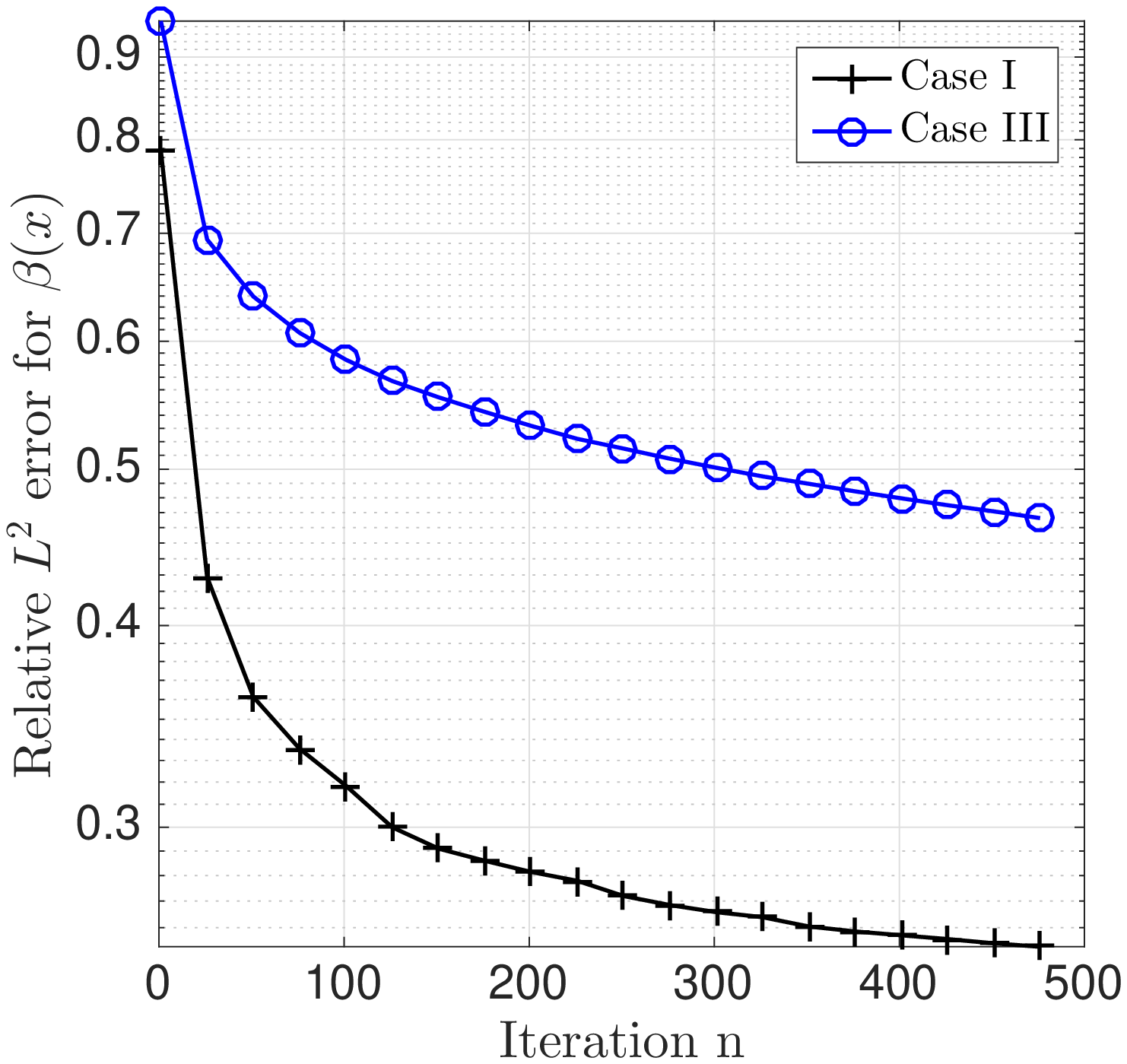}} 
\caption{Observation points starting before or after the support of bathymetry. (c) and (d) show results for the configuration given in (b) for case \rom1 and case \rom2. Note amplitudes and fluid depth are not to scale. }
\label{obs_place}
\end{figure}

With that in mind, in fig \ref{obs_place} we analyse the placement of the first observation points for case \rom1 and \rom2. All previous computations were conducted with the first point within the support of the initial condition and before the bathymetry is observed with the first sensor $y_1^{(O)}$ at $x = L/10$ . Here we consider the case where the first sensor $y_1^{(o)}$ is placed after the bathymetry has already been observed at $x = L/2$, and outside the support of the initial condition. A visual comparison of the two configurations for case \rom 1 is given in fig \ref{obs_place}(a) and (b). Fig \ref{obs_place}(c) and \ref{obs_place}(d) give the results for cases \rom1 (Gaussian $\beta$ and Gaussian $\phi$) and \rom2 (Gaussian $\beta$ and sinusoidal $\phi$). If we compare the convergence to results in fig \ref{results_filt}, we see that for both cases, the convergence is worse for $y_1^{(o)}$ at $x = L/2$ (fig \ref{obs_place}), than when $y_1^{(o)}$ is at $x = L/10$ (fig \ref{results_filt}).

From an observability perspective, this is not so surprising. We go back to the hypothesis presented in section \ref{sec4}  suggesting a lack of observability  of bathymetry at some point $x_0$ by a set of measurements taken at $\{x_j\}$, $j = 1, ..., N_{obs}$, where $x_0 + \delta <x_j$ for some distance $\delta >0$. If we place our first sensor at a point after the bathymetry is observed, then due to the non-linear effects of the bathymetry perturbation $\beta$ from $z=0$ on the evolution of the surface wave, any observability of what the bathymetry was before this perturbation is effectively lost, and the wave propagating at the sensor $y_1^{(o)}$ does not convey what the shape of the sea floor was prior to the perturbation. This is because the evolution up to that point of the surface wave is not unique; multiple forms of the propagating wavefront could result in the solution $\eta(x,t)$ at the first measurement point, because the bathymetry perturbation causes the propagating wave front to no longer retain the initial form. 

It was observed in Kevlahan et al. (2019) \cite{khan_2019} that for a flat bottom and a Guassian initial condition, the solution of the linear shallow water equations is that of the wave equation, and given some initial condition $\phi(x)$ centred at $x=0$, for $x >0$ the solution $\eta(x,t) = (1/2) (\phi(x-ct))$ is the transport of of the initial condition. Additionally for low amplitudes of $\phi(x)$, it was shown that the non-linear wave equations have a solution which is similar in form to the linear case, with slight steepening at the peaks at $t>0$. And so in the present analysis, we see that before the the bathymetry perturbation is observed, the solution $\eta(x,t)$ propagates with a form close to the initial condition, $\phi(x)$, however once the bathymetry is observed this assumption is no longer true, and subsequently observability of the bathymetry prior to the perturbation is no longer possible. 

And thus, we can stipulate that placing the first observation point within the support of $\phi(x)$ and before the support of $\beta(x)$ (the domain in which we wish to find a reconstruction) gives better convergence than placing the sensors after the bathymetry is already observed by the surface wave. In realistic settings, this could translate to placing an ``initial condition sensor'' to give the form at some simulation time $t=0$, and then clustering the remaining sensors appropriately around the bathymetry to be observed.

\subsection{Sensitivity of propagating surface waves to perturbation from flat bottom assumption \& to bathymetry reconstruction error}

Our final analysis in this study concerns the sensitivity of the surface wave to the bathymetry. We consider this in two ways; first we analyse the sensitivity of the surface wave to any perturbations in bathymetry from a flat bottom. Secondly, given some optimal reconstruction $\beta^{(b)}$, we wish to gauge the sensitivity of the propagating surface wave $\eta(x,t)$ to the errors in the reconstruction.

\begin{figure}[H]
\centering
\subfigure[Case \rom1 surface wave]{\includegraphics[width=0.25\textwidth]{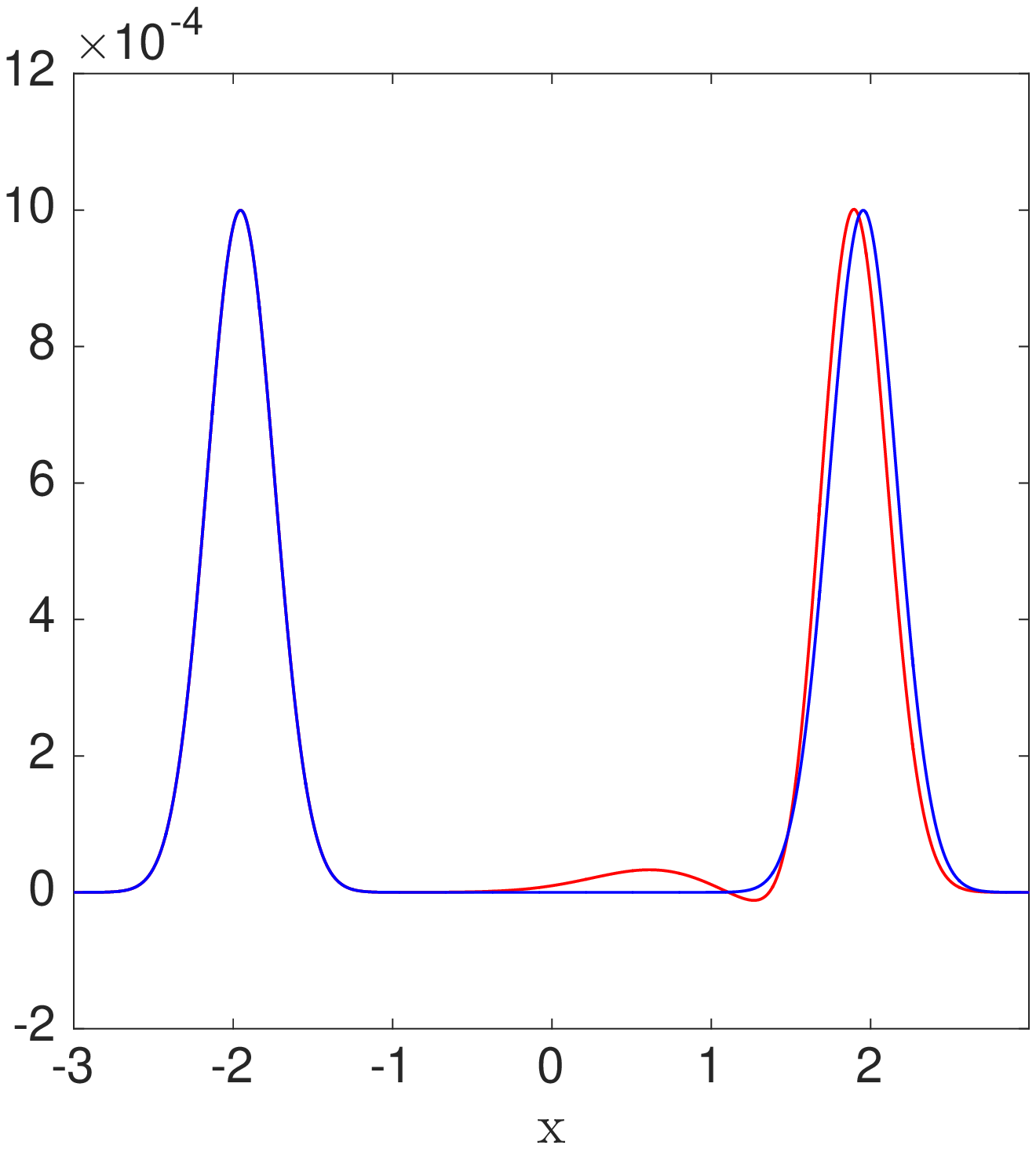}}\hspace{40pt}
\subfigure[Case \rom1 absolute error]{\includegraphics[width=0.25\textwidth]{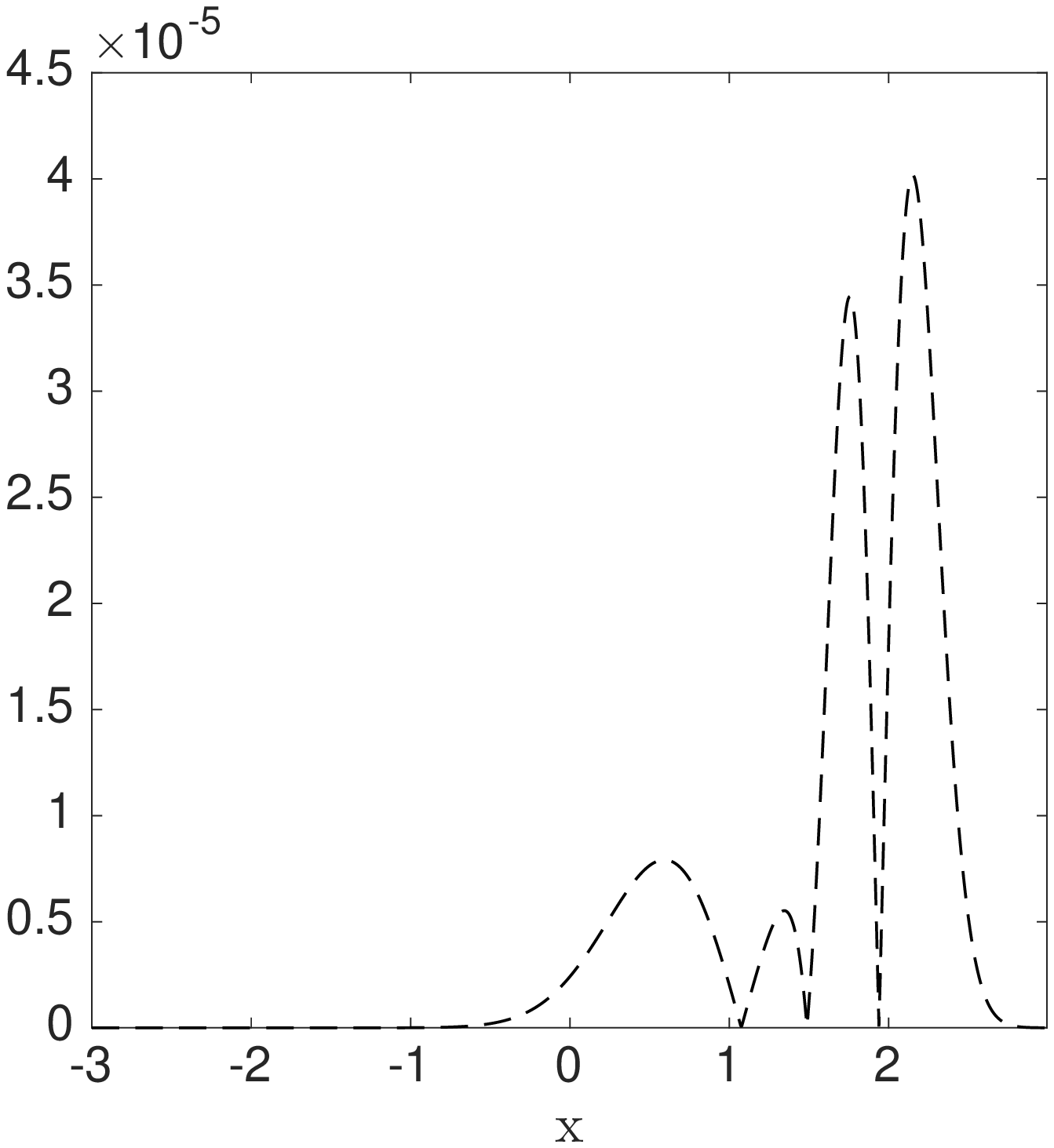}} \hspace{40pt}
\subfigure[Case \rom1 error spectrum]{\includegraphics[width=0.25\textwidth]{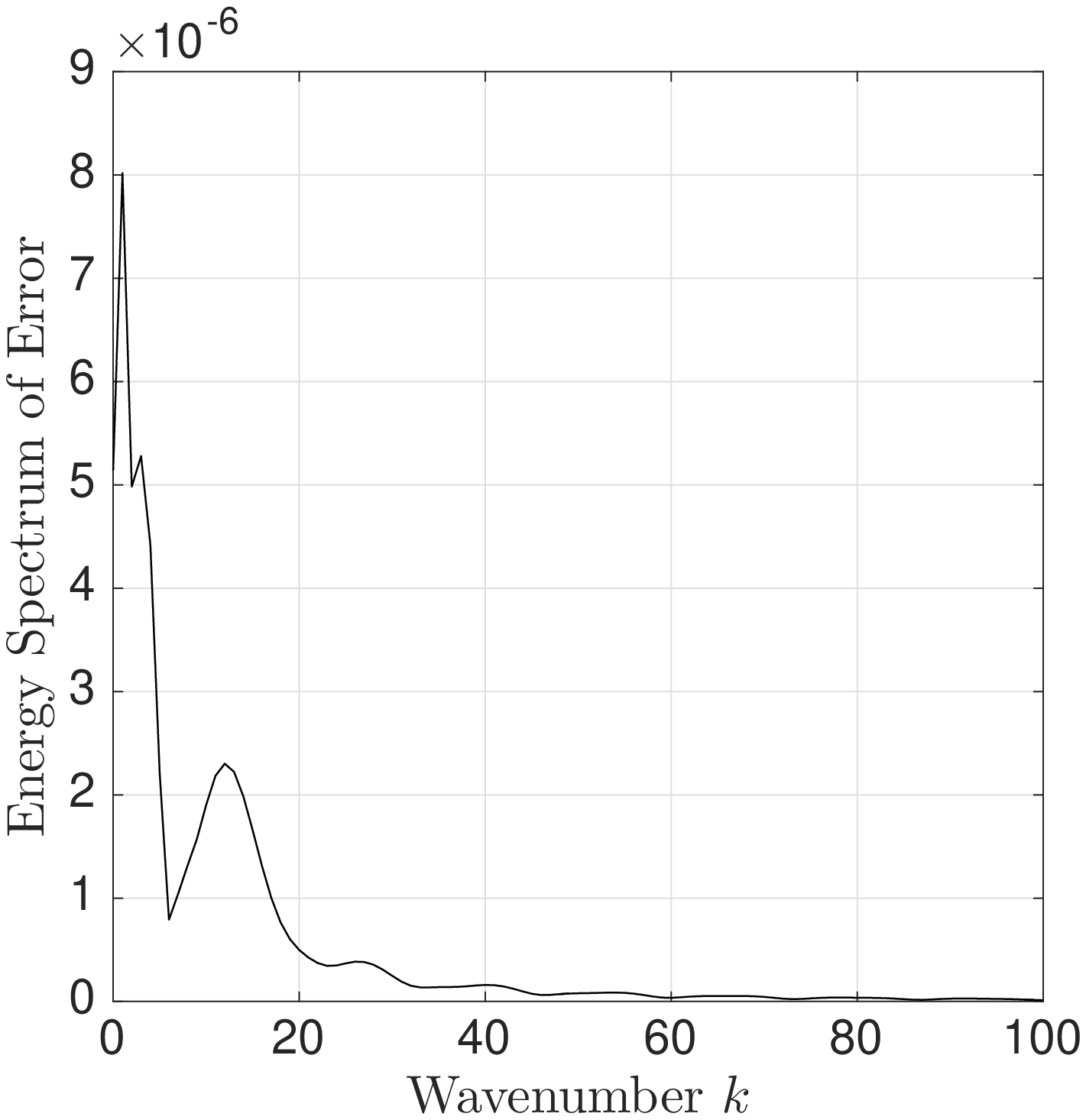}} 

\subfigure[Case \rom2 surface wave]{\includegraphics[width=0.25\textwidth]{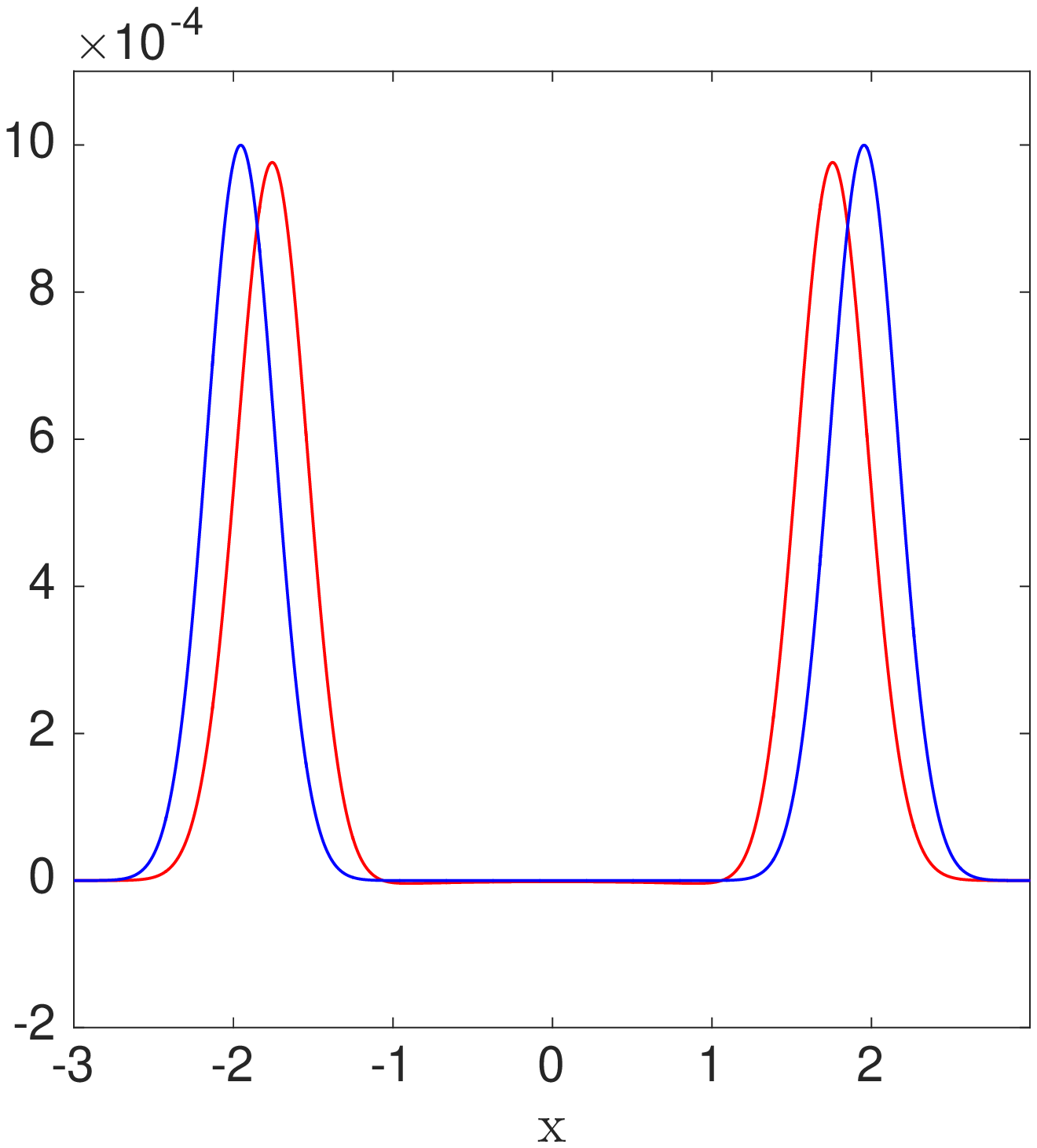}} \hspace{40pt}
\subfigure[Case \rom2 absolute error]{\includegraphics[width=0.25\textwidth]{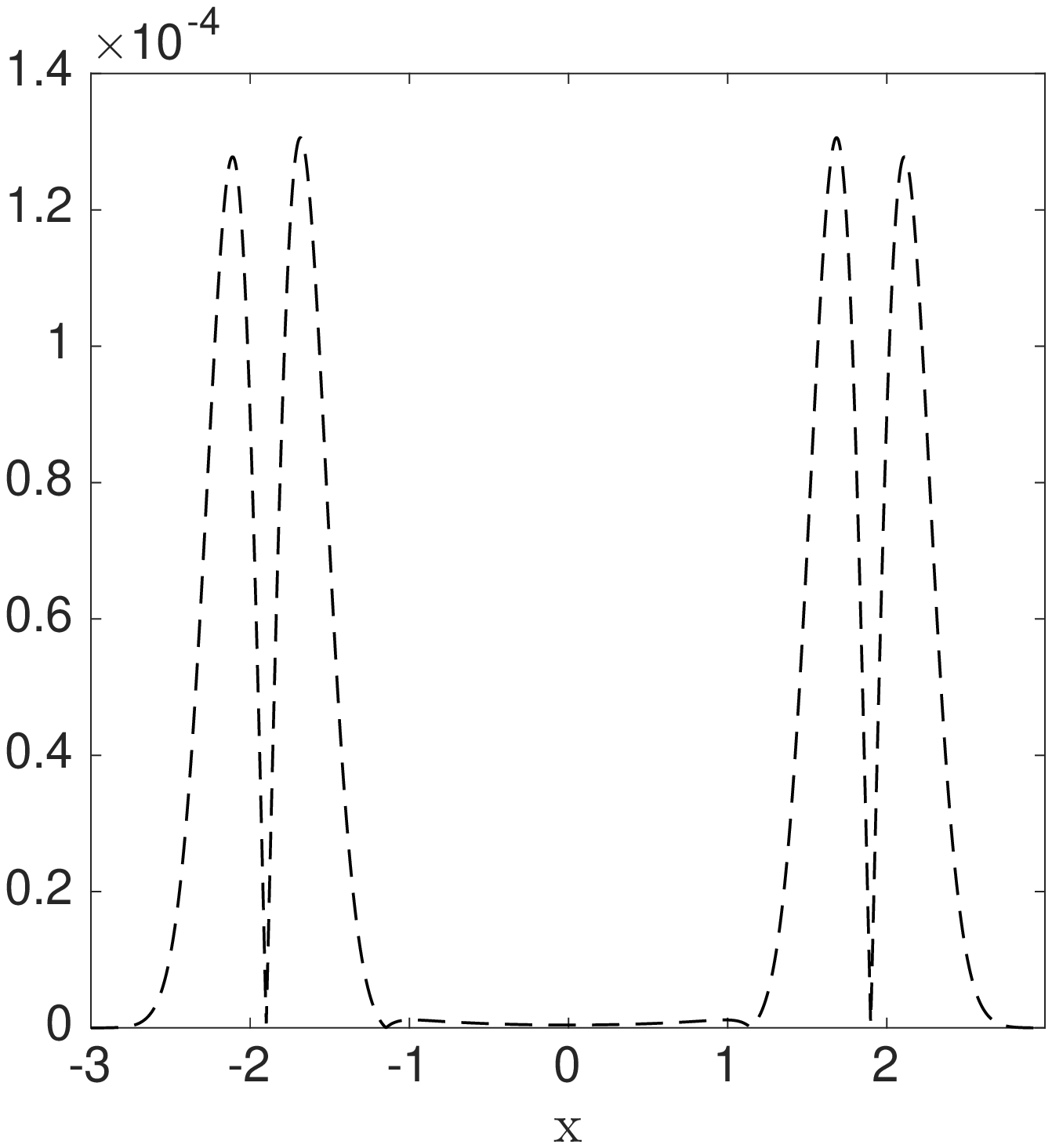}} \hspace{40pt}
\subfigure[Case \rom2 error spectrum]{\includegraphics[width=0.25\textwidth]{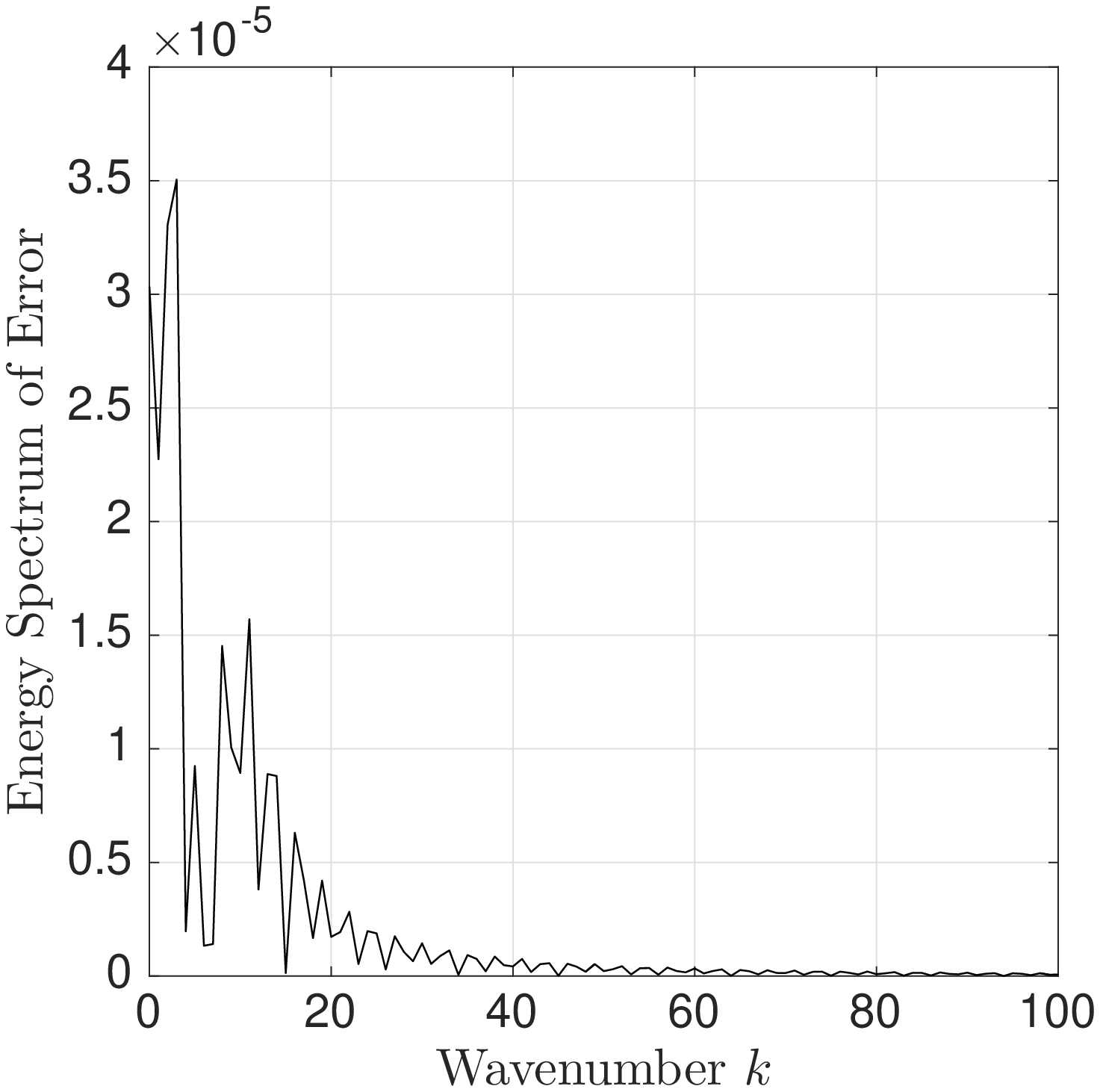}} 

\subfigure[Case \rom3 surface wave]{\includegraphics[width=0.25\textwidth]{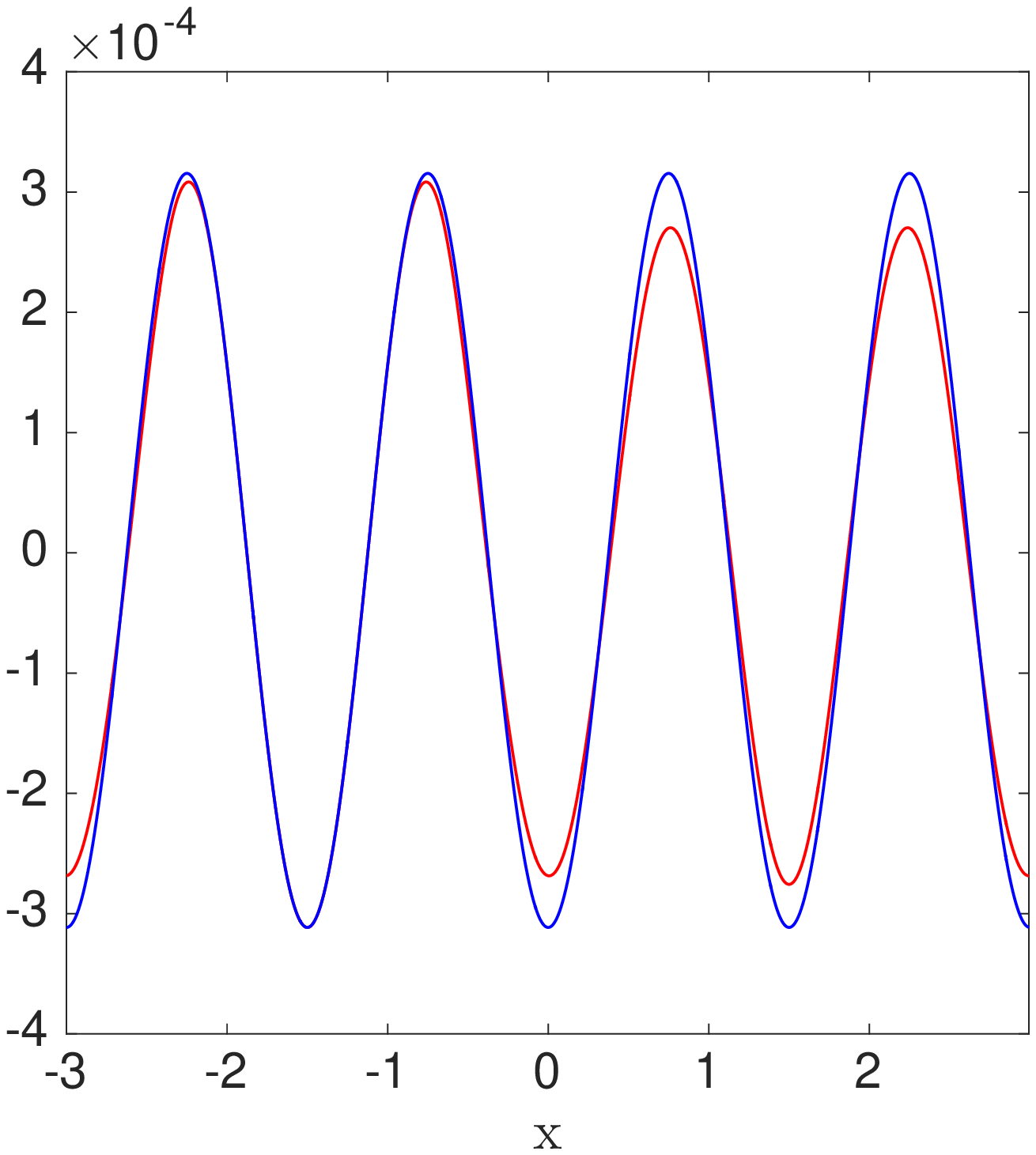}} \hspace{40pt}
\subfigure[Case \rom3 absolute error]{\includegraphics[width=0.25\textwidth]{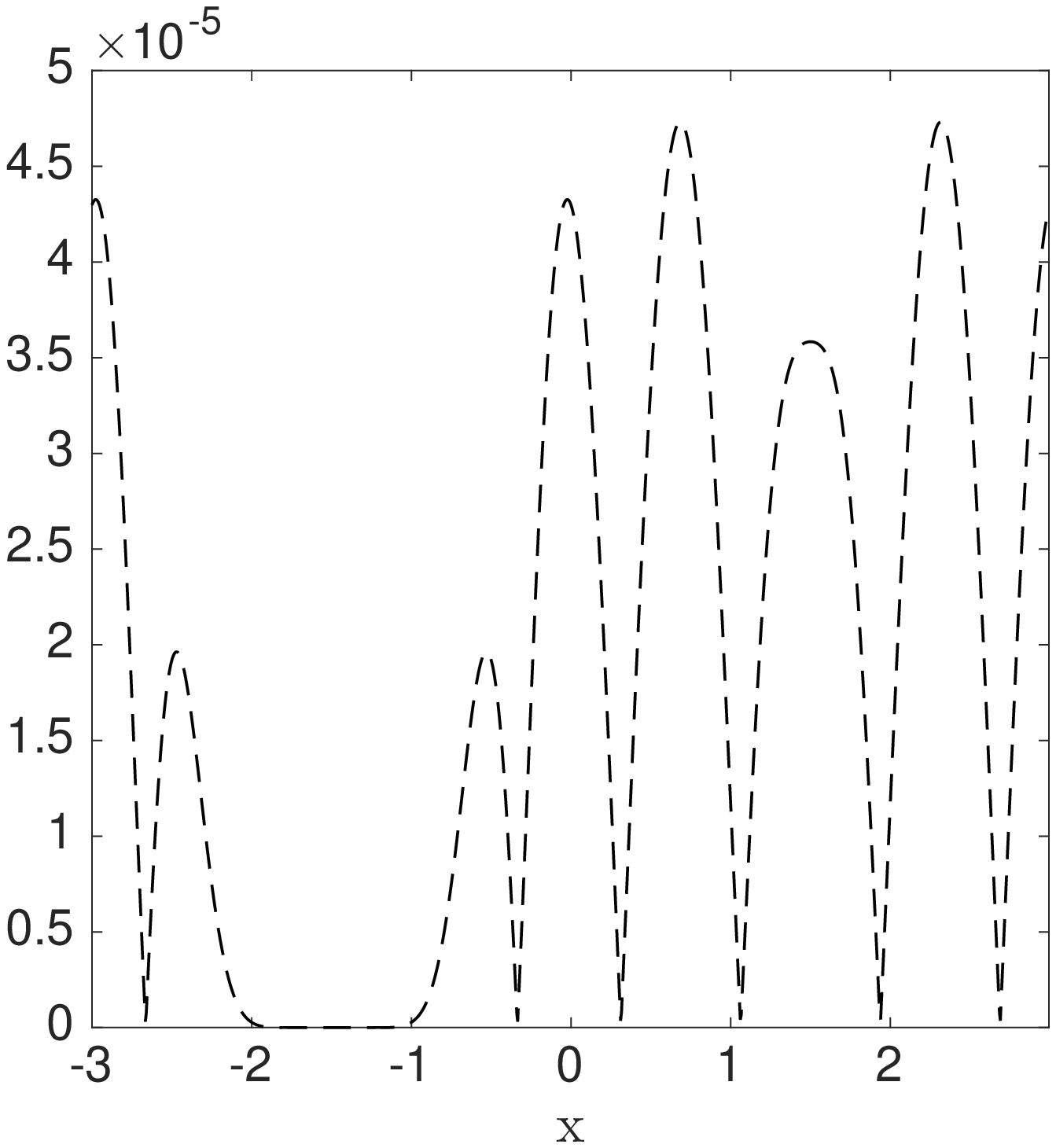}} \hspace{40pt}
\subfigure[Case \rom3 error spectrum]{\includegraphics[width=0.25\textwidth]{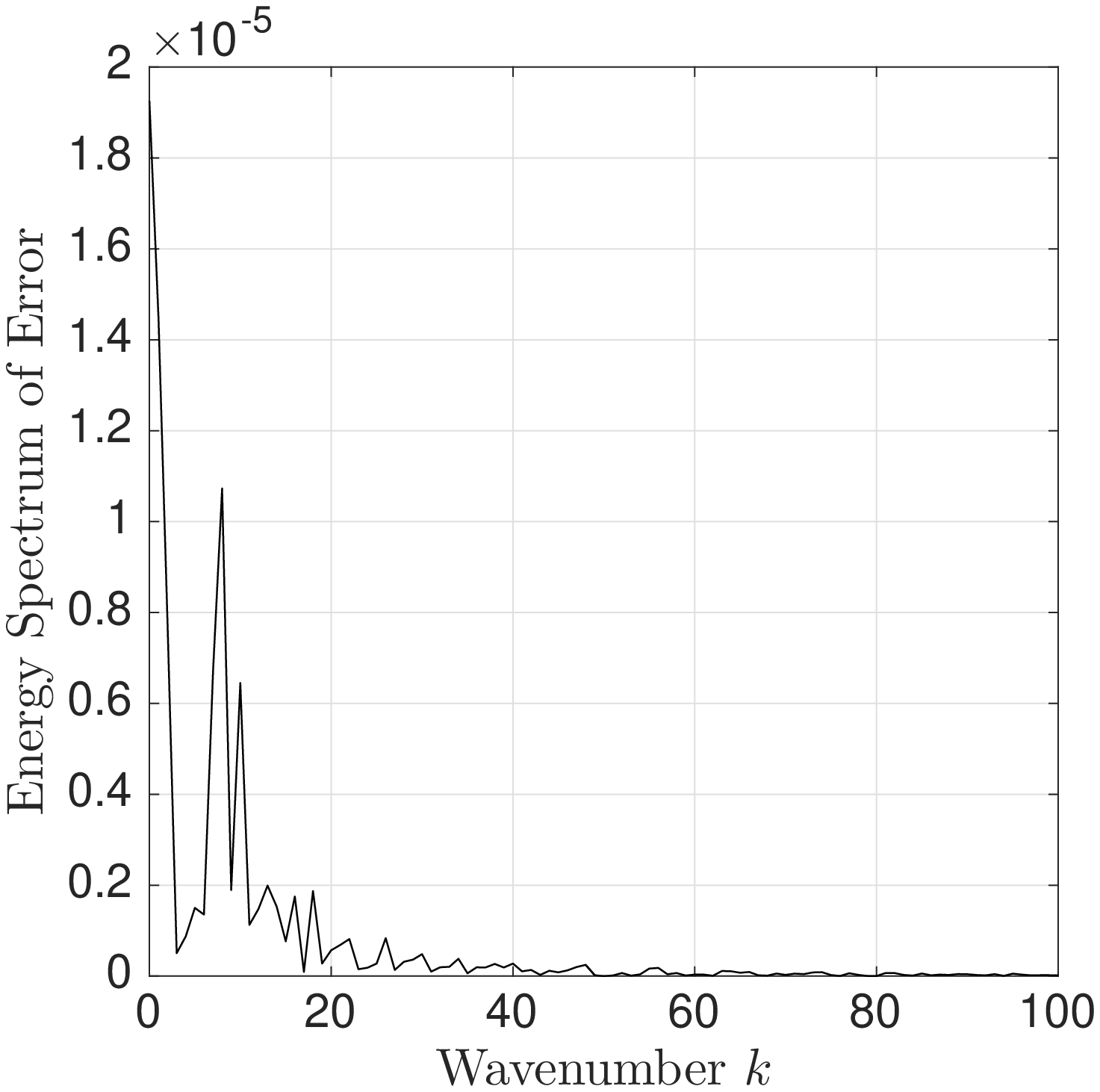}}
\caption{(a), (d) \& (g) show the propagating surface wave at time $t = 1.95$ (final time $T=2$) with flat bottom (blue) and bathymetry (red) for each case \rom1, \rom2 and \rom3 respectively. (b), (e) \& (h) shows the absolute error for each case, and (c), (f) \& (i) give the spectrum for the absolute error.}
\label{pert_fb}
\end{figure}

For the former, we solve the nonlinear shallow water equations with $\beta =0$, and with bathymetry as outlined for cases \rom1, \rom2 and \rom3. In figure \ref{pert_fb}, we compare the propagation of the surface wave with and without bathymetry, and denote the solutions $\eta_o(x,t)$ and $\eta_{\beta}(x,t)$ respectively. fig \ref{pert_fb} (a), \ref{pert_fb} (d) and \ref{pert_fb}(g) show plots of $\eta_o(x)$ (blue) and $\eta_{\beta}(x)$ (red) at the time $t= 1.95$. We can see for case \rom1, we see a slight deviation between the two solutions at the base $z \approx H$, but otherwise they remain very similar. For case \rom2, the two solutions retain a similar shape, but it is evident that the wave propagation speed $c$ has been effected by the bathymetry- as $c$ is dependent on the depth, this is to be expected, and we see that $\eta_{\beta}(x,t)$ is travelling more slowly. In case \rom3 we have a sinusoidal initial condition and Gaussian bathymetry centred at $x=1.5$. The effects of bathymetry are clear by the slightly damped amplitudes of $\eta_{\beta}(x)$. We can also see this is only for $x>0$, and so the bathymetry does not effect the left hand side propagation (the slight deviation on the left side in fig \ref{pert_fb}(g) is due to the periodic boundary effects in the numerical implementation which for a sinusoidal $\phi(x)$ can already be observed at $t=1.95$).

In general, we can see that propagation with or without bathymetry, though discernible from each other, does not significantly change qualitatively other than some slight damping, minor perturbations and small propagation speed differences in the different cases considered here. We can see this in the plots of the absolute error as a function of $x$ and energy spectrum of the error in fig \ref{pert_fb}. The highest error we see is in case \rom2, at $\ord(1)$, and this is due to the differences in $c$ for $\eta_{\beta}(x,t)$ and $\eta_o(x,t)$. 

The purpose of this analysis is to then pose the question: if the surface wave sensitivity to bathymetry perturbations from a flat bottom assumption is relatively low, then how significant is the effect of the error in bathymetry reconstruction, if our purpose is to use it to get forecast solutions for $\eta(x,t)$? After all, if our priority is to utilise these reconstructions of the bathymetry to create more accurate predictions of tsunami waves, then the significant consideration is not the error between the optimal reconstruction and the true bathymetry, but the error resulting error in $\eta(x,t)$.

\begin{figure}[H]
\centering
\subfigure[Case \rom 1]{\includegraphics[width=0.26\textwidth]{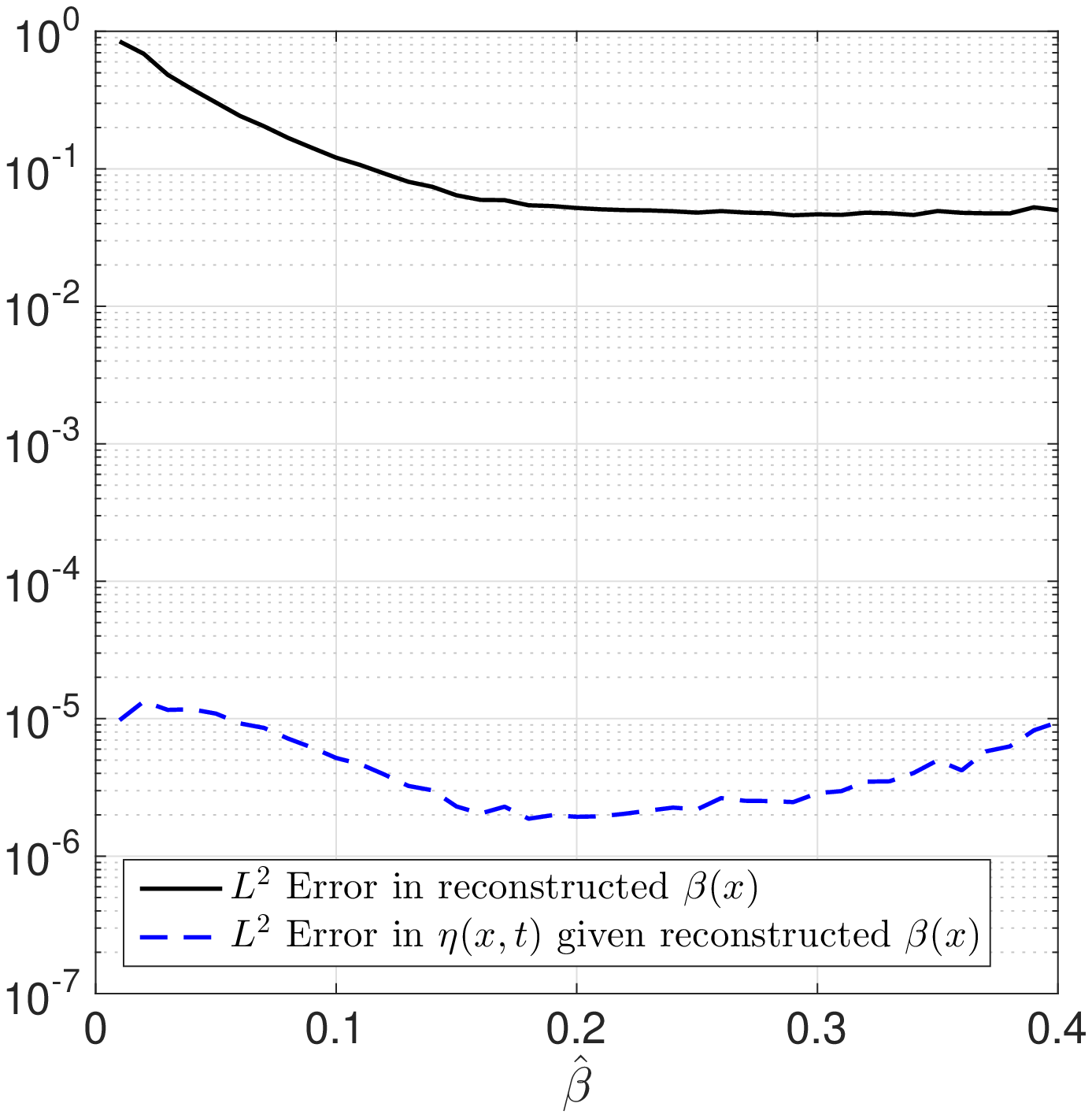}} \hspace{40pt}
\subfigure[Case \rom 2]{\includegraphics[width=0.26\textwidth]{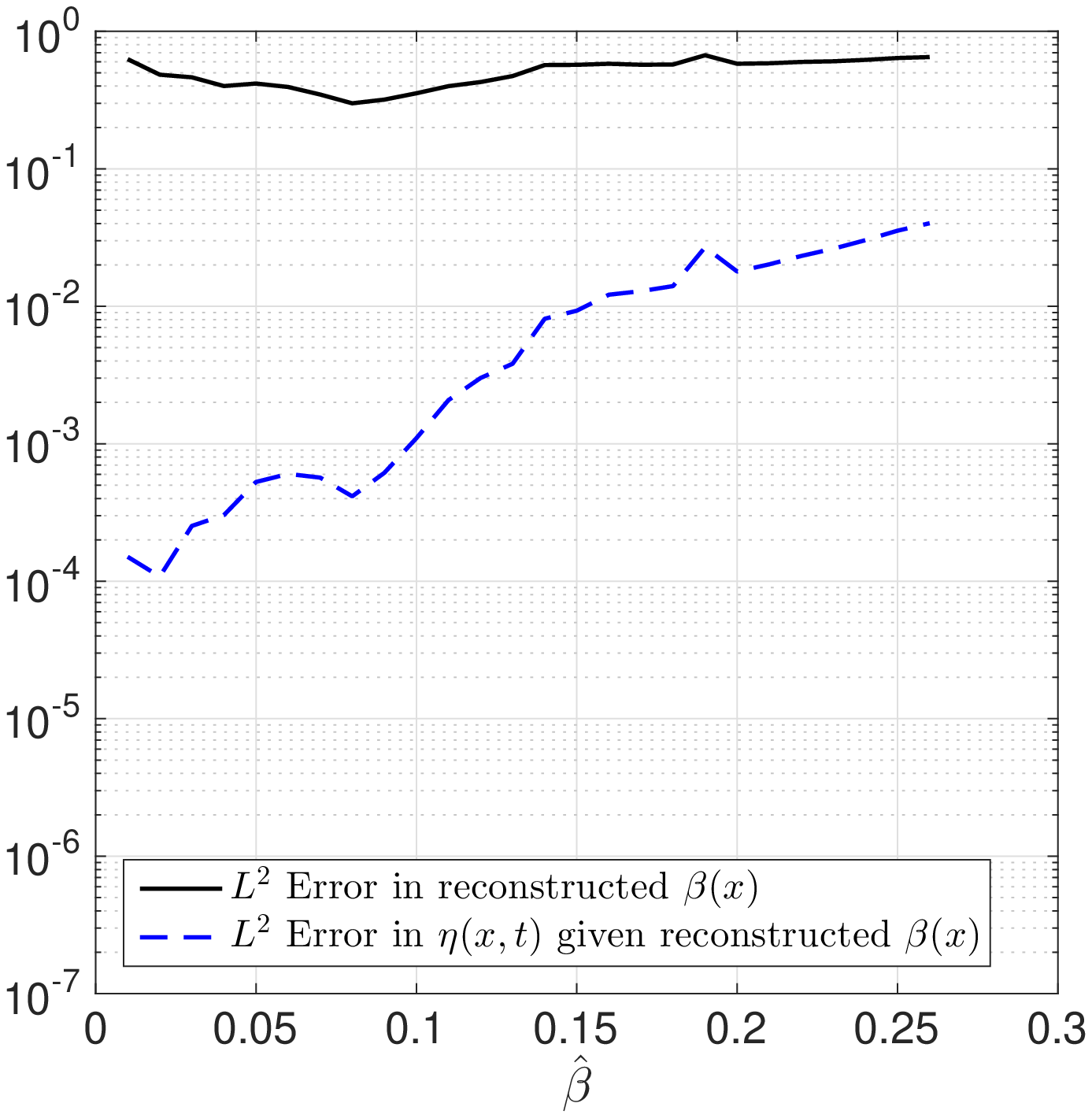}} \hspace{40pt}
\subfigure[Case \rom 3]{\includegraphics[width=0.26\textwidth]{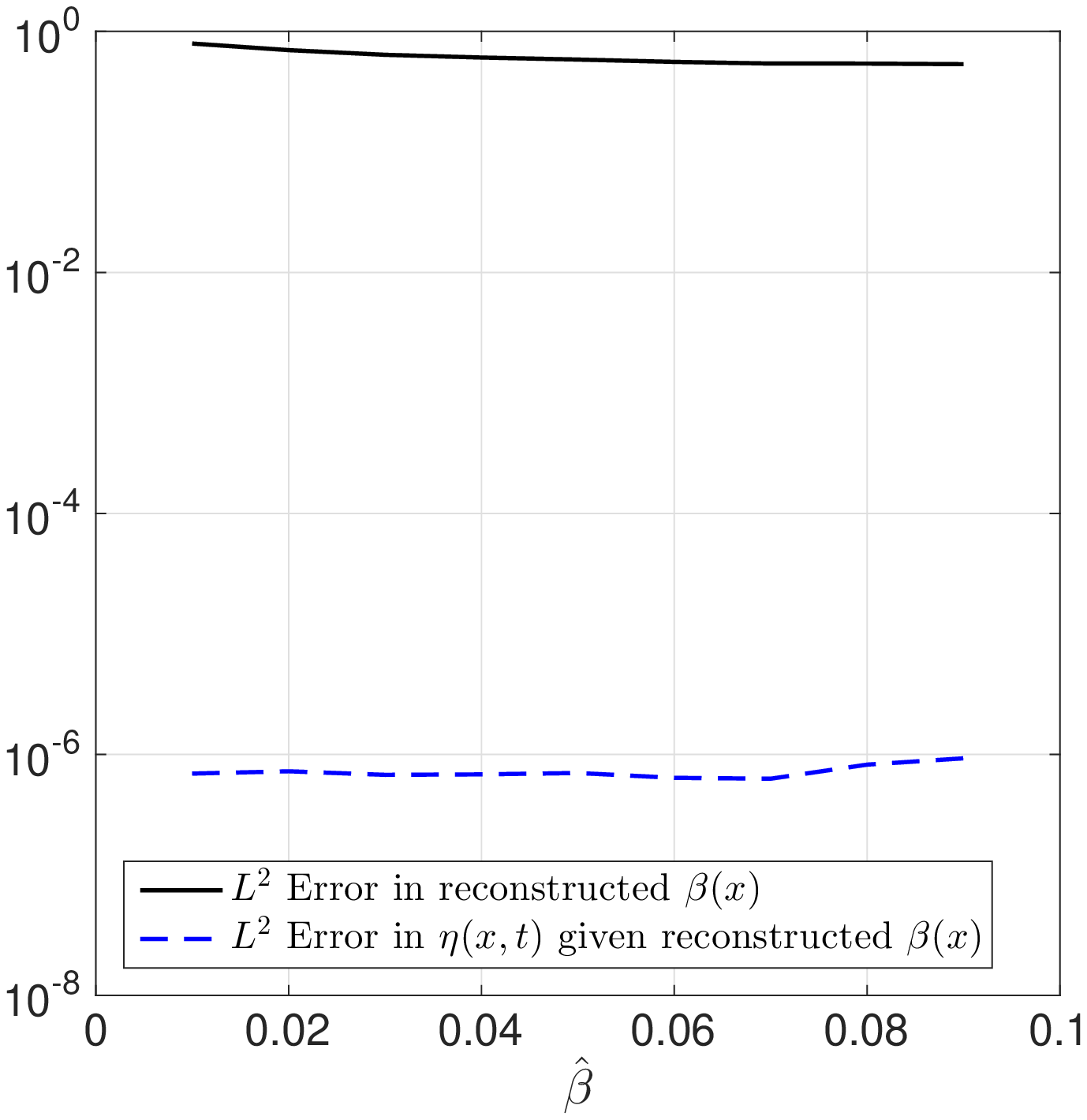}}
\caption{The $L^2$ in the bathymetry reconstruction relative to the exact bathymetry, shown for different amplitudes $\hat{\beta}$, and the resulting relative $L^2$ error in the propagating surface wave $\eta(x,t)$ given the optimally reconstructed bathymetry.Amplitude of the initial condition is fixed to be $1\%$ of $\hat{\beta}$}
\label{sens_surf}
\end{figure}

The results of this subsequent analysis are summarised in fig \ref{sens_surf}. The $L^2$ error in the bathymetry is plotted alongside the resulting $L^2$ error (in space and time) in the surface wave, as a function of bathymetry amplitude. Our intuition of low sensitivity is proven correct; In each of the three cases, the error in the surface wave is orders of magnitude lower than the error in the bathymetry estimation. In fact for case \rom1 and case \rom3 where we have a Gaussian bathymetry, the error is approximately six orders of magnitude less. In case \rom2 the error increases with the bathymetry amplitude, but if we impose the aforementioned condition of restricting the bathymetry to $10\%$ of H for accurate convergence, the error in the surface wave is still three orders of magnitude smaller than the bathymetry error.

The consequences of this low sensitivity are significant. It allows us to set less strict tolerance levels for the convergence scheme. In fact, if we wish to restrict the relative error in the surface wave to a fixed percentage, then we can derive intervals for which the bathymetry reconstruction error is permissible. additionally, we also do not require our optimal reconstruction $\beta^{(b)}$ to be unique. 

\section{Conclusion}

In conclusion, we can stipulate that this study provides a first step for better understanding the role of observations and model parameters in bathymetry assimilation. We are limited somewhat by the lack of analytical solution for the shallow water system with non-zero bathymetry, but these computational results should provide a solid platform for further analyses. 

The concepts illustrated here extend to the full three dimensional case as well, and the next step in our work is to account for the extra factors that arise, as well as additional degrees of freedom in the observation parameter.  The Sobolev gradient smoothing is a particularly powerful tool that can effectively be used to regularise noise in optimisation schemes, and it should prove a valuable tool when extending this analysis to more realistic ocean models. 

The key conclusions we reached regarding the system parameters were the necessary conditions for convergence, on the amplitudes of the initial condition and bathymetry relative to the average fluid depth. We also concluded that observation operators that provide measurements of the surface wave before the bathymetry is observed as well as after provide a better estimate of the bathymetry shape. 

Perhaps the most significant conclusion of this work is that the evolution of the surface wave $\eta(x,t)$ has low sensitivity to errors in the reconstructed bathymetry. If this conclusion can be confirmed in higher dimensions and with inclusion of additional complexities such as turbulence, Coriolis effect and multiple fluid layers of varying density, then this can greatly enhance tsunami forecast models by quantifying exact tolerance levels for the prediction, and relaxes the resolution criteria on smaller scales in our reconstruction and system. Our objective is to extend this analysis to include real bathymetry data, as found in the ETOPO2 global relief database \cite{etopo2}.  
In addition to extension to higher dimensions, our current work involves a more rigorous sensitivity analysis, and the objective is to derive analytical formulations that let us gauge the sensitivity of arbitrary response functions (such as the error in the surface wave given the reconstructed bathymetry) to perturbations in the observation operator, or to parameters in the system such as the bathymetry and initial condition amplitudes. Works such as Le Dimet et al. (1992) \cite{ld_92} provide a keen insight on the use of second order adjoint methodologies and utilising the Hessian of our cost function $\mathcal{J}$ to derive the sensitivity of such response functions. In addition, variance based analysis that allows the decomposition of the variance of a response function to individual components in the system \cite{sobol} would also be an effective tool to enhance such sensitivity quantifications.

\bibliographystyle{plain}
\bibliography{bibliography}

\begin{thebibliography}{10}

\bibitem{etopo2}
2-minute gridded global relief data (etopo2) v2, 2006.

\bibitem{situ_obs}
Luca~R. Centurioni, Jon Turton, Rick Lumpkin, Lancelot Braasch, Gary
  Brassington, Yi~Chao, Etienne Charpentier, Zhaohui Chen, Gary Corlett,
  Kathleen Dohan, Craig Donlon, Champika Gallage, Verena Hormann, Alexander
  Ignatov, Bruce Ingleby, Robert Jensen, Boris~A. Kelly-Gerreyn, Inga~M.
  Koszalka, Xiaopei Lin, Eric Lindstrom, Nikolai Maximenko, Christopher~J.
  Merchant, Peter Minnett, Anne O’Carroll, Theresa Paluszkiewicz, Paul Poli,
  Pierre-Marie Poulain, Gilles Reverdin, Xiujun Sun, Val Swail, Sidney
  Thurston, Lixin Wu, Lisan Yu, Bin Wang, and Dongxiao Zhang.
\newblock Global in situ observations of essential climate and ocean variables
  at the air–sea interface.
\newblock {\em Frontiers in Marine Science}, 6:419, 2019.

\bibitem{marsden93}
A.J. Chorin and J.~E. Marsden.
\newblock {\em A Mathematical Introduction to Fluid Mechanics}.
\newblock Springer Verlag, 1993.

\bibitem{cobelli_petitjeans_maurel_pagneux_2018}
Pablo~J. Cobelli, Philippe Petitjeans, Agnès Maurel, and Vincent Pagneux.
\newblock Determination of the bottom deformation from space- and time-resolved
  water wave measurements.
\newblock {\em Journal of Fluid Mechanics}, 835:301–326, 2018.

\bibitem{craig_sulem_99}
Walter Craig and Catherine Sulem.
\newblock Asymptotics of surface waves over random bathymetry.
\newblock {\em Quarterly of Applied Mathematics}, 68(1):91--112, 2010.

\bibitem{tsunami_ch}
Michiel Damen, Paul van Dijk, Job Duim, Harald van~der Werff, Bart Krol, Benno
  Masselink, and Frank van Ruitenbeek.
\newblock Characteristics of tsunamis.
\newblock Accessed: 2020-01-24.

\bibitem{dugan_97}
J.~P. Dugan.
\newblock Bathymetry measurements from long range airborne imaging systems.
\newblock {\em Proc. 4th Int. Conf. Remote Sensing for Marine and Coastal
  Environments. Orlando, FL: ERIM}, 1:451–457, March 1997.

\bibitem{nichols94}
A.~K. Griffith and N.~K. Nichols.
\newblock Data assimilation using optimal control theory.
\newblock {\em Numerical Analysis Report}, 10(2):101--109, 1994.

\bibitem{Grill_Sub_96}
S.T. Grilli and R.~Subramanya.
\newblock Numerical modeling of wave breaking induced by fixed or moving
  boundaries.
\newblock {\em Computational Mechanics}, 17:374--391, 1996.

\bibitem{grilli_98}
Stéphan~T. Grilli.
\newblock Depth inversion in shallow water based on nonlinear properties of
  shoaling periodic waves.
\newblock {\em Coastal Engineering}, 35:185--209, March 1998.

\bibitem{jang_park_12}
T.~S. Jang, Hong~Gun Sung, and Jinsoo Park.
\newblock A determination of an abrupt motion of the sea bottom by using
  snapshot data of water waves.
\newblock {\em Mathematical Problems in Engineering}, 2012, 2012.

\bibitem{JANG2010146}
T.S. Jang, S.L. Han, and T.~Kinoshita.
\newblock An inverse measurement of the sudden underwater movement of the
  sea-floor by using the time-history record of the water-wave elevation.
\newblock {\em Wave Motion}, 47(3):146 -- 155, 2010.

\bibitem{khan_2019}
N.K. Kevlahan, R.~Khan, and B~Protas.
\newblock On the convergence of data assimilation for the one-dimensional
  shallow water equations with sparse observations.
\newblock {\em Adv Comput Math}, 45:3195–3216, 2019.

\bibitem{losch_wunsch_03}
Martin Losch and Carl Wunsch.
\newblock Bottom topography as a control variable in an ocean model.
\newblock {\em Journal of Atmospheric and Oceanic Technology},
  20(11):1685--1696, 2003.

\bibitem{lubard_80}
S.~C. Lubard, J.~E. Krimmel, L.~R. Thebaud, D.~D. Evans, and O.~H. Shemdin.
\newblock Optical image and laser slope meter intercomparison of high-frequency
  waves.
\newblock {\em J. Geophys. Res}, 85:4996 -- 5002, 1980.

\bibitem{matharu2019optimal}
Pritpal Matharu and Bartosz Protas.
\newblock Optimal closures in a simple model for turbulent flows, 2019.

\bibitem{seabed2030}
Larry Mayer, Martin Jakobsson, Graham Allen, Boris Dorschel, Robin Falconer,
  Vicki Ferrini, Geoffroy Lamarche, Helen Snaith, and Pauline Weatherall.
\newblock The nippon foundation—gebco seabed 2030 project: The quest to see
  the world’s oceans completely mapped by 2030.
\newblock {\em Geosciences}, 8(2), 2018.

\bibitem{MINZHANG201441}
Hu~Minzhang, Li~Jiancheng, Li~Hui, and Xin Lelin.
\newblock Bathymetry predicted from vertical gravity gradient anomalies and
  ship soundings.
\newblock {\em Geodesy and Geodynamics}, 5(1):41 -- 46, 2014.

\bibitem{nakamura06}
K.~Nakamura~et al.
\newblock Sequential data assimilation: Information fusion of a numerical
  simulation and large scale observation data.
\newblock {\em Journal of Universal Computing}, 12(6):608--626, 2006.

\bibitem{Nichols_09}
D.P. Nicholls and M.~Taber.
\newblock Detection of ocean bathymetry from surface wave measurements.
\newblock {\em European Journal of Mechanics B/Fluids}, 28:224 -- 233, March
  2009.

\bibitem{ozisik}
M.N. Ozisik and H.~R.~B. Orlande.
\newblock {\em Inverse Heat Transfer: Fundamentals And Applications}.
\newblock Taylor \& Francis, 2000.

\bibitem{dugan_02}
C.~C. Piotrowski and J.~P. Dugan.
\newblock Accuracy of bathymetry and current retrievals from airborne optical
  time-series imaging of shoaling waves.
\newblock {\em IEEE Transactions on Geoscience and Remote Sensing},
  40:2606--2618, Dec 2002.

\bibitem{sobol}
I.M Sobol.
\newblock Global sensitivity indices for nonlinear mathematical models and
  their monte carlo estimates.
\newblock {\em Mathematics and Computers in Simulation}, 55(1):271 -- 280,
  2001.
\newblock The Second IMACS Seminar on Monte Carlo Methods.

\bibitem{ruuth_spiteri}
Raymond~J. Spiteri and Steven~J. Ruuth.
\newblock A new class of optimal high-order strong-stability-preserving time
  discretization methods.
\newblock {\em SIAM Journal on Numerical Analysis}, 40(2):469--491, 2002.

\bibitem{Tsai_Yu_96}
W~Tsai and D~K~P Yue.
\newblock Computation of nonlinear free-surface flows.
\newblock {\em Annual Review of Fluid Mechanics}, 28(1):249--278, 1996.

\bibitem{ld_92}
Zhi Wang, Ionel Navon, François-Xavier Le~Dimet, and X~Zou.
\newblock The second order adjoint analysis: Theory and applications.
\newblock {\em Meteorology and Atmospheric Physics}, 50:3--20, 01 1992.

\bibitem{dugan_williams_97}
J.Z. Williams and J.~Dugan.
\newblock Bathymetry measurements using electro-optical remote sensing.
\newblock {\em Proc. 4th Int. Conf. on Remote Sensing for Marine and Coastal
  Environments, Orlando, FL}, 1997.

\bibitem{wunsch_1996}
Carl Wunsch.
\newblock {\em The Ocean Circulation Inverse Problem}.
\newblock Cambridge University Press, 1996.

\end{thebibliography}
\nocite{*}

\end{document}